\documentclass[a4paper,aps,prd,superscriptaddress,nofootinbib,preprintnumbers]{revtex4}
\usepackage{amsmath,multirow}
\usepackage{graphicx,tabularx,slashed}
\usepackage{url}
\usepackage{hyperref}

\newcommand{\sulr}{\ensuremath{\tilde{u}_{L,R}}}
\newcommand{\sdlr}{\ensuremath{\tilde{d}_{L,R}}}
\newcommand{\sslr}{\ensuremath{\tilde{s}_{L,R}}}
\newcommand{\sclr}{\ensuremath{\tilde{c}_{L,R}}}
\newcommand{\sul}{\ensuremath{\tilde{u}_L}}
\newcommand{\sdl}{\ensuremath{\tilde{d}_L}}

\newcommand{\sur}{\ensuremath{\tilde{u}_R}}
\newcommand{\sdr}{\ensuremath{\tilde{d}_R}}

\newcommand{\sql}{\ensuremath{\tilde{q}_L}}
\newcommand{\sqr}{\ensuremath{\tilde{q}_R}}
\newcommand{\sq}{\ensuremath{\tilde{q}}}
\newcommand{\su}{\ensuremath{\tilde{u}}}
\newcommand{\sd}{\ensuremath{\tilde{d}}}

\newcommand{\myrbox}[1]{\parbox{4.0cm}{#1}}
\providecommand{\msqs}{m_{\tilde{q}_s}}

\newcommand\one{\leavevmode\hbox{\small1\normalsize\kern-.33em1}}

\newcommand{\qqquad}{\qquad \qquad}
\newcommand{\qqqquad}{\qquad \qquad \qquad}

\newcommand{\msbar}{\ensuremath{\overline{\text{MS}}}}

\newcommand{\go}{\tilde{g}}

\newcommand{\nz}[1]{\tilde{\chi}_{#1}^0}

\providecommand{\mgo}{m_{\tilde{g}}}

\providecommand{\msq}{m_{\tilde{q}}}

\newcommand{\tev}{{\ensuremath\rm TeV}}

\newcommand{\mg}{{\sc MadGolem\,}}

\def\slashchar#1{\setbox0=\hbox{$#1$}           % set a box for #1
   \dimen0=\wd0                                 % and get its size
   \setbox1=\hbox{/} \dimen1=\wd1               % get size of /
   \ifdim\dimen0>\dimen1                        % #1 is bigger
      \rlap{\hbox to \dimen0{\hfil/\hfil}}      % so center / in box
      #1                                        % and print #1
   \else                                        % / is bigger
      \rlap{\hbox to \dimen1{\hfil$#1$\hfil}}   % so center #1
      /                                         % and print /
   \fi}

\def\eg{{\sl e.g.} \,}
\def\ie{{\sl i.e.} \,}

\begin{document}

\date{\today}

\title{Automated Squark and Gluino Production to Next-to-Leading Order}

\author{Dorival Gon\c{c}alves-Netto}
\affiliation{Institut f\"ur Theoretische Physik, Universit\"at Heidelberg, Germany}

\author{David L\'opez-Val}
\affiliation{Institut f\"ur Theoretische Physik, Universit\"at Heidelberg, Germany}

\author{Kentarou Mawatari}
\affiliation{Theoretische Natuurkunde and IIHE/ELEM, Vrije Universiteit Brussel, Belgium \\
             and International Solvay Institutes, Brussels, Belgium}

\author{Tilman Plehn}
\affiliation{Institut f\"ur Theoretische Physik, Universit\"at Heidelberg, Germany}

\author{Ioan Wigmore}
\affiliation{SUPA, School of Physics \& Astronomy, The University of Edinburgh, UK}

\begin{abstract}
 We present completely general next-to-leading order predictions for
 squark and gluino production at the LHC, based on the fully automated
 {\sc MadGolem} tool. Without any assumptions on the mass spectrum we
 predict production rates and examine the structure of the massless
 and massive quantum corrections. This allows us to quantify theory
 uncertainties induced by the spectrum assumptions commonly
 made. Going beyond total rates we compare general fixed-order
 distributions to resummed predictions from jet merging. As part of
 this comprehensive study we present the {\sc MadGolem} treatment of
 ultraviolet, infrared and on-shell divergences.
\end{abstract}

\maketitle

\tableofcontents

\newpage

%%%%%%%%%%%%%%%%%%%%%%%%%%%%%%%%%%%%%%%%%%%%%%%%%%%%%%%%%%%%%%%%%%%%%%%%
\section{Introduction}
\label{sec:intro}

With the LHC close to completing its 8 TeV run models predicting heavy
new particles~\cite{reviews} are under intense scrutiny. Experimental
searches~\cite{atlas,cms} are probing vast parameter regions in the
minimal supersymmetric Standard Model (MSSM), most notably those parts
of the squark--gluino mass plane which can be described in terms of
gravity mediation~\cite{massplane}. Inclusive searches for the
production and decay of squarks and gluinos plays a leading role and
require an accurate as well as flexible framework for theory
predictions.  Understanding decay jets as well as QCD jet
radiation~\cite{qcd_radiation} is a crucial aspect, affecting
triggering, rate measurements~\cite{autofocus}, or kinematic
reconstruction. Advanced analysis tools like subjet
structures~\cite{subjets} increase the need for precise QCD and jets
predictions.\bigskip

Squark and gluino production to leading order was studied 30
years ago~\cite{squarkpairLO}. Next-to-leading order (NLO) QCD corrections
were first computed almost 20 years
ago~\cite{squarkpairNLO,gluinopairNLO,squarkgluinoNLO} and made public
in the {\sc Prospino} package~\cite{prospino}\footnote{All results
  shown in this paper have been checked to agree with {\sc Prospino2}
  wherever applicable.}. These calculations substantially reduce the
theoretical uncertainties to the $20-30\%$ level. More recently,
electroweak corrections~\cite{squarkEW}, resummed
predictions~\cite{squarkgluinoResummed}, and (approximate) NNLO
predictions~\cite{squarkgluinoNNLO} have been made available, further
decreasing the theoretical uncertainties. Essentially all of these
precision studies make simplifying assumptions about the squark mass
spectrum and focus on improving total cross section predictions.

In this paper we present numerical results as well as the underlying
structure of the completely automized \mg approach to NLO
predictions~\cite{madgolem_sqn,madgolem_sgluons,madgolem_talk}. It
allows us to go beyond current limitations, like assumptions on the
supersymmetric mass spectrum or the focus on total rates.  We compute
the total and differential NLO rates going through all squark and
gluino pair production channels for several benchmark parameter
points. We study in detail the structure and numerical impact of the
real and virtual QCD and SUSY-QCD effects for each of these
channels. Particular emphasis we devote to illustrating the reduction
of the theoretical uncertainties in total rates and kinematic
distributions as a key improvement of NLO predictions.  Finally, we
conduct a comprehensive comparison of the fixed-order differential
cross sections with those obtained by multi-jet matrix element
merging, including a variation of the renormalization and
factorization scales. Many details on the computation and its
numerical validation are included in the four appendices.\bigskip

This study, alongside with its earlier
squark--neutralino~\cite{madgolem_sqn} and
sgluon-pair~\cite{madgolem_sgluons} counter parts, are examples of
fully automized NLO computations in TeV-scale new physics models. This
kind of automation will significantly enhance the availability of
precision predictions for LHC observables in and beyond the Standard
Model~\cite{automation}, at a time when standard new physics scenarios
for the LHC are becoming less and less likely.  \mg is an independent,
highly modular add-on to {\sc MadGraph}~\cite{mg4,mg5}, benefiting
from its event simulation and analysis features.  It generates all
tree-level diagrams and helicity amplitudes in the {\sc MadGraph} v4.5
framework~\cite{mg4} and relies on the {\sc Helas}~\cite{helas}
library for the numerical evaluation.  The one-loop amplitudes are
generated by {\sc Qgraf}~\cite{qgraf} and {\sc
  Golem}~\cite{golem,golem_lib}. Supersymmetric counterterms and
Catani-Seymour dipoles~\cite{catani_seymour} are part of our model
implementation and can easily be adapted for other new physics models.
The subtraction of infrared and
on-shell~\cite{squarkgluinoNLO,on-shell} divergences is completely
automized.  \mg is currently undergoing final tests and will be
released to the LHC community soon.

%%%%%%%%%%%%%%%%%%%%%%%%%%%%%%%%%%%%%%%%%%%%%%%%%%%%%%%%%%%%
\section{Rates}
\label{sec:rates}

As a starting point we systematically analyze squark and gluino
production rates at next-to-leading order. We entertain all possible
production channels at the LHC involving pairs of squarks and gluinos
in the final state:
\begin{alignat}{5}
 pp \to \sq\sq, \sq\sq^*, \sq\go, \go\go \; .
 \label{eq:channels}
\end{alignat}
Following the typical decay signature we focus on the dominant first
and second generation squarks $\sq = \sulr, \sdlr, \sslr, \sclr$.  The
associated quarks we can safely assume to be massless. Moreover, we
disregard flavor-mixing, \ie the SUSY-QCD couplings are
flavor-diagonal. Further removing this latter assumption is foreseen in
the \mg setup.  In our numerical analysis we use the CTEQ6L1 and
CTEQ6M parton densities~\cite{cteq}. Unless stated otherwise, we fix
both the central renormalization and factorization scales to the
average final-state mass $\mu_R = \mu_F = \mu^0 =(m_1+m_2)/2$. From
previous studies, this choice is known to lead to perturbatively
stable results~\cite{squarkgluinoNLO}.\bigskip

As real corrections we include all channels with a three-particle
final state, in which a light parton accompanies the heavy
superpartners. The associated infrared divergences we subtract using
Catani-Seymour dipoles~\cite{catani_seymour}, generalized to include
the massive colored SUSY particles. Details on this implementation are
included in Appendix~\ref{sec:cs}.  With the help of an FKS-like cutoff
$\alpha$~\cite{alpha} we can select the phase space regions covered
by the dipole subtraction to include more ($\alpha \to 1$) or less
($\alpha \ll 1$) of the non-divergent phase space. Our default choice
is $\alpha = 1$, but the total rates must not change with varying
$\alpha$.

Virtual corrections include the one-loop exchange of virtual gluons
and gluinos.  The standard 't Hooft-Feynman gauge is employed for
internal gluons to avoid higher rank loop integrals.  Accordingly,
Fadeev-Popov ghosts appear in the gluon self-energy and in the
three-gluon vertex corrections.  Ultraviolet divergences are cancelled
by renormalizing the strong coupling constant and all
masses. Supersymmetry identifies the strong gauge coupling constant
$g_s$ and the Yukawa coupling of the quark--squark--gluino interaction,
$\hat{g}_s$.  At the one-loop level dimensional regularization induces
an explicit breaking of this symmetry via the mismatch between the 2
fermionic gluino components and the $(2-2\epsilon)$ degrees of freedom
of the transverse gluon field. We restore the underlying supersymmetry
with an appropriate finite counter
term~\cite{squarkgluinoNLO,Martin:1993yx}. Details on the
renormalization procedure can be found in
Appendix~\ref{sec:cts}.\bigskip

Finally, we have to remove potential divergences in the three-body
phase space due to intermediate resonant states. An example is the
appearance of on-shell gluinos as part of the correction to
squark--antisquark production, $qg \to \sq\go\to\sq\sq^*
q$~\cite{squarkgluinoNLO,on-shell}.  In addition to the technical
complication of a divergent rate these on-shell states introduce a
double counting once we sum all squark and gluino production rates to
next-to-leading order.  In the Standard Model a similar problem
appears in $Wt$ single top production which requires a separation from
top pair production, so our \mg implementation should benefit Standard
Model processes as well.

Following the {\sc Prospino} scheme we remove on-shell divergences
locally through a point-by-point subtraction over the entire phase
space. Off-shell pieces in the limit of vanishing particle width are
genuine parts of the NLO real emission and hence left untouched.  This
procedure preserves the gauge invariance of the entire matrix element
as well as the spin correlations between the intermediate particles
and the final state. The subtraction terms have a Breit-Wigner shape
and are automatically generated.  

Note that for an actual observable we of course need to combine the
pair production and associated production channels.  Initial-state jet
radiation at the LHC may be as hard as decay jets~\cite{qcd_radiation}
and thus cannot be distinguished on an event-by-event basis.  A
detailed account of our on-shell subtraction is provided in
Appendix~\ref{sec:os}.

%%%%%%%%%%%%%%%%%%%%%%%%%%%%%%%%%%%%%%%%%%%%%%%%%%%%%%%%%%%%%%%%%%%%%%%%
\subsection{Parameter space}
\label{sec:mssm}

%------------------------------------------------
\begin{table}[b]
\begin{center}
\begin{small}
\begin{tabular}{l|rrrrr|c}\hline\hline
 & $m_{\su_L}$ &\hspace*{2mm} $m_{\su_R}$ 
 &\hspace*{2mm} $m_{\sd_L}$ &\hspace*{2mm} $m_{\sd_R}$ 
 &\hspace*{2mm} $m_{\go}$ & mass hierarchy \\\hline  
 CMSSM 10.2.2 & 1162 & 1120 & 1165 & 1116 & 1255 & $\sq_R<\sq_L<\go$ \\
 CMSSM 40.2.2 & 1200 & 1168 & 1202 & 1165 & 1170 & $\sq_R<\go<\sq_L$ \\
 CMSSM 40.3.2 & 1299 & 1284 & 1301 & 1284 &  932 & $\go<\sq_R<\sq_L$ \\
 mGMSB 1.2    &  899 &  868 &  902 &  867 &  946 & $\sq_R<\sq_L<\go$ \\  
 mGMSB 2.1.2  &  933 &  897 &  936 &  895 &  786 & $\go<\sq_R<\sq_L$ \\ 
 mAMSB 1.3    & 1274 & 1280 & 1276 & 1289 & 1282 &
   
 \; $\su_L<\su_R<\go,\,\sd_L<\go<\sd_R$ \\ \hline\hline 
\end{tabular}
\end{small}
\end{center}
\caption{Squark and gluino masses (in GeV) for different benchmark points.}
\label{tab:sps}
\end{table}
%------------------------------------------------

%------------------------------------------------
\begin{table}[t]
\begin{small}
\begin{tabular}{l|rrcc|rrcc|rrcc|rrcc|rrcc} \hline\hline
 & \multicolumn{4}{c|}{$\su_L\su_L$} 
 & \multicolumn{4}{c|}{$\su_R\su_R$} 
 & \multicolumn{4}{c|}{$\su_L\su_R$} 
 & \multicolumn{4}{c|}{$\su\sd$} \\ 
 & $\sigma^\text{LO}$ & $\sigma^\text{NLO}$ && $K$  
 & $\sigma^\text{LO}$ & $\sigma^\text{NLO}$ && $K$ 
 & $\sigma^\text{LO}$ & $\sigma^\text{NLO}$ && $K$ 
 & $\sigma^\text{LO}$ & $\sigma^\text{NLO}$ && $K$ \\ \hline 
 CMSSM 10.2.2
  & 26.2 &  29.2 && 1.11 & 31.0 & 34.3 && 1.11 & 26.2 & 30.7 && 1.17
  & 87.7 & 104.8 && 1.19 \\ 
 CMSSM 40.2.2
  & 22.8 &  26.0 && 1.14 & 26.0 & 29.4 && 1.13 & 25.2 & 30.2 && 1.20 
  & 75.2 &  91.2 && 1.21 \\ 
 CMSSM 40.3.2
  & 14.8 & 18.1 && 1.22 & 15.8 & 19.1 && 1.21 & 23.1 & 29.9 && 1.29 
  & 49.8 & 63.6 && 1.28 \\ 
 mGMSB 1.2
  &  85.3 &  97.0 && 1.14 & 98.1 & 110.7 && 1.13 & 99.7 & 120.4 && 1.21 
  & 316.6 & 387.8 && 1.22 \\
 mGMSB 2.1.2
  &  73.9 &  88.7 && 1.20 & 87.6 & 104.5 && 1.19 & 113.9 & 144.5 && 1.27
  & 293.3 & 372.6 && 1.27 \\
 mAMSB 1.3
  & 16.8 & 18.9 && 1.13 & 16.4 & 18.4 && 1.12 & 16.1 & 19.1 && 1.19
  & 48.3 & 58.1 && 1.20 \\ \hline\hline
 & \multicolumn{4}{c|}{$\su_L\su_L^*$} 
 & \multicolumn{4}{c|}{$\su_R\su_R^*$} 
 & \multicolumn{4}{c|}{$\su_L\su_R^*,\su_R\su_L^*$} 
 & \multicolumn{4}{c|}{$\su\sd^*$} \\ 
 \hspace*{18mm} 
 & $\sigma^\text{LO}$ & $\sigma^\text{NLO}$ && $K$  
 & $\sigma^\text{LO}$ & $\sigma^\text{NLO}$ && $K$ 
 & $\sigma^\text{LO}$ & $\sigma^\text{NLO}$ && $K$ 
 & $\sigma^\text{LO}$ & $\sigma^\text{NLO}$ && $K$ \\ \hline 
 CMSSM 10.2.2
  &  3.0 &  4.6 && 1.54 & 3.8 & 5.8 && 1.53 & 4.6 & 6.0 && 1.30
  & 16.0 & 19.3 && 1.21 \\ 
 CMSSM 40.2.2
  &  2.5 &  3.8 && 1.49 & 3.0 & 4.6 && 1.53 & 3.7 & 4.9 && 1.32 
  & 13.1 & 15.8 && 1.21 \\ 
 CMSSM 40.3.2
  & 1.7 &  2.5 && 1.44 & 1.9 & 2.7 && 1.44 & 1.9 & 2.6 && 1.33
  & 7.7 &  9.3 && 1.20 \\
 mGMSB 1.2
  & 17.8 &  27.5 && 1.54 & 21.9 & 33.7 && 1.54 & 21.1 & 27.8 && 1.32 
  & 74.1 &  92.8 && 1.25 \\
 mGMSB 2.1.2
  & 16.0 & 23.0 && 1.44 & 20.2 & 29.2 && 1.45 & 17.1 & 22.5 && 1.32
  & 66.0 & 81.6 && 1.24 \\
 mAMSB 1.3
  & 1.6 &  2.4 && 1.54 & 1.5 & 2.3 && 1.53 & 2.2 & 3.0 && 1.32
  & 7.7 &  9.2 && 1.20 \\ \hline\hline
 & \multicolumn{4}{c|}{$\su_L\go$} 
 & \multicolumn{4}{c|}{$\su_R\go$} 
 & \multicolumn{4}{c|}{$\su_L^*\go$} 
 & \multicolumn{4}{c|}{$\sd\go$} 
 & \multicolumn{4}{c}{$\go\go$} \\ 
 \hspace*{18mm}   
 & $\sigma^\text{LO}$ & $\sigma^\text{NLO}$ && $K$  
 & $\sigma^\text{LO}$ & $\sigma^\text{NLO}$ && $K$ 
 & $\sigma^\text{LO}$ & $\sigma^\text{NLO}$ && $K$ 
 & $\sigma^\text{LO}$ & $\sigma^\text{NLO}$ && $K$ 
 & $\sigma^\text{LO}$ & $\sigma^\text{NLO}$ && $K$ \\ \hline 
 CMSSM 10.2.2
  & 78.7 & 108.6 && 1.38 & 87.7 & 120.3 && 1.37 & 2.3 & 3.8 && 1.63
  & 58.2 &  83.6 && 1.44 & 23.3 & 53.4 && 2.29 \\
 CMSSM 40.2.2
  & 93.5 & 131.3 && 1.40 & 101.7 & 142.3 && 1.40 & 2.8 & 4.6 && 1.65
  & 68.7 & 100.5 && 1.46 &  41.1 & 94.5 && 2.30 \\
 CMSSM 40.3.2
  & 159.4 & 239.5 && 1.50 & 165.6 & 248.2 && 1.50 & 5.2 & 9.0 && 1.73
  & 116.3 & 182.0 && 1.57 & 249.2 & 552.9 && 2.22 \\
 mGMSB 1.2
  & 467.0 & 610.6 && 1.31 & 511.4 & 665.4 && 1.30 & 18.7 & 28.3 && 1.52
  & 371.2 & 503.3 && 1.36 & 222.8 & 453.4 && 2.03 \\
 mGMSB 2.1.2
  & 777.0 & 1077.6 && 1.39 & 868.0 & 1193.9 && 1.38 & 33.6 & 52.5 && 1.56
  & 638.1 &  914.6 && 1.43 & 849.6 & 1755.0 && 2.07 \\
 mAMSB 1.3
  & 54.4 & 78.1 && 1.44 & 53.5 & 77.0 && 1.44 & 1.5 & 2.6 && 1.71
  & 36.3 & 54.5 && 1.50 & 19.0 & 46.1 && 2.42 \\ \hline\hline
\end{tabular}
\end{small}
\caption{Total cross sections (in fb) and corresponding $K$ factors
  for squark and gluino production at $\sqrt{S}=14$~TeV. The renormalization
  and factorization scales
  are chosen as the average final state mass.  The notation $\su\sd$
  indicates the summation over all possible final-state chiralities
  $\su\sd = \sul \sdl + \sul \sdr + \sur \sdl + \sur \sdr$.
  Symmetry factors $1/2$ are automatically included, when applicable.}
\label{tab:14tev}
\end{table}
%------------------------------------------------

%------------------------------------------------
\begin{table}
\begin{small}
\begin{tabular}{l|rrcc|rrcc|rrcc|rrcc} 
\hline\hline & \multicolumn{4}{c|}{$\su_L\su_L$} &
\multicolumn{4}{c|}{$\su_L\su_R$} &
\multicolumn{4}{c|}{$\su_L\su_L^*$} &
\multicolumn{4}{c}{$\su_L\su_R^*,\su_R\su_L^*$} \\ 
& $\sigma^\text{LO}$ & $\sigma^\text{NLO}$ && $K$ 
& $\sigma^\text{LO}$ & $\sigma^\text{NLO}$ && $K$ 
& $\sigma^\text{LO}$ & $\sigma^\text{NLO}$ && $K$ 
& $\sigma^\text{LO}$ & $\sigma^\text{NLO}$ && $K$ \\ \hline 
 CMSSM 10.2.2 
  & 3.6 & 3.8 && 1.06 & 2.5 & 2.8 && 1.12 & 0.09 & 0.16 && 1.68 & 0.27 & 0.36 && 1.36 \\ 
 CMSSM 40.2.2 
  & 2.9 & 3.2 && 1.09 & 2.2 & 2.6 && 1.15 & 0.07 & 0.12 && 1.63 & 0.19 &0.27 && 1.38 \\ 
 CMSSM 40.3.2 
  & 1.5 & 1.7 && 1.16 & 1.6 & 2.0 && 1.24 & 0.04 & 0.06 && 1.59 & 0.08 & 0.11 && 1.42 \\ 
 mGMSB 1.2 
  & 19.7 & 21.6 && 1.10 & 17.1 & 19.9 && 1.16 & 1.1 & 1.8 && 1.62 & 2.2 & 2.9 && 1.33 \\ 
 mGMSB 2.1.2 
  & 15.9 & 18.5 && 1.16 & 18.8 & 23.2 && 1.23 & 0.9 & 1.4 && 1.51 & 1.7 & 2.2 && 1.34 \\ 
 mAMSB 1.3 
  & 1.8 & 1.9 && 1.08 & 1.1 & 1.3 && 1.14 & 0.04 & 0.06 && 1.72 & 0.09 & 0.13 && 1.40 \\ \hline\hline
 & \multicolumn{4}{c|}{$\su_L\go$} 
 & \multicolumn{4}{c|}{$\su_L^*\go$} 
 %& \multicolumn{4}{c|}{$\su_L\su_L^*$} 
 & \multicolumn{4}{c|}{$\go\go$} \\ 
 & $\sigma^\text{LO}$ & $\sigma^\text{NLO}$ && $K$  
 & $\sigma^\text{LO}$ & $\sigma^\text{NLO}$ && $K$ 
 %& $\sigma^\text{LO}$ & $\sigma^\text{NLO}$ && $K$ 
 & $\sigma^\text{LO}$ & $\sigma^\text{NLO}$ && $K$ \\ \hline 
 CMSSM 10.2.2 
  & 3.0 & 5.2 && 1.74 & 0.04 & 0.09 && 2.25 & 0.34 & 1.19 && 3.51 \\
 CMSSM 40.2.2 
  & 3.8 & 6.7 && 1.76 & 0.05 & 0.12 && 2.26 & 0.74 & 2.57 && 3.49 \\ 
 CMSSM 40.3.2 
  & 7.8 & 14.5 && 1.85 & 0.12 & 0.28 && 2.33 & 8.68 & 26.4 && 3.04  \\ 
 mGMSB 1.2 
  & 34.9 & 53.8 && 1.54 & 0.7 & 1.3 && 1.91 & 7.29 &  20.6 && 2.83 \\
 mGMSB 2.1.2 
  & 67.3 & 108.4 && 1.61 & 1.5 & 2.9 && 1.93 & 40.85 & 112.0 && 2.74 \\ 
 mAMSB 1.3 
  & 1.8 & 3.2 && 1.84 & 0.02 & 0.05 && 2.42 & 0.25 & 0.95 && 3.81 \\ 
  \hline\hline
\end{tabular}
\end{small}
\caption{Total cross sections (in fb) and corresponding $K$ factors
  for squark and gluino production at $\sqrt{S}=8$~TeV. All scales are
  chosen at the average final state mass.}
\label{tab:8tev}
\end{table}
%------------------------------------------------

The effect of NLO corrections on LHC cross sections varies from
production channel to production channel and from one mass spectrum to
another. In particular the hierarchy between squark and gluino
masses affects the behavior of QCD corrections. In Tab.~\ref{tab:sps}
we list a set of conventional mass spectra~\cite{benchmarks} which in the following
we will use to study squark and gluino pair production in some detail.
Scenarios labelled CMSSM-\# (constrained MSSM) are derived from
GUT-scale universality conditions with squark and gluino masses
above 1~TeV. Each of them exhibits a different squark--gluino mass
hierarchy.  The benchmark points denoted as mGMSB-\# and mAMSB-\#
represent gauge-mediated and anomaly-mediated supersymmetry
breaking. While the entire spectrum including the weak gauginos is
very different compared to CMSSM-type scenarios, they are considerably
less distinctive when we limit ourselves to light-flavor squark and
gluino masses. The position of each of the benchmark points in the
squark--gluino mass plane defines a maximal mediation scale for SUSY
breaking, but this argument cannot simply be turned
around~\cite{massplane}.\bigskip

In Tabs.~\ref{tab:14tev} and \ref{tab:8tev} we document a comprehensive
numerical survey over these MSSM parameter for LHC center-of-mass
energies of $\sqrt{S} = 14$~TeV and $\sqrt{S} = 8$~TeV. These cross
sections can be used to test \mg once it is publicly available. Using
the general \mg setup it is possible to separate the squark flavor and
chirality in squark pair production and in associated squark--gluino
production.  The size of the NLO QCD effects we express through the
consistent ratio $K \equiv \sigma^\text{NLO}/\sigma^\text{LO}$, in
spite of some well-known problems with the convergence of the LO
parton densities we will notice below.

Already at the level of total cross sections we confirm a number of
well known general trends which are essentially common to all
production mechanisms. First, the significance of the QCD quantum
effects manifests itself as sizable $K$ factors spanning the range
$K\sim 1.1 - 2.4$ for 14~TeV center-of-mass energy. For 8~TeV we
observe uncomfortably large $K$ factors which have nothing to do with
real or virtual QCD corrections. Instead, they indicate poor
perturbative behavior of the CTEQ parton densities.  These suppress LO
production rates of heavy particles with $\mathcal{O}(\tev)$ masses,
while the NLO predictions are perturbatively stable. This can be
checked by comparing the CTEQ rate predictions to other parton
densities.

Second, the different color charges of squarks and gluinos as well as
their spin are clearly reflected in the production rates.
Interactions among color octets will give larger rates than color
triplets. Similarly, fermion pairs yield larger cross sections than
scalar pairs. This effect is not only observed in the LO and NLO rates
but also in the relative $K$ factor.

Third, both the total LO and NLO cross sections decrease with
increasing superpartner masses. Non-trivial effects can for example be
understood from the threshold behavior of virtual corrections and the
real emission, which may in part overcome the phase space suppression
of the NLO diagrams~\cite{squarkgluinoNLO}.  We will expand on all
these aspects later in this Section.

%%%%%%%%%%%%%%%%%%%%%%%%%%%%%%%%%%%%%%%%%%%%%%%%%%%%%%%%%%%%%%%%%%%%%%%%
\subsection{Squark pair production}

%------------------------------------------------
\begin{figure}[b]
\includegraphics[width=0.65\textwidth]{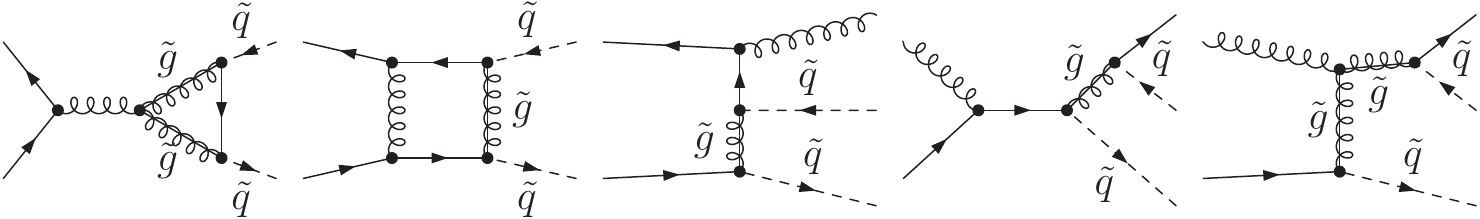} 
\caption{Sample Feynman diagrams for squark--antisquark production to
  NLO.  Virtual corrections involve the exchange of gluons, gluinos
  and squarks. Real corrections denote the emission of one quark or
  gluon.}
\label{fig:diags}
\end{figure}
%------------------------------------------------

Squark pair production can lead to a multitude of final states, which
we first classify into two basic categories:
\begin{enumerate}
\item squark--squark pairs $\sq\sq$, to leading order mediated by
  $t$-channel gluino interchange between colliding quarks. This
  mechanism is flavor-locked, so first generation squarks will
  dominate. In particular in proton-proton collisions at large
  parton-$x$ values this channel will contribute large cross sections
  because it links incoming valence quarks.
 \item squark--antisquark pairs $\sq\sq^*$ with three distinct
   sub-channels: $q\bar{q}$ annihilation through an $s$-channel gluon;
   $q\bar{q}$ scattering via a $t$-channel gluino, and $gg$ fusion
   with $s$-channel and $t$-channel diagrams.  Due to the large
   adjoint color charge and the higher spin representations involved
   the $gg$ initial-state dominates at the LHC up to moderate
   parton-$x$ values.  In the absence of flavor mixing, the
   gluino-induced sub-channel is flavor-locked to the initial state
   while the other two are flavor-locked within the final state.
   First and second generation squarks will therefore contribute with
   similar rates. All but the gluino-induced channels will also lead
   to sbottom and stop pair production~\cite{squarkpairNLO}.
\end{enumerate}

The predicted LO and NLO rates alongside their $K$ factors we document
in Tables~\ref{tab:14tev} and \ref{tab:8tev}.  The production of
squark pairs $\sq\sq$ yields cross sections of 10 to 100~fb for squark
and gluino masses around 1 TeV. The squark--antisquark rates for this
mass range are roughly one order of magnitude smaller.  These cross
sections are highly sensitive to the strongly interacting superpartner
masses. This is largely due to kinematics, \ie the different squark
masses in each benchmark point.  For instance, the production of the
lighter right-handed squarks comes with larger production rates
than that of their left-handed counterparts. According to
Tab.~\ref{tab:sps} this is true for all benchmark points except for
mAMSB~1.3.  This means that in a squark--(anti)squark sample
right-handed squarks will be overrepresented.  This can be a problem
if the NLO computation does not keep track of the different masses of
left-handed and right-handed quarks.

In contrast, we see that the $K$ factors barely change
between benchmark points, because the bulk of the NLO effects are
genuine QCD effects. However, all $K$ factors range around $K\sim
1.2$ for squark-squark production -- correspondingly, for squark-antisquark production
they render $K \sim 1.2-1.5$ depending
on the specific channel.
This effect is used by {\sc Prospino2.1}, where the different
squark masses only enter at leading order, while the NLO corrections
are computed with a universal squark mass. Some sample Feynman
diagrams we show in Fig.~\ref{fig:diags}. The supersymmetric QCD
corrections including one-loop squark and gluino loops are power
suppressed by the heavy particle masses.\bigskip

An interesting observation we make for squark pairs with different
chiralities, \eg $\su_L\su_R$. As mentioned above, all $\sq\sq$
channels proceed via a $t$-channel gluino.  For identical final-state
chiralities, the gluino propagator corresponds to a mass insertion --
enhancing the LO rates for heavy gluinos. This is not true for
$\su_L\su_R$ production, where we probe the $\slashed{p}$ term in the
gluino propagator.  This difference can be read off
Tab.~\ref{tab:14tev}. The $\sul\sul$ and $\sur\sur$
channels are suppressed from CMSSM~10.2.2 to CMSSM~40.3.2, following a
decrease in the gluino mass.  On the other hand, the $\su_L\su_R$ rate
remains quite constant. This different behavior is also visible from
their $K$ factors, which are ordered as $K_{LL} \sim K_{RR} <
K_{LR}$.\bigskip

%------------------------------------------------
\begin{figure}[t]
\begin{center}
\begin{tabular}{ccc}
\includegraphics[scale=0.45]{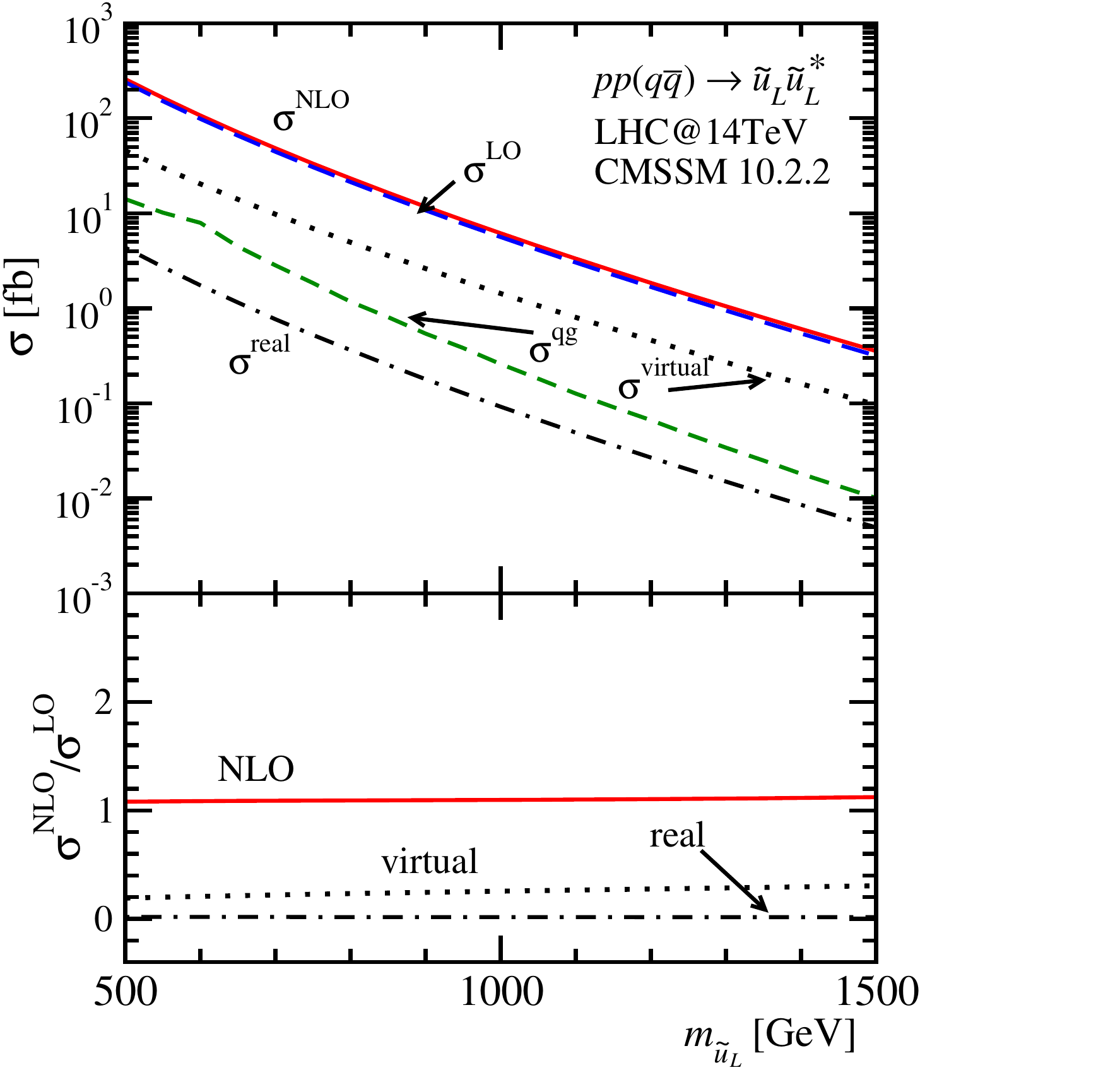} & \,&
\includegraphics[scale=0.45]{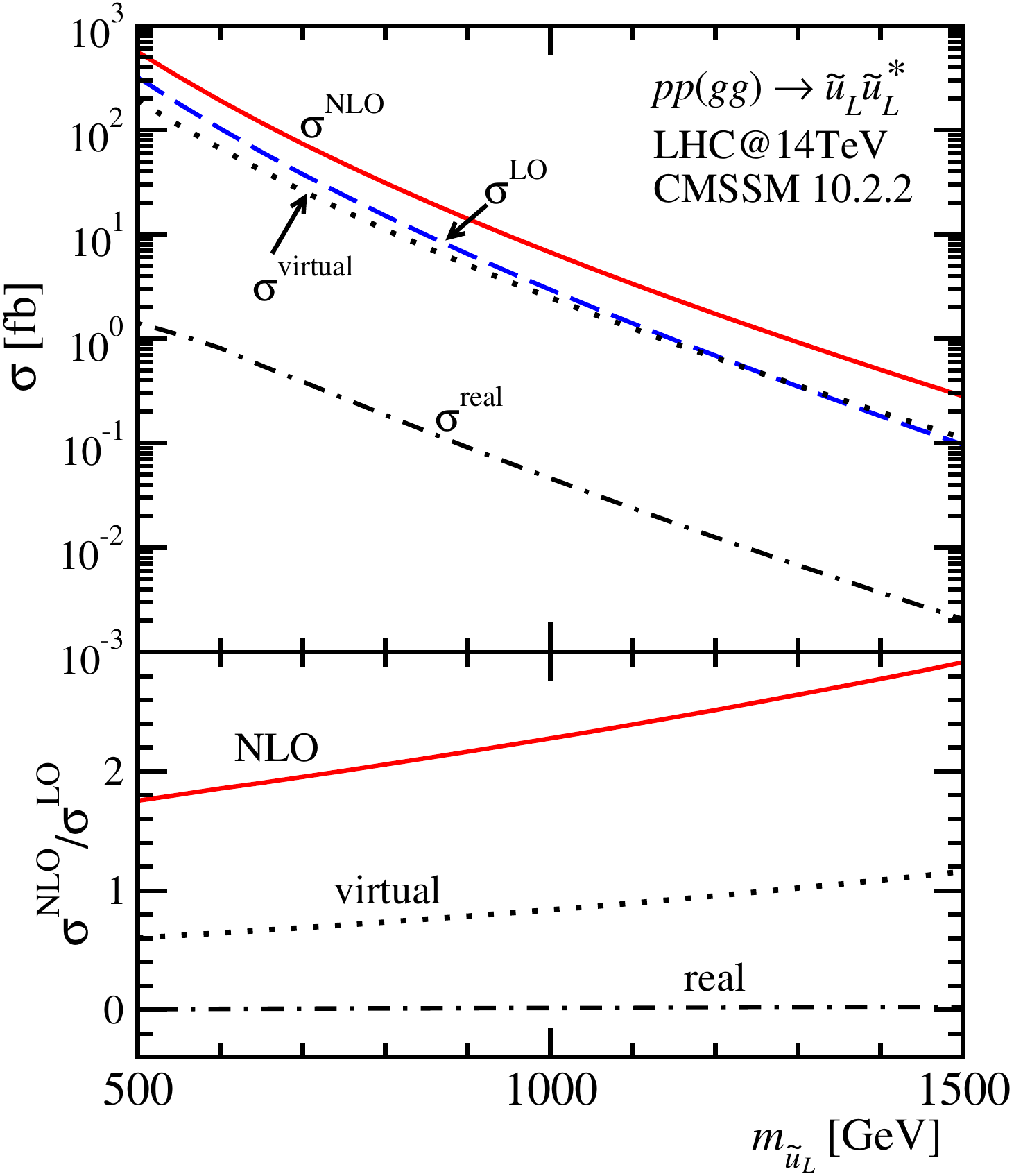}
\end{tabular}
\end{center}
\caption{Cross sections for $\sul\sul^*$ production for the different
  initial states as a function of the squark and the gluino
  masses. The $q\bar{q}$ process (left) includes also the $qg$
  crossed-channels. Together with $m_{\sul}$ we vary all squark and
  gluino masses such that the mass splittings of the CMSSM~10.2.2
  benchmark point are kept. In the lower panels we evaluate the
  relative size of the NLO cross section with respect to the total LO
  rate for each sub-channel.}
\label{fig:virtuala}
\end{figure}
%------------------------------------------------

%------------------------------------------------
\begin{figure}[b]
\begin{center}
\includegraphics[width=0.45\textwidth]{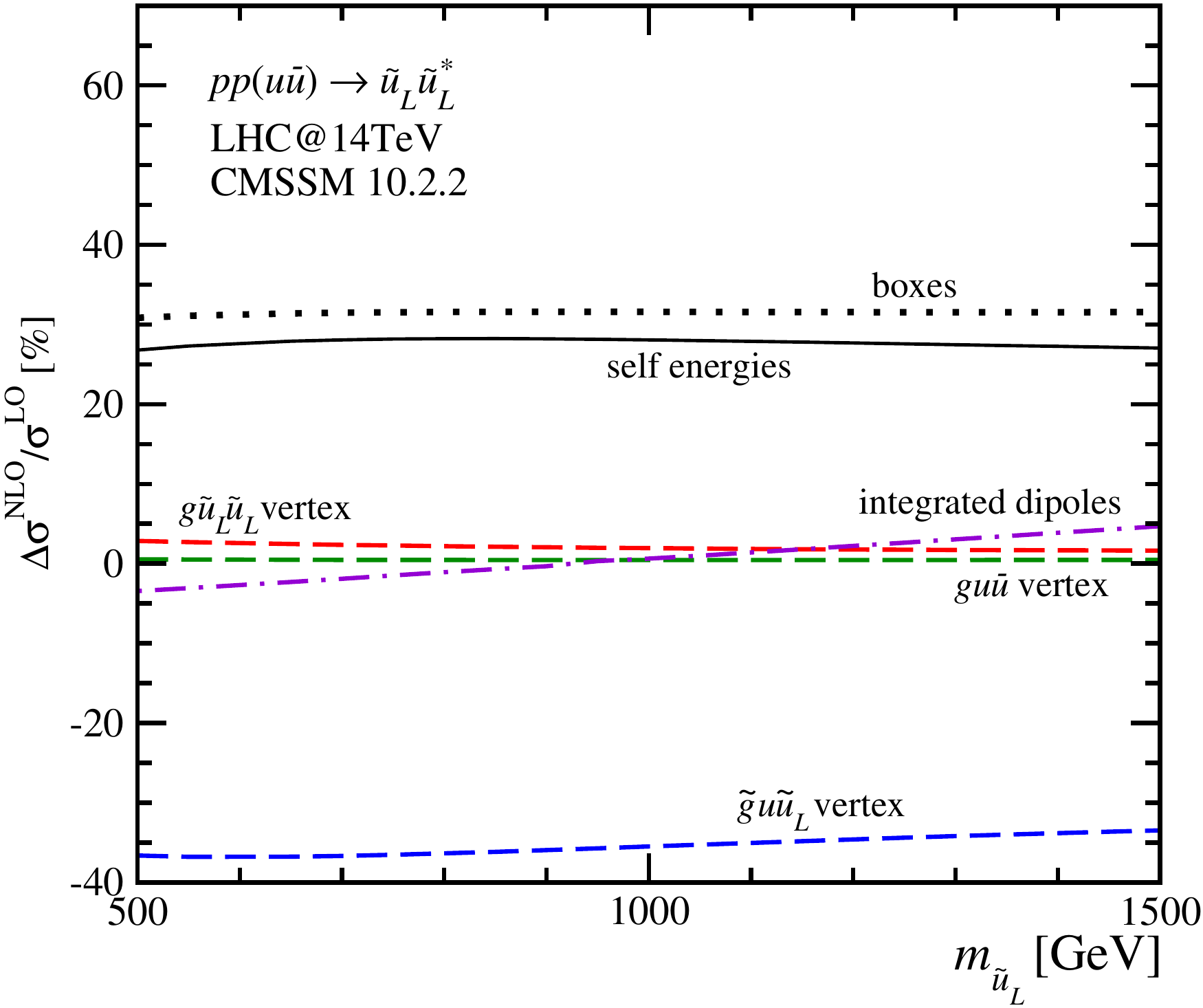}
\hspace*{0.05\textwidth}
\includegraphics[width=0.45\textwidth]{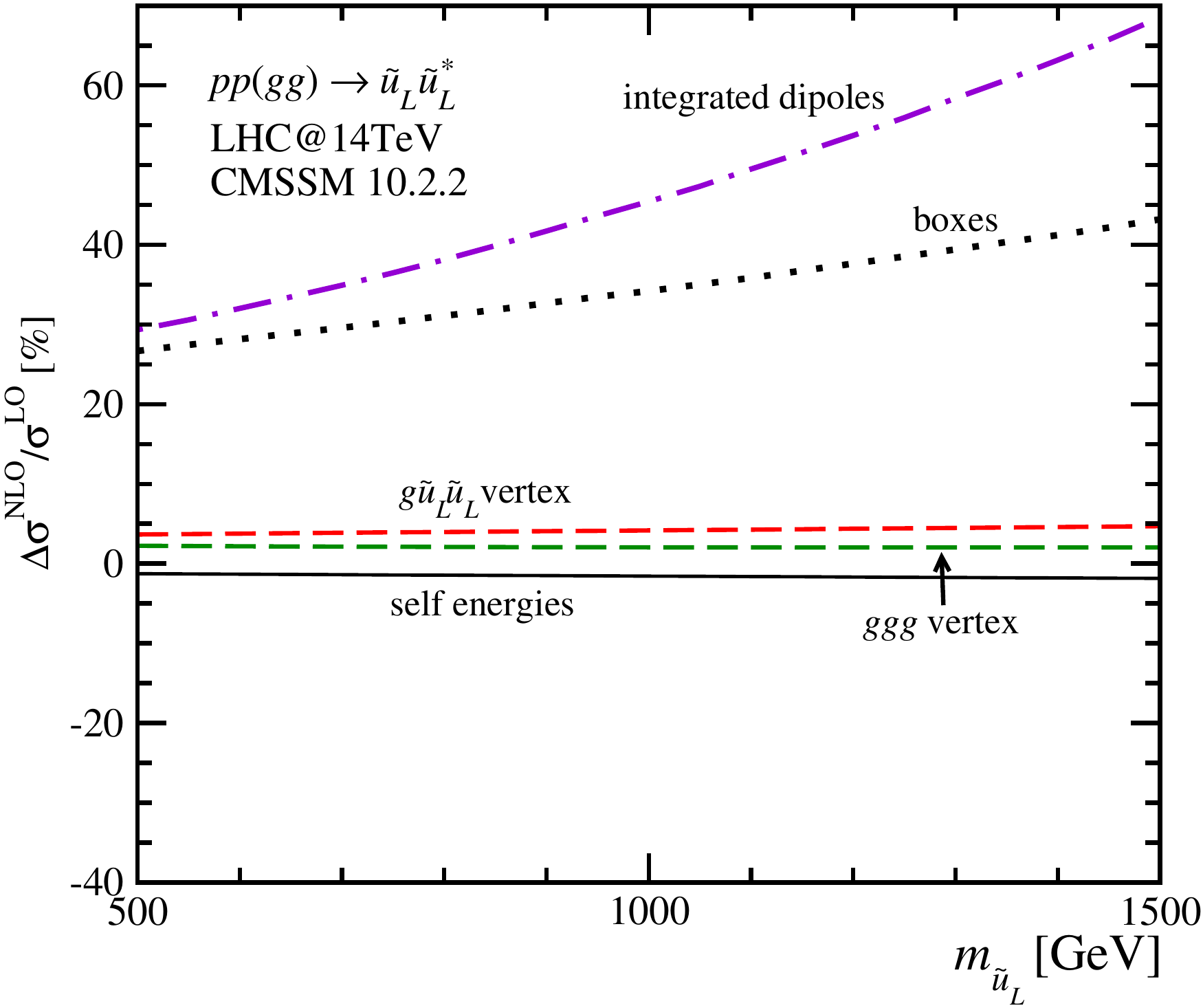}
\end{center}
\caption{Relative shift $\Delta\sigma^\text{NLO}/\sigma^{\text{LO}}$ 
 for the different parts of the
  virtual corrections to $q\bar{q}/gg \to \sul \sul^*$ production.
  All squark and gluino masses we vary in parallel, just like in
  Fig.~\ref{fig:virtuala}.}
\label{fig:virtualb}
\end{figure}
%------------------------------------------------

In Fig.~\ref{fig:virtuala} we separate the real and virtual QCD and
SUSY-QCD corrections for $\sul \sul^*$ production as a function of
the final state mass $m_{\su_L}$. All the other heavy masses we vary
simultaneously, keeping the absolute mass splittings of the
CMSSM~10.2.2 benchmark point shown in Tab.~\ref{tab:sps}.  The two
main partonic subprocesses contributing to the process we show
separately.  The separation into real and virtual corrections we
define through Catani-Seymour dipoles with a FKS-like cutoff $\alpha =
1$. The integrated dipoles count towards the virtual corrections while
only the hard gluon radiation counts towards the real
corrections. This is the reason why the real corrections appear
negligible.  The cross sections for both
the gluon fusion $gg$ and the quark-antiquark $q\bar{q}$ subprocesses
are essentially determined by the squark masses
and the corresponding phase space suppression.  The gluon fusion dominates
in the lower squark mass range, contributing with rates of roughly
a factor $2$ above the $q\bar{q}$ mechanism. Conversely, the $gg$
channel depletes slightly faster than the $q\bar{q}$, especially for large squark masses. This
can be traced back to the respective scaling behavior of the cross
sections~\cite{squarkpairNLO} as a function of the partonic energy,
and its correlation to the parton luminosities. Indeed, heavier
final-states probe larger parton-$x$ values --- this being the region
where the quark parton densities become more competitive, while the
gluon luminosity depletes.\bigskip

The lower panels of Fig.~\ref{fig:virtuala} show the relative size of
the NLO contributions with respect to the total LO rate.  While
$\sigma^\text{virtual}/\sigma^\text{LO}$ grows with
increasing squark masses, specially for the $gg$ initial state,
$\sigma^\text{real}/\sigma^\text{LO}$ stays constant.  This effect is
related to threshold enhancements: first, a long-range gluon exchange
between slowly moving squarks in the $gg \to \su\su^*$ channel gives
rise to a Coulomb singularity $\sigma \sim \pi \alpha_s/\beta$, where
$\beta$ denotes the relative squark velocity in the center-of-mass
frame, $\beta \equiv \sqrt{1-4m^2_{\su}/\hat{S}}$.  This is nothing but the
well-known Sommerfeld enhancement~\cite{sommerfeld}. The associated
threshold singularity cancels the leading $\sigma \sim \beta$
dependence from the phase space and leads to finite rates but
divergent $K$ factors~\cite{squarkgluinoNLO}. In addition, there
exists a logarithmic enhancement $\sigma \sim [A\log^2(\beta) +
  B\log(8\beta^2)]$ from initial-state soft gluon radiation. This
second effect is common to the $gg$ and $q\bar{q}$ initial
states. Threshold effects can be re-summed to improve the precision of
the cross section prediction~\cite{squarkgluinoResummed}.\bigskip

%------------------------------------------------
\begin{figure}[t]
\includegraphics[width=0.8\textwidth]{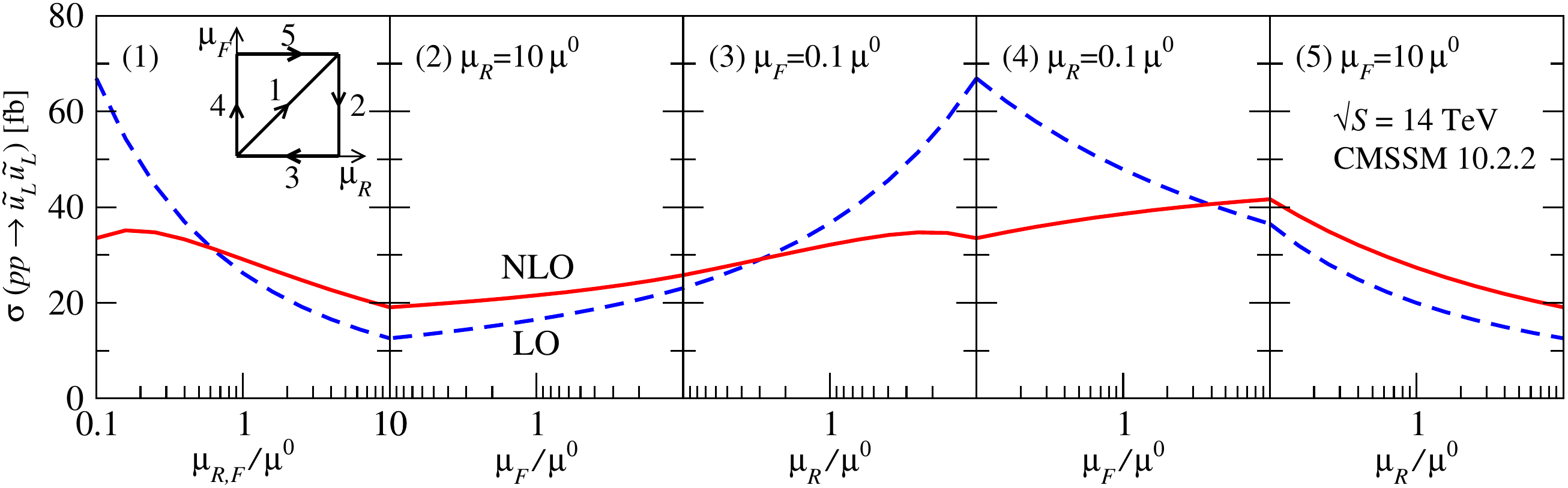}
\includegraphics[width=0.8\textwidth]{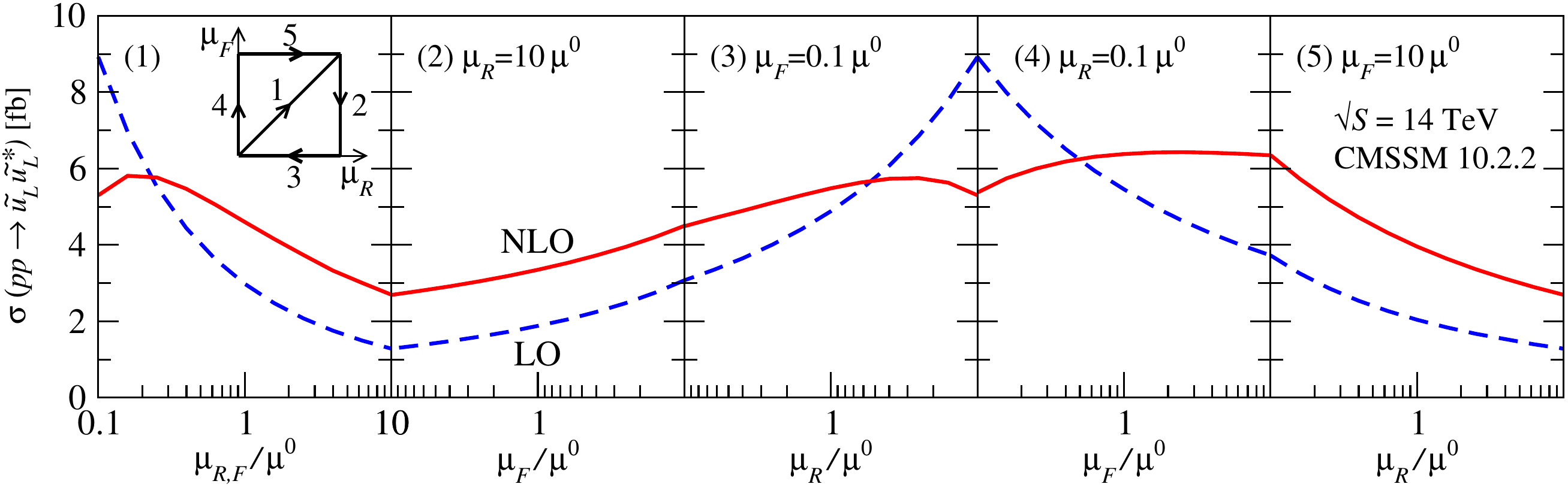}
\caption{Renormalization and factorization scale dependence for squark
  pair production $pp \to \sul\sul$ (upper) and $pp \to \sul \sul^*$
  (lower).  The plots trace a contour in the $\mu_R$-$\mu_F$ plane in
  the range $\mu = (0.1 - 10) \times \mu^0$ with $\mu^0 = m_{\sul}$.
  All MSSM parameters follow the CMSSM~10.2.2 benchmark point in
  Tab.~\ref{tab:sps}.}
\label{fig:scale-sqsq}
\end{figure}
%------------------------------------------------ 

The internal architecture of the virtual corrections we analyze in
Fig.~\ref{fig:virtualb}.  Virtual diagrams come in different one-loop
topologies: self-energy and wave-function corrections, three-point 
vertex corrections, and box corrections.  The box diagrams
also include the one-loop corrections to the quartic $gg\sq\sq$
vertex.  Again, we assume the specific flavor/chirality final state
$\su_L\su^*_L$ with the CMSSM~10.2.2 parameter point. Just
like in Fig.~\ref{fig:virtuala} the masses vary in parallel, keeping
the splitting constant.  The threshold effects discussed in the
previous paragraph are nicely visible in the increasing ratio
$\Delta\sigma^\text{NLO}/\sigma^\text{LO}$ for the boxes and the integrated dipoles,
where the quantity $\Delta\sigma^\text{NLO}/\sigma^{\rm NLO}$
accounts for the genuine $\mathcal{O}(\alpha_s)$ NLO contributions. 
This enhancement leads to sizable
quantum effects in the $30\% - 70\%$ range for the $gg$ initial state.

For the $q\bar{q}$-initiated subprocess the integrated dipoles are
numerically far smaller. The bulk of the virtual corrections is driven
by the boxes, the gluino self-energy, and the negative
quark--squark--gluino vertex correction.  Their remarkable size we can
trace back to mass insertions in the gluino-mediated
diagrams.  Barring these dominant sources, Fig.~\ref{fig:virtualb}
illustrates that all remaining NLO contributions stay at the $\sim
5\%$ level or below.  In the absence of threshold effects, all these
pieces are insensitive to the squark mass. As a consequence, both the
LO and the NLO cross sections undergo essentially the same phase space
suppression as a function of the final state mass.  Because we vary
all masses in parallel this is also indicative of the dominance of the
gluon-mediated QCD effects as compared to SUSY-QCD corrections.  In
the large-mass regime the latter have to be power suppressed, matching
on to the decoupling regime.\bigskip

%------------------------------------------------
\begin{figure}[t]
\includegraphics[width=0.40\textwidth]{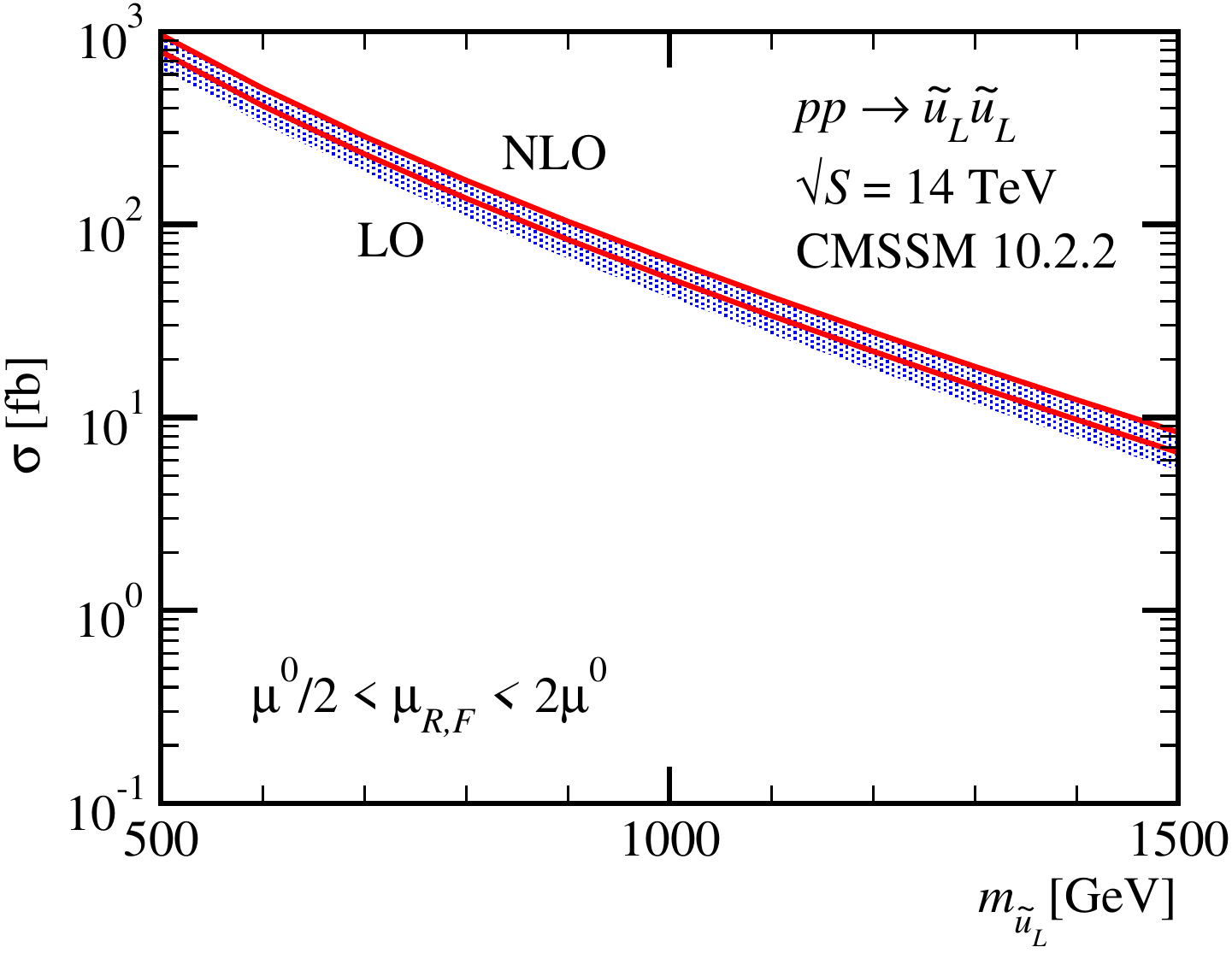}
\hspace*{0.10\textwidth}
\includegraphics[width=0.40\textwidth]{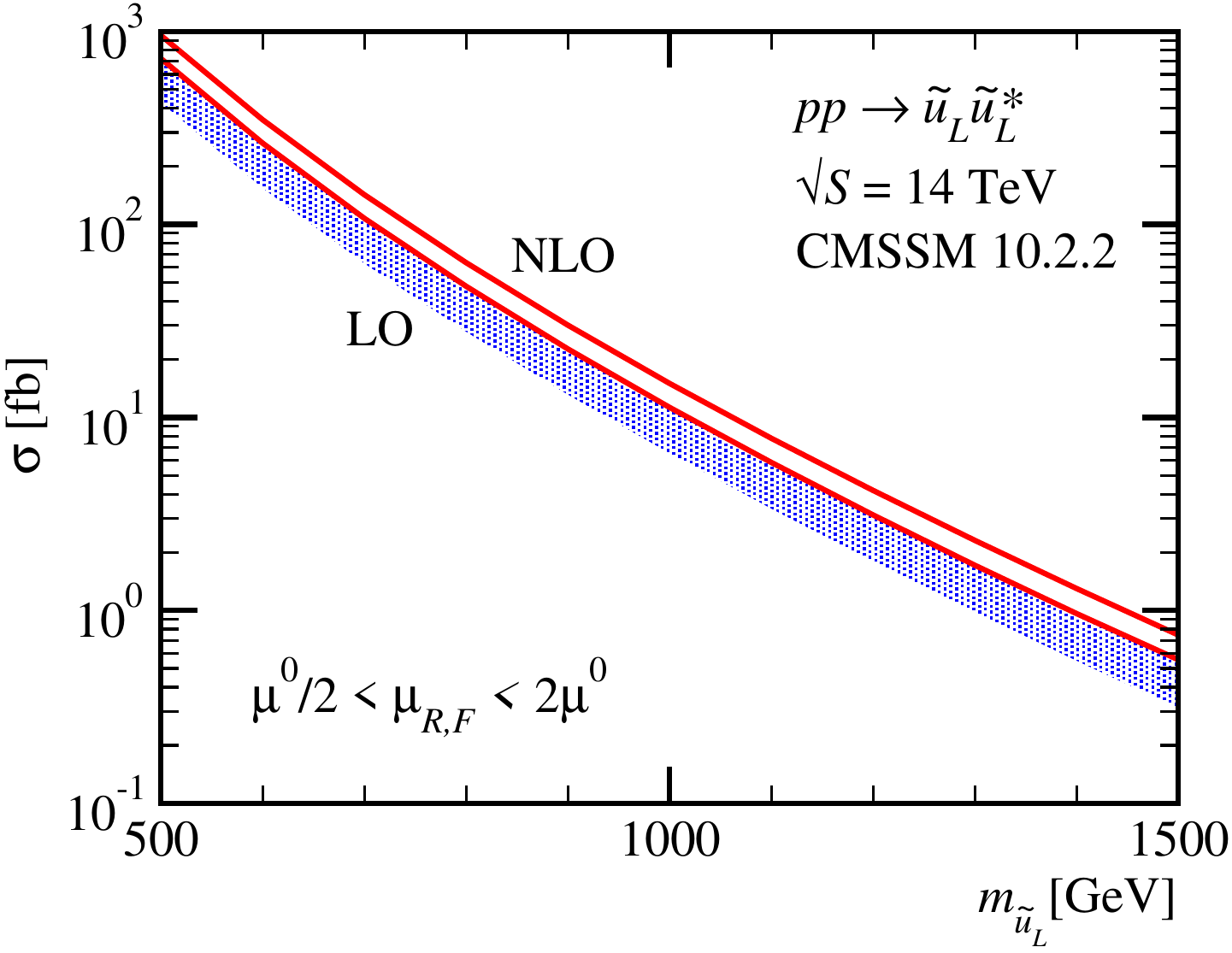}
\caption{Cross sections $\sigma(pp \to \tilde{u}_L\tilde{u}_L)$ (left)
  and $\sigma(pp \to \tilde{u}_L\tilde{u}_L^*)$ (right) as a function
  of the squark mass.  The band corresponds to the scale variation
  envelope $\mu^0/2 < \mu_{R,F} < 2\mu^0$, where $\mu^0=m_{\sul}$. 
  The central MSSM parameters are given by the CMSSM~10.2.2 benchmark
  point. The squark and gluino masses we vary in parallel, just like
  in Fig.~\ref{fig:virtuala}.}
\label{fig:overmass}
\end{figure}
%------------------------------------------------

The fact that cross section predictions increase, \ie exclusion limits
become stronger once we include NLO cross sections is only a
superficial effect of the improved QCD predictions. The main reason
for higher order calculations is the increased precision, reflected in
the renormalization and factorization scale dependence. As is well
know, these scale dependences do not have to be an accurate measure
of the theoretical uncertainty. This can be seen for example in
Drell-Yan-type processes at the LHC where the LO factorization scale
dependence hugely undershoots the known NLO corrections. For the pair
production of heavy states mediated by the strong interaction the
detailed studies of top pairs give us hope that the scale dependence
can be used as a reasonable error estimate.

In Fig.~\ref{fig:scale-sqsq} we trace the scale dependences of
squark--squark and squark--antisquark production. Note that such a
separate scale variation is not possible in {\sc Prospino}, where both
scales are identified in the analytic expressions. We profile the
behavior of $\sigma^\text{LO}(\mu)$ and $\sigma^\text{NLO}(\mu)$ for
an independent variation of the renormalization and the factorization
scales in the range $\mu^0/10 < \mu_{R,F} < 10\mu^0$. As usual, the
central scale choice is $\mu^0 = m_{\sul}$.  The path across the
$\mu_R$ -$ \mu_F$ plane we illustrate in the little square in the left
panel. The numerical results are again given for the CMSSM~10.2.2
parameter point and $\sqrt{S} = 14$~TeV. As expected, the
renormalization scale dependence dominates the leading order scale
dependence. Unlike in other cases there is no cancellation between the
renormalization and the factorization scale dependences.  The
stabilization of the scale dependence manifests itself as smoother NLO
slope. While the LO scale variation covers an $\mathcal{O}(100\%)$
band, the improved NLO uncertainty is limited to
$\mathcal{O}(30\%)$. Interestingly, the NLO plateau at small scales is
not generated by a combination of the two scale dependences, but is
visible for a variation of the renormalization scale alone at fixed
small values of the factorization scale.\bigskip

%------------------------------------------------
\begin{figure}[t]
\begin{center}
\includegraphics[width=0.49\textwidth]{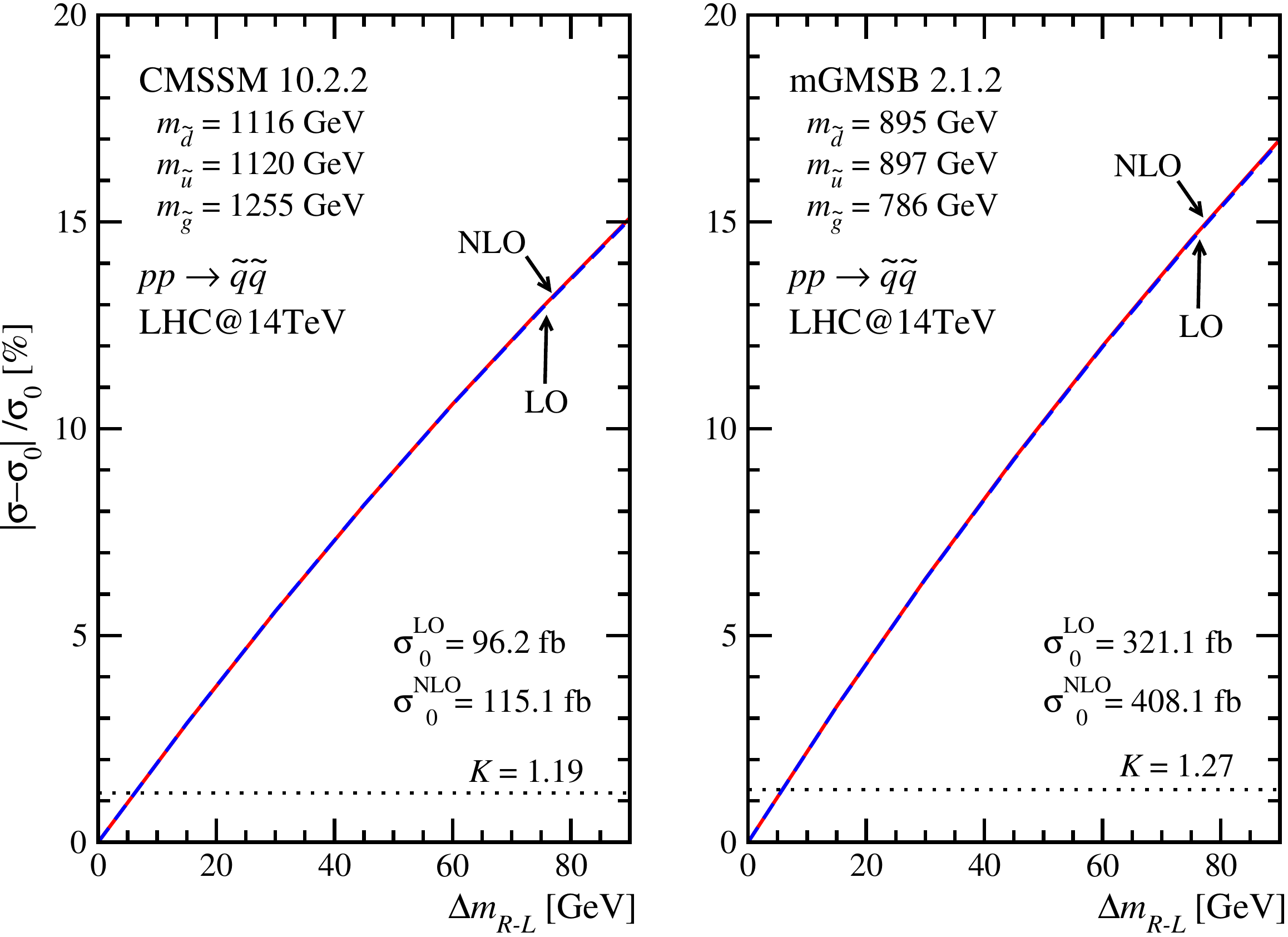}
\hspace*{0.00\textwidth}
\includegraphics[width=0.49\textwidth]{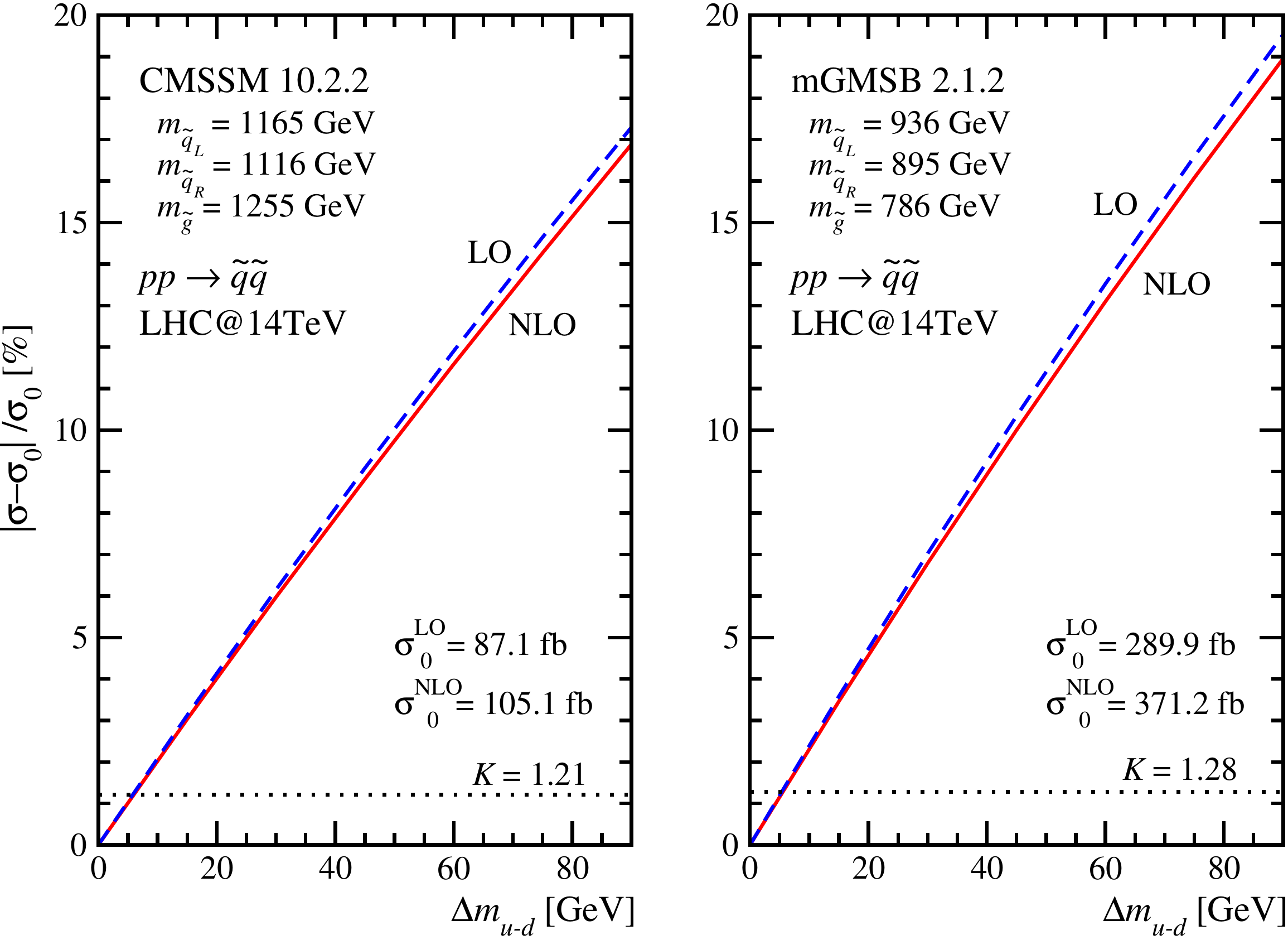}
\end{center}
\caption{Cross sections for squark pair production $pp\to \sq\sq$
  ($\sq = \sulr,\sdlr$) as a function of mass splittings. In the left
  panels we vary the right-left splitting keeping the flavor splitting
  constant.  In the right panels we vary the $\su$-$\sd$ flavor
  splitting fixing the right-left splitting. We show the shift with
  respect to the degenerate spectrum with the masses and the total
  rates $\sigma_0 \equiv \sigma(\Delta m = 0)$ given in each
  panel.}
\label{fig:prospino}
\end{figure}
%------------------------------------------------

In Fig.~\ref{fig:overmass} we show the usual LO and NLO cross
sections as a function of the final-state mass $m_{\sul}$. The error
bar around the central values represents a simultaneous scale
variation $[\mu^0/2, 2\mu^0]$. Both error bands nicely overlap and 
reflect, for $\tilde{u}_L\tilde{u}_L$, a reduction of the theoretical uncertainties
from
$\mathcal{O}(50\%)$ at LO
down to $\mathcal{O}(20\%)$ at NLO -- similarly, 
from $\mathcal{O}(60\%)$ down to $\mathcal{O}(30\%)$ for $\tilde{u}_L\tilde{u}_L^*$.\bigskip

The most significant upgrade of the \mg automated framework compared
to previous calculations is that we do not have to assume any
simplifying relations between the supersymmetric masses.  We can
freely sweep over the entire parameter space of a given model, varying
each input parameters independently. This differs from {\sc Prospino}
or other precision tools which rely on a single mass scale for all
light-flavor squarks for all next-to-leading order effects. A fully
general scan as shown in Tabs.~\ref{tab:14tev}-\ref{tab:8tev}, is thus
beyond the reach of these tools. 

Figure~\ref{fig:prospino} shows quantitative results for this
generalized NLO computation. As an example we focus on the (partially
inclusive) production of all first-generation squark pairs $pp \to
\sq\sq$, with $\sq = \sul,\sur,\sdl,\sdr$ and examine an independent
variation of the different squark masses.  In the left two panels we
study the effect of a right-left mass separation while identifying
s-up and s-charm masses as well as s-down and s-strange masses to
the CMSSM~10.2.2 and mGMSB~2.1.2 values shown in
Tab.~\ref{tab:sps}. We show the change in the total squark pair cross
sections with a growing mass splitting $\Delta m_{R-L} \equiv m_{\sqr}
- m_{\sql}$, where $\sq = \su, \sd, \tilde{c}, \tilde{s}$.  The
results we evaluate in terms of $|\sigma-\sigma_0|/\sigma_0$, where
$\sigma_0$ denotes the cross section for $\Delta m_{R-L}=0$. The
associated $K$ factor is displayed at the bottom of each panel. In the
right panels we show the same analysis for an up-versus-down squark mass
splitting $\Delta m_{u-d} \equiv m_{\su}-m_{\sd}$, with $\su =
\su,\tilde{c}$ and $\sd = \sd,\tilde{s}$. The masses of their
respective chiral components are separated as in the corresponding
benchmark scenarios.

The results highlight, first of all, that a fully flexible mass
spectrum leaves a measurable footprint in the total cross
sections. The rates change by $\mathcal{O}(5-20\%)$ for a squark mass
splitting of $10-100$ GeV, as commonly featured by MSSM benchmark
points. These effects lie roughly in the same ball park as higher
order corrections beyond the fixed next-to-leading order predictions.
At the same time, the $K$ factors stay essentially constant with a
varying mass splitting. This follows from the fact that while
kinematic effects change the cross sections significantly, the NLO
corrections are mostly sensitive to mass splittings through SUSY-QCD
effects which are typically mass suppressed. For squark pair
production, in particular, the NLO rate dependence on $\Delta m_{R-L}$
and on $\Delta m_{u-d}$ is affected by the gluino self-energy
corrections. Our \mg results confirm that taking into account the full
mass spectrum in the LO rate predictions and re-weighting them with a
$K$ factor computed most efficiently for a mass degenerate spectrum
gives an accurate estimate of the full NLO rates.

%------------------------------------------------
\begin{figure}[b]
\includegraphics[width=0.38\textwidth]{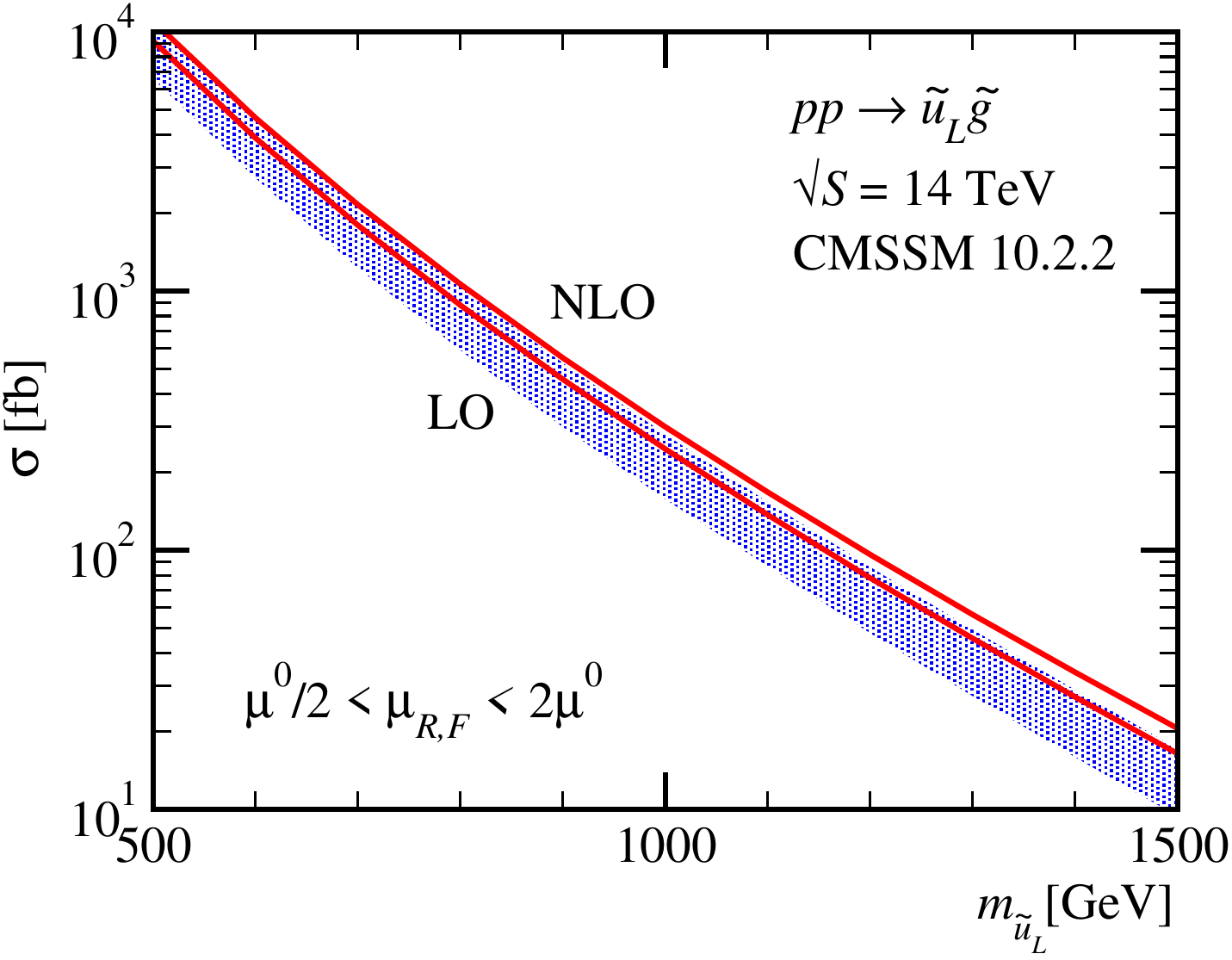}
\caption{Cross sections for $\sigma(pp \to \sul \go)$ as a function of
  the squark mass $m_{\sul}$.  The band corresponds to a scale
  variation $\mu^0/2 < \mu_{R,F} < 2\mu^0$, where
  $\mu^0=(m_{\sul}+m_{\go})/2$. 
  The MSSM parameters are given by the
  CMSSM~10.2.2 benchmark point. The squark and gluino masses we vary
  in parallel, just like in Fig.~\ref{fig:virtuala}.}
\label{fig:overmass_ulgo}
\end{figure}
%------------------------------------------------

%%%%%%%%%%%%%%%%%%%%%%%%%%%%%%%%%%%%%%%%%%%%%%%%%%%%%%%%%%%%%%%%%%%%%%%%
\subsection{Squark--gluino production}

Unlike squark or gluino pair production the associated production
process does not have a QCD-only component and is always flavor
locked, $qg \to \sq\go$. This makes it the most model dependent
signature.  First generation squarks, mostly $\sulr$, will be
copiously produced, and some of the structures will be reminiscent of
the electroweak production of squarks with electroweak gauginos, $pp
\to \sq\tilde{\chi}$~\cite{madgolem_sqn}.

Moreover, in this particular channel on-shell divergences can have a
twofold origin: they can either stem from an on-shell gluino or an
on-shell squark, depending on which of these particles is
heavier. This makes associated squark--gluino production the key
channel to test our numerical \mg implementation of automized on-shell
subtraction.\bigskip

Several qualitative expectations we can nicely confirm from
Tabs.~\ref{tab:14tev} and \ref{tab:8tev}. For instance, we see how
$\sul\go$ production dominates over the charge conjugated channel
$\sul^*\go$, simply due to the valence $u$ quark. This is also the
reason why the QCD corrections are larger for the $\sul^*\go$ process,
because $gg$-initiated NLO contributions are not suppressed by the
relative size of the underlying parton luminosities.

%------------------------------------------------
\begin{figure}[t]
\includegraphics[width=0.8\textwidth]{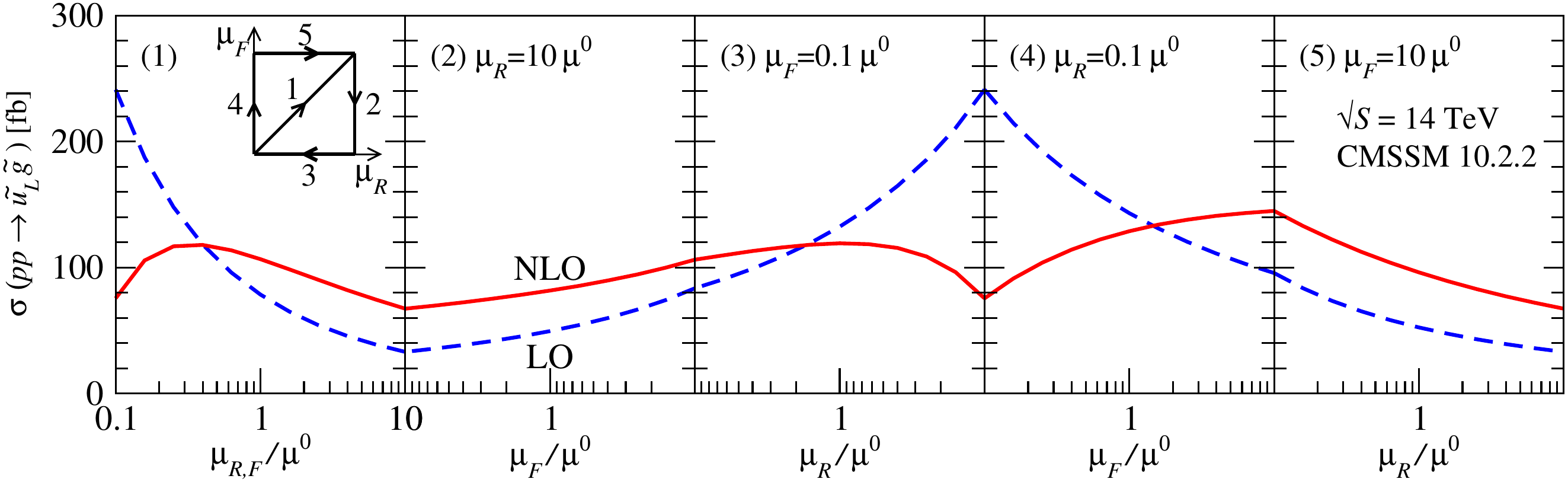}
\caption{Renormalization and factorization scale dependence for
  $\tilde{u}_L \go$ associated production. The plot traces a contour in the
  $\mu_R$-$\mu_F$ plane in the range $\mu = (0.1 - 10) \times \mu^0$
  with $\mu^0 = (m_{\sul}+\mgo)/2$. All parameters are the same as for
  Fig.~\ref{fig:scale-sqsq}, with mass values $m_{\sul}= 1162$~GeV and
  $m_{\go}= 1255$~GeV.}
\label{fig:scale-sqgo}
\end{figure}
%------------------------------------------------ 

%------------------------------------------------
\begin{figure}[t]
\begin{center}
\includegraphics[width=0.49\textwidth]{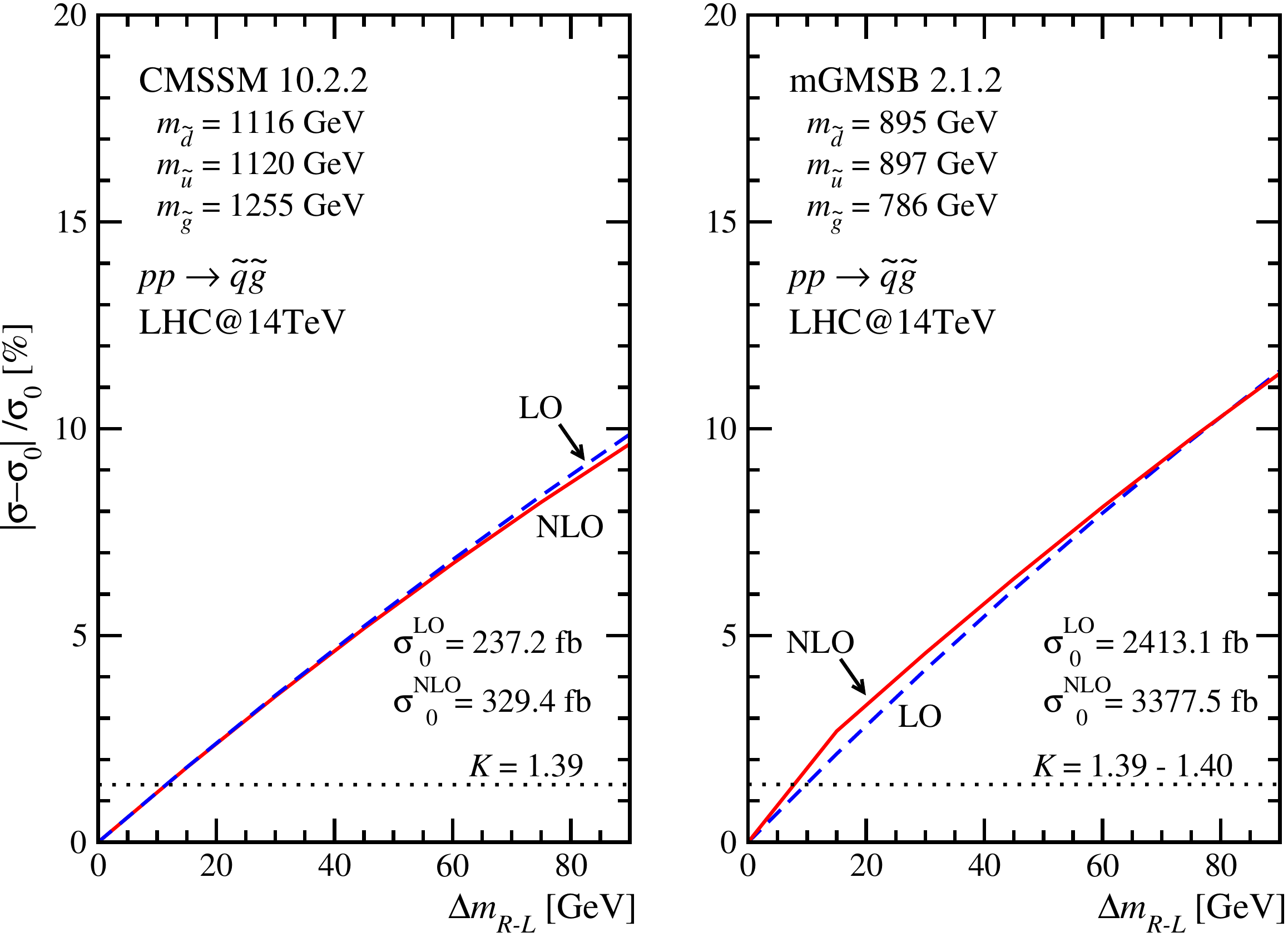}
\hspace*{0.00\textwidth}
\includegraphics[width=0.49\textwidth]{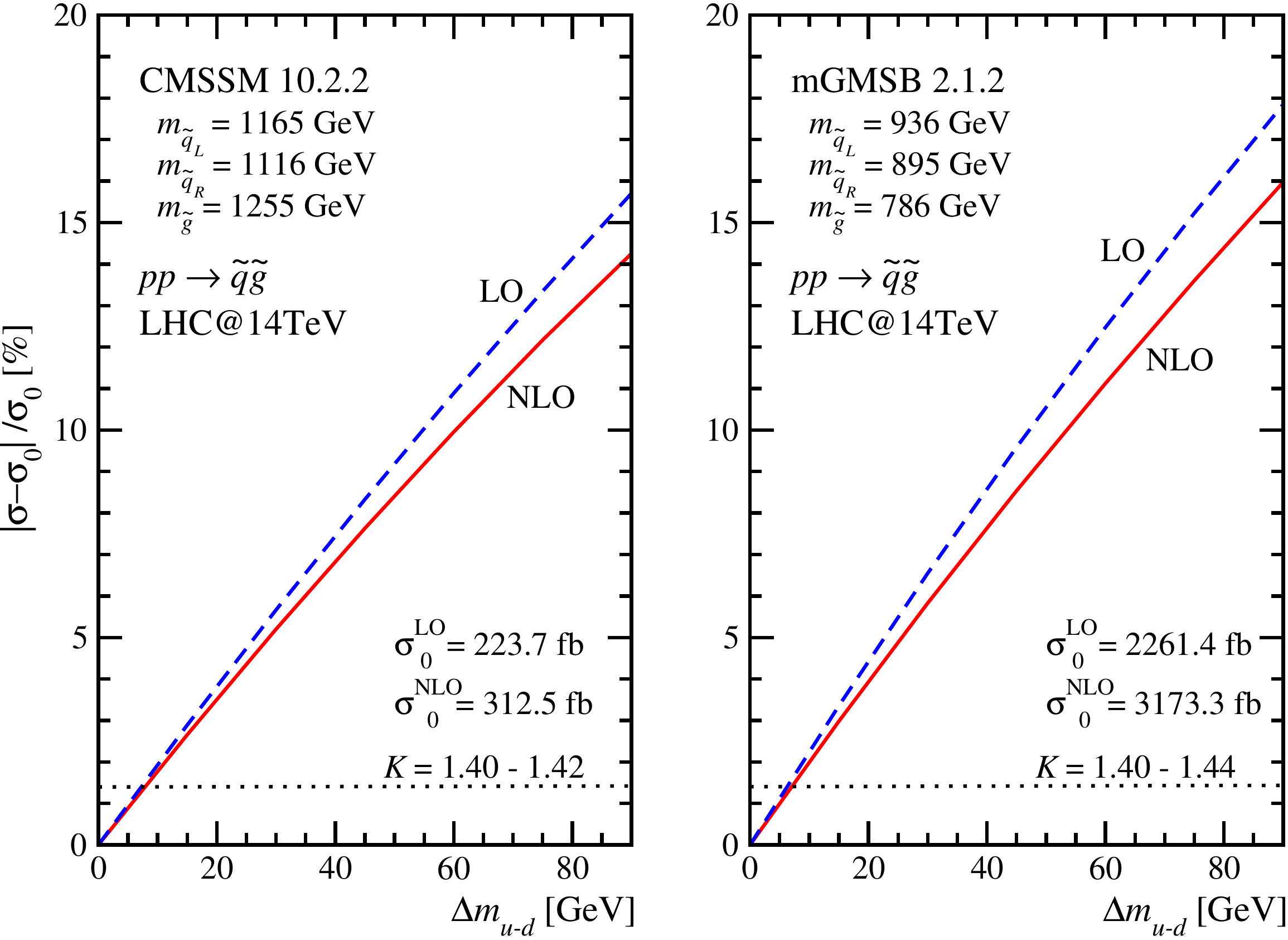}
\end{center}
\caption{Cross sections for squark--gluino production $pp\to \sq\go$
  ($\sq = \sulr,\sdlr$) as a function of mass splittings, within the
  same setup as in Fig.~\ref{fig:prospino}.}
\label{fig:masses-sqgo-rates}
\end{figure}
%------------------------------------------------

The dependence on the final state masses we show in
Fig.~\ref{fig:overmass_ulgo}, where we display the total cross
sections $\sigma(pp \to \sul\go)$ as a function of the final-state
squark mass $m_{\sul}$, noting that the gluino mass is changed together
with the squark mass. The total cross section is pulled down by
roughly three orders of magnitude when the final-state mass increases by
a factor of three. We find cross sections as large as $\sigma(\sul\go)
\sim \mathcal{O}(10)$ pb for $m_{\tilde{u}_L} \lesssim 500$ GeV, which fall
down to $\mathcal{O}(10)$ fb for $m_{\tilde{u_L}} \lesssim 1.5$~TeV. A
remarkable reduction 
in scale dependence 
of $\mathcal{O}(60)\%$ down to $\mathcal{O}(20)\%$
can be assessed by
contrasting the LO and NLO uncertainty bands.  A complementary
viewpoint we provide in Fig.~\ref{fig:scale-sqgo}, where we probe
scale variations of the total cross section as usually in the
two-dimensional renormalization vs factorization scale plane.\bigskip

Finally, we address the effect of a general squark mass pattern on the
total rates. Our analysis follows
Fig.~\ref{fig:prospino}, now for the process $pp \to \sq\go$ with $\sq
= \sulr, \sdlr$. Again, we study the different CMSSM 10.2.2 and
mGMSB 2.1.2 scenarios.  For each of them, we explore the relative
change in the total rate $|\sigma-\sigma^0|/\sigma^0$ when we increase
mass splittings from zero ($\sigma_0$).  We separately examine i) fixing
all left-handed and right-handed squarks at one common mass value
and increasing the right-left mass splitting $\Delta m_{R-L}$;
and ii) setting a common mass for up-type and down-type squarks and
increasing $\Delta m_{u-d}$. Similarly to the squark
pair case, in Fig.~\ref{fig:masses-sqgo-rates} we find variations up to $20\%$ for mass splittings up to
$\mathcal{O}(100)$~GeV. The LO and NLO cross sections scale in
parallel, with minor differences at the per-cent level. As expected
from the squark pair case the footprint of a non-degenerate squark
spectrum factorizes from the QCD corrections, with remaining
non-factorizing effects the level of a few per-cent.

%------------------------------------------------
\begin{figure}[t]
\includegraphics[width=0.8\textwidth]{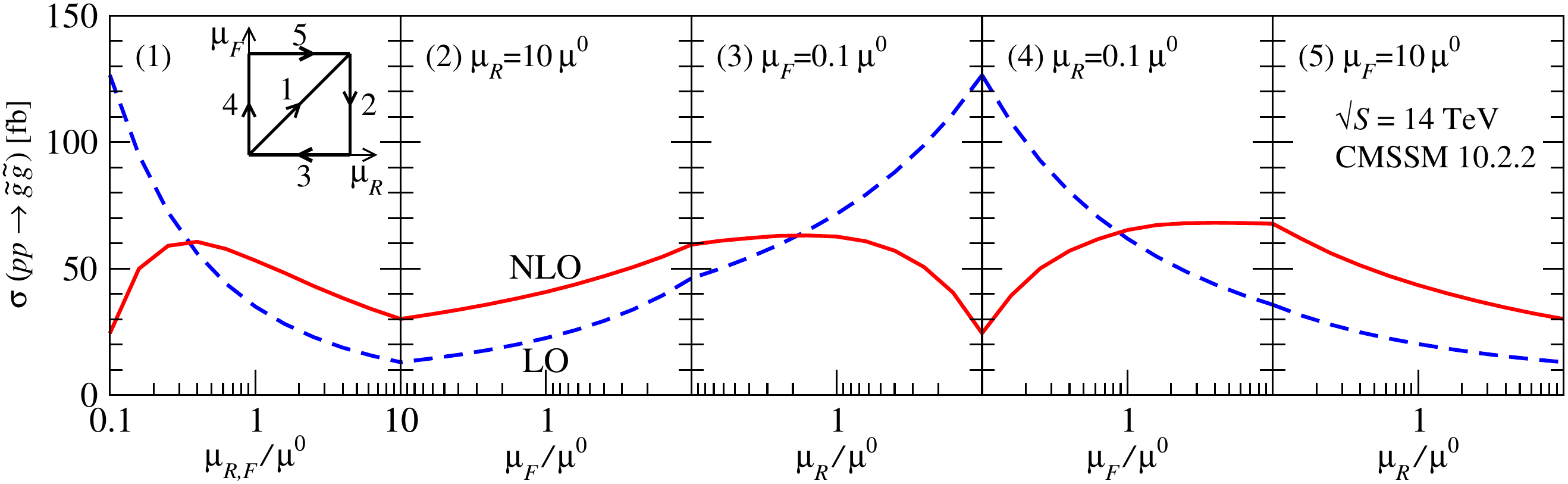}
\caption{Renormalization and factorization scale dependence for gluino
  pair production. The plot traces a contour in the $\mu_R$-$\mu_F$
  plane in the range $\mu = (0.1 - 10) \times \mu^0$ with 
  $\mu^0 = \mgo$. All parameters
  are the same as for Fig.~\ref{fig:scale-sqsq}, with $m_{\go}=
  1255$~GeV.}
\label{fig:scale-gogo}
\end{figure}
%------------------------------------------------ 

%------------------------------------------------
\begin{figure}[b]
\includegraphics[width=0.38\textwidth]{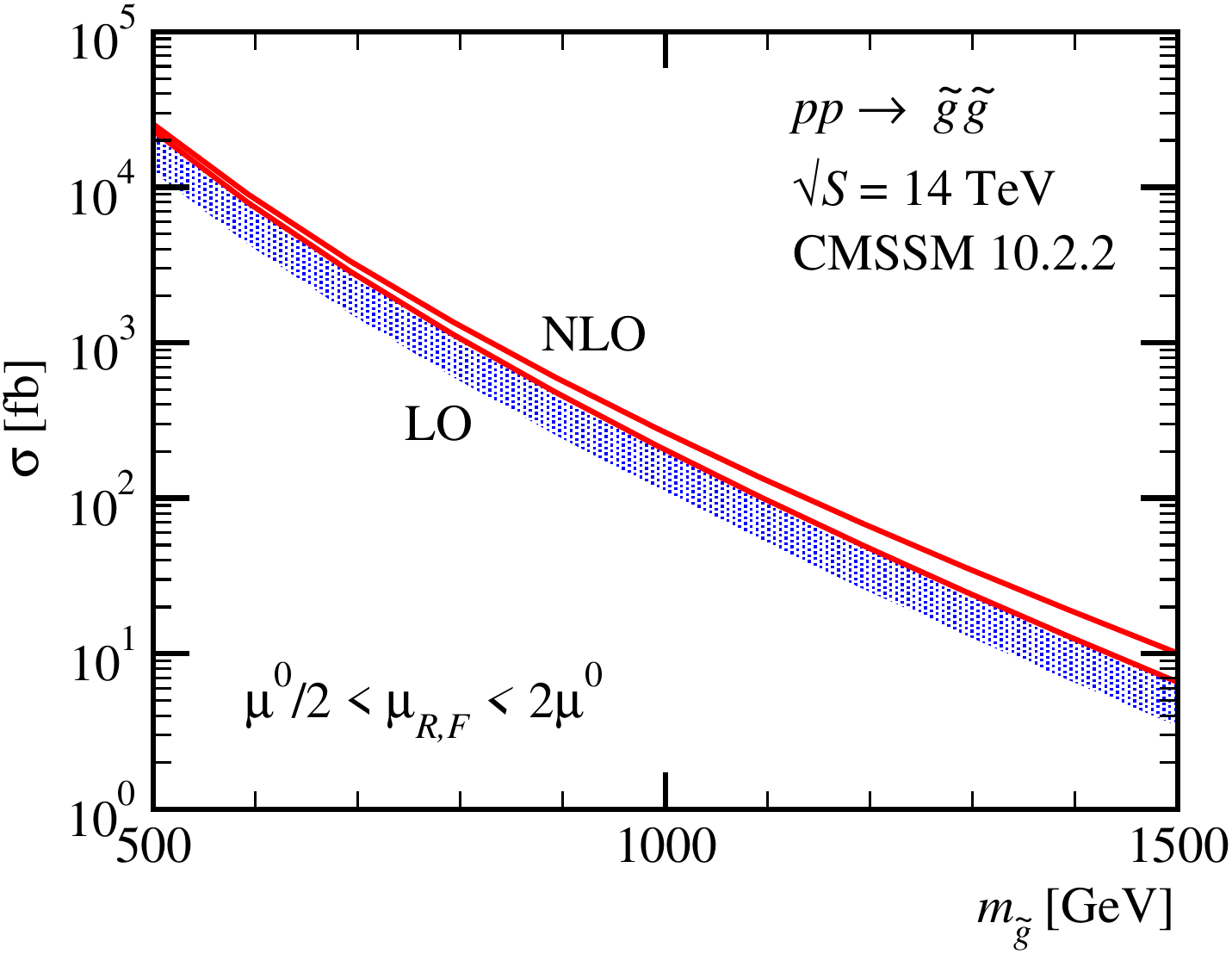}
\caption{Cross sections for $\sigma(pp \to \go \go)$ as a function of
  the gluino mass $m_{\go}$.  The band corresponds to a scale
  variation $\mu^0/2 < \mu_{R,F} < 2 \mu^0$ with $\mu^0 = \mgo$.
  The MSSM parameters are
  given by the CMSSM~10.2.2 benchmark point. The squark and gluino
  masses we vary in parallel, just like in Fig.~\ref{fig:virtuala}.}
\label{fig:overmass_gogo}
\end{figure}
%------------------------------------------------

%%%%%%%%%%%%%%%%%%%%%%%%%%%%%%%%%%%%%%%%%%%%%%%%%%%%%%%%%%%%%%%%%%%%%%%%
\subsection{Gluino pair production}

Finally, similar phenomenological
trends we identify for gluino-pair final-states. The NLO effects are
particularly sizable (cf. Tables~\ref{tab:14tev} and \ref{tab:8tev})
with $K$ factors in the ball-park of $\sim 2$ for $\sqrt{S}= 14$
TeV, and even surpassing $K\sim 2.5$ for the lower nominal LHC energy
$\sqrt{S}= 8$ TeV.  These results essentially reproduce what is
included in {\sc Prospino}.

The separate dependence on the factorization and renormalization
scales we display in Fig.~\ref{fig:scale-gogo}. As before, the
simultaneous scale variation captures the complete theoretical
uncertainty well. Including the NLO corrections significantly reduces
the dependence on the renormalization as well as on the factorization
scale. 
In Fig.~\ref{fig:overmass_gogo} we show the envelope of the
scale variation together with the central LO and NLO rate predictions
as a function of the gluino mass. 
It reflects a reduction of the theoretical uncertainties from
$\mathcal{O}(70\%)$ at LO down to $\mathcal{O}(30\%)$ at NLO.
In spite of the large $K$ factor
triggered by the LO parton densities the two bands nicely
overlap.\bigskip

From the discussions of the squark--pair and squark--gluino channels
we expect the effect of the non-degenerate squark spectrum on gluino
pair production to be small. The leading phase space effects which
appear as factorizing corrections in the other production channels are
absent for gluino pairs. Only the LO squark mass dependence through
the $t$-channel exchange diagram and the specific NLO loop effects
remain. For the mGMSB 2.1.2 benchmark point we compute these effects and find
a deviation of roughly 2\% when we allow for a right-left mass splitting of 100~GeV.

%------------------------------------------------
\begin{figure}[t!]
\begin{center}
\begin{tabular}{ccc}
\includegraphics[width=0.4\textwidth]{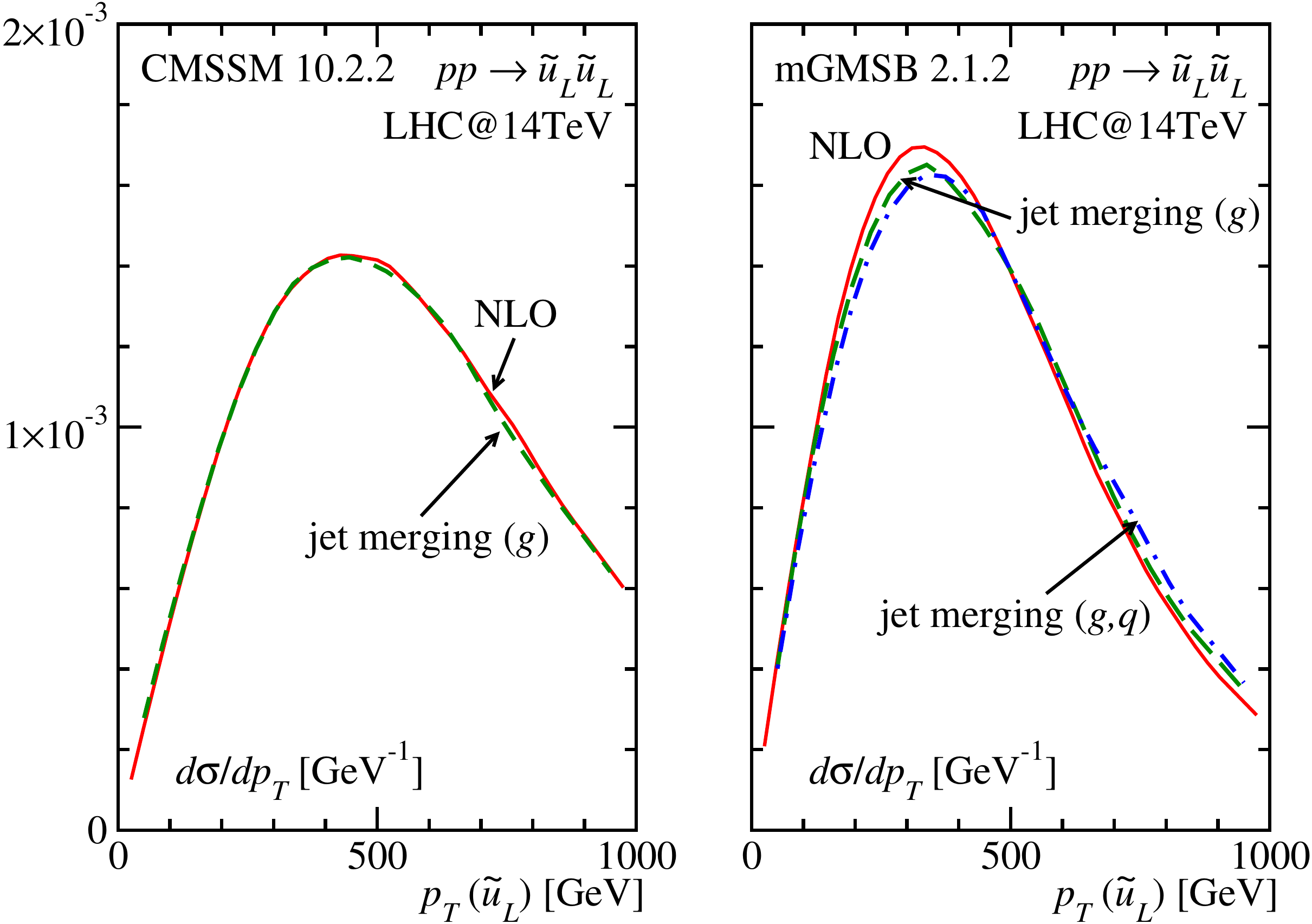} & &
\includegraphics[width=0.4\textwidth]{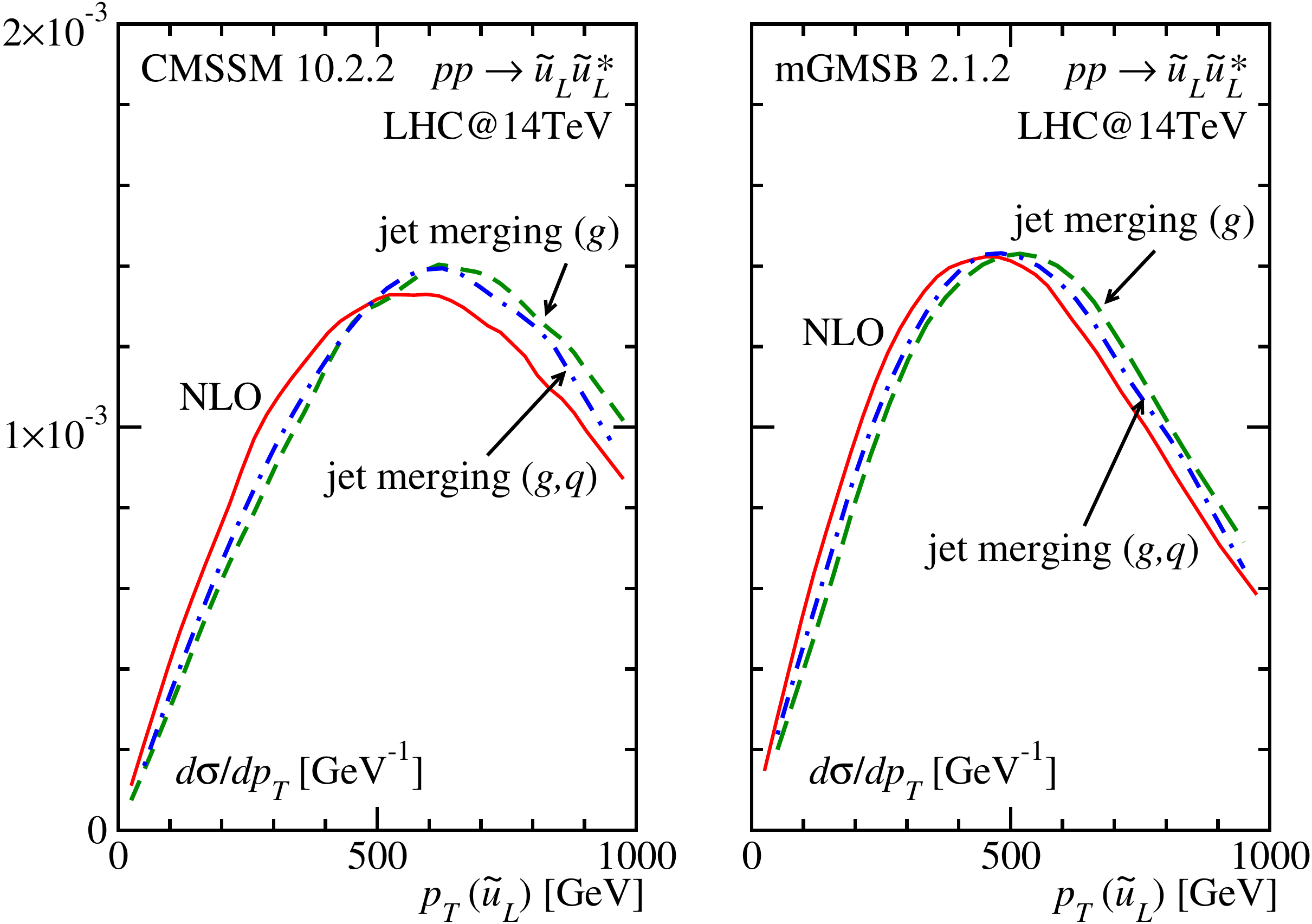} 
\\ & \hspace{0.2cm} & \\
\includegraphics[width=0.4\textwidth]{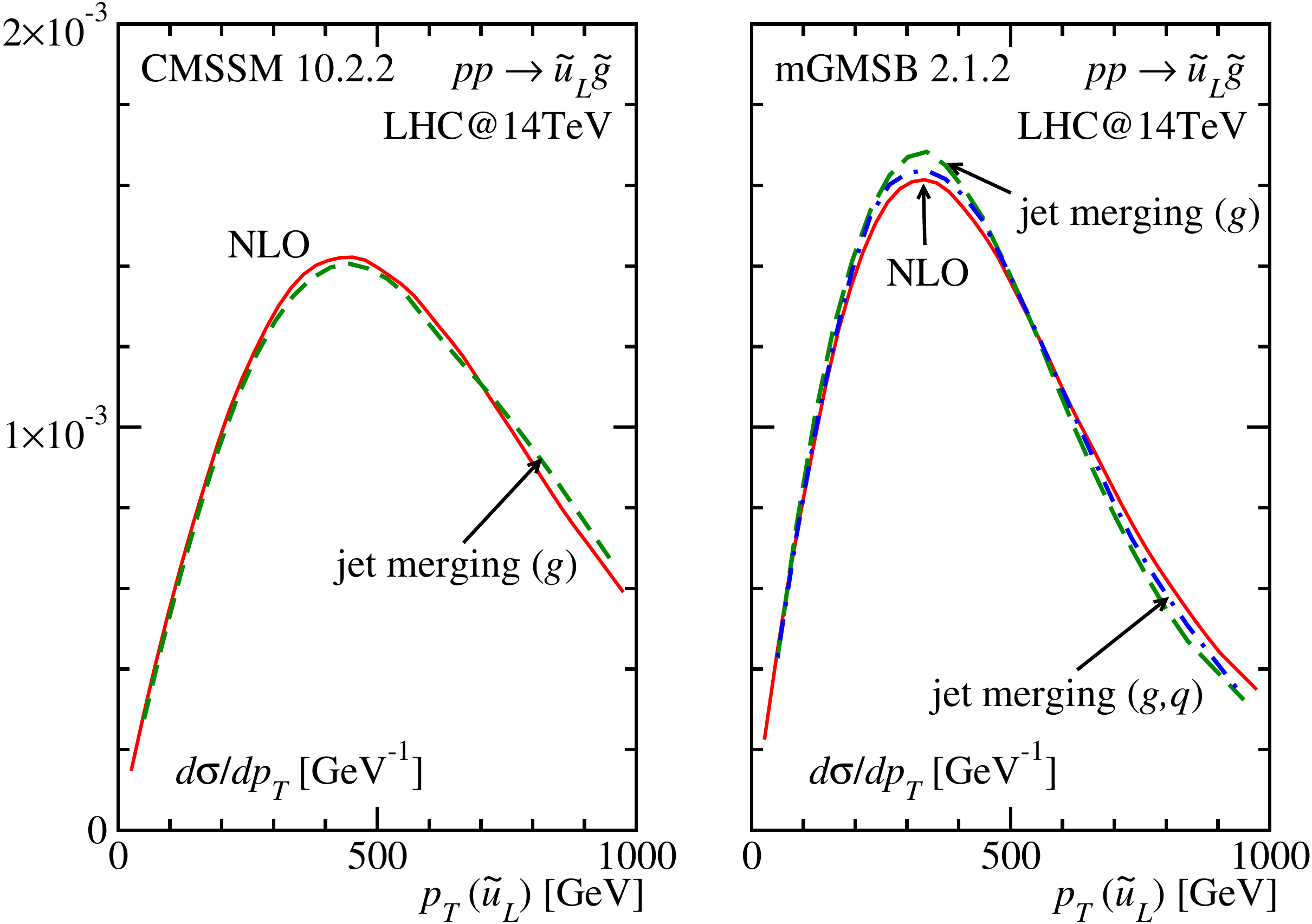} & & 
\includegraphics[width=0.4\textwidth]{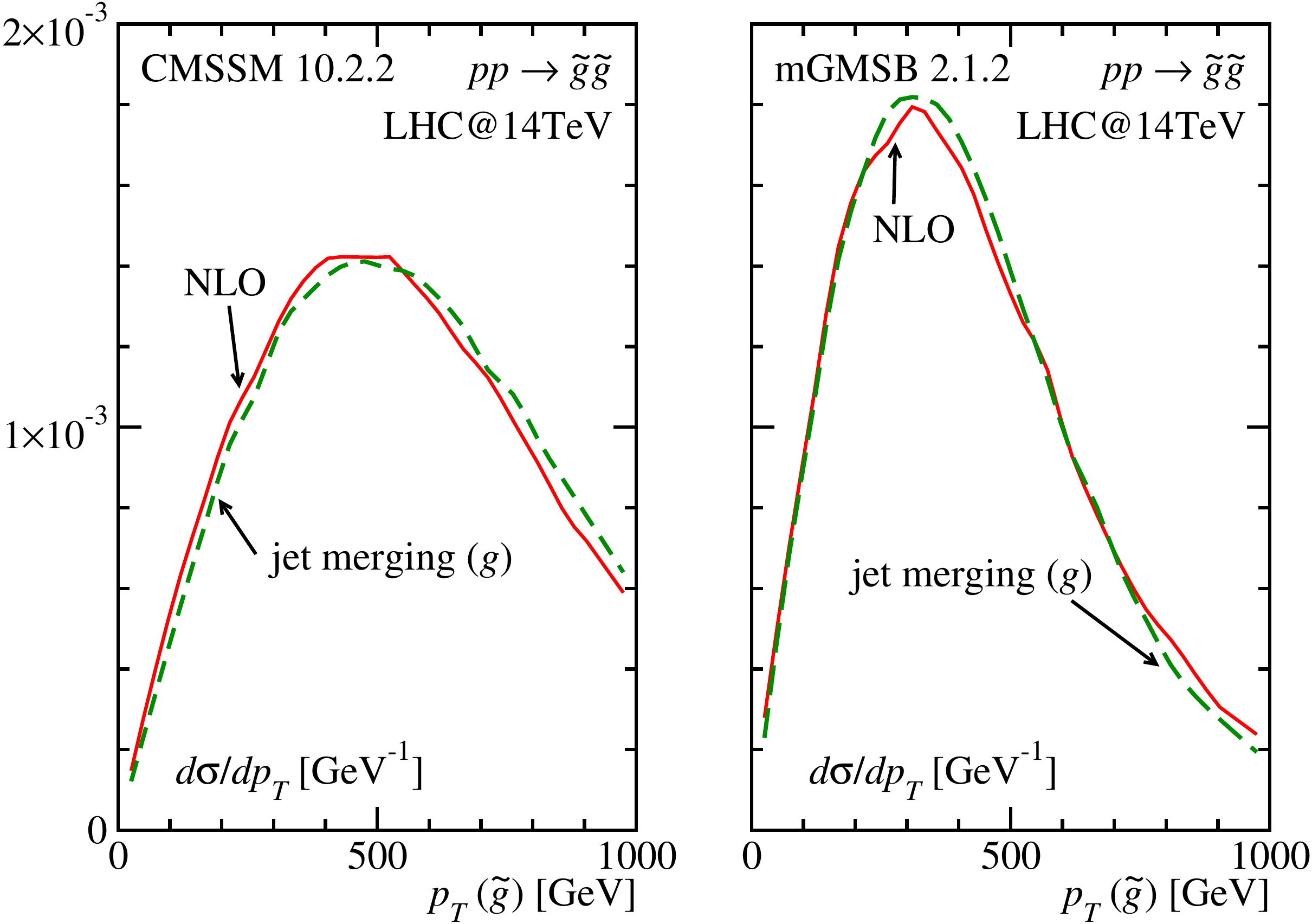} 
 \end{tabular}
\end{center}
\caption{Normalized transverse momentum distributions for different
  processes for the benchmark points CMSSM~10.2.2 and mGMSB~2.1.2.  We
  compare NLO predictions to LO jet merging~\cite{mlm} with three
  different setups: up to one hard gluon; up to two hard gluons; up to
  one hard quark or gluon jet. The latter two we only display when
  differences are visible.}
\label{fig:distrib}
\end{figure}
%------------------------------------------------

%%%%%%%%%%%%%%%%%%%%%%%%%%%%%%%%%%%%%%%%%%%%%%%%%%%%%%%%%%%%%%%%%%%%%%%%
\section{Distributions}
\label{sec:merging}

All through Sec.~\ref{sec:rates} we have limited our discussions to
total cross sections. This corresponds to the way higher-order
corrections to new physics processes are usually implemented in
experimental analyses. Event simulation including all differential
cross section is performed by any of the parton-shower Monte
Carlos. Because the hard process scale is given by the heavy particle
masses it is usually well above the typical jet momenta required by
inclusive searches. This means that the parton shower approximation is
justified~\cite{qcd_radiation} while the total cross sections have to
be corrected for higher-order effects. In the original {\sc Prospino}
calculations transverse momentum and rapidity distributions for the
heavy squarks and gluinos were
studied~\cite{squarkpairNLO,squarkgluinoNLO,gluinopairNLO}, indicating
that no large NLO effects should be expected. 
\mg allows us to include a comprehensive study
of distributions in this paper.

%%%%%%%%%%%%%%%%%%%%%%%%%%%%%%%%%%%%%%%%%%%%%%%%%%%%%%%%%%%%%%%%%%%%%%%%
\subsection{Fixed order vs jet merging}

To make quantitative statements beyond total cross sections we use \mg
to compute NLO distributions for different squark and gluino final
states. Because the \mg output is weighted events for the regularized
virtual and real emission channels we can plot any distribution which
makes sense in perturbative QCD. The only limitation is the validity
of fixed-order QCD, reaching its limitations for example when studying
the jet recoil against the heavy squark--gluino system.\bigskip

For comparison we do not rely on the usual parton shower simulations,
but employ more modern matrix element and parton shower
merging~\cite{ckkw,mlm,lecture}. We generate tree-level matrix element
events with zero, one, or two hard jets with the help of {\sc
  MadGraph}5~\cite{mg5} and combine them with each other and with the {\sc
  Pythia}~\cite{pythia} shower using the MLM procedure~\cite{mlm} as
implemented in {\sc MadGraph}. When defining the hard matrix element
corrections we follow three different approaches. First, we include up
to one additional hard gluon in the matrix elements. This
automatically excludes all topologies which could lead to on-shell
divergences. Second, we instead allow for two additional hard gluons
in the matrix elements. As before, we avoid any issues with on-shell
singularities. Finally, we generate samples with one additional quark
or gluon.  In this case, on-shell divergences will appear just like
for the real emission contributing to the NLO rate.  These
singularities we remove using the numerical prescription implemented
in {\sc MadGraph}~\cite{mg5}. It subtracts all events with phase space
configurations close to the on-shell poles. While this subtraction is
not equivalent to the consistent {\sc Prospino} scheme and does not
have a well-defined zero-width limit we have checked that it gives
numerically similar results as long as we only compare normalized
distributions.\bigskip

Our results for the transverse momentum distributions of squarks and
gluinos we present in Fig.~\ref{fig:distrib}.  We focus on the
CMSSM~10.2.2 and mGMSB~2.1.2 benchmark points as representative MSSM
scenarios. They exemplify both possible squark--gluino mass
hierarchies.  As described above, we show the NLO predictions and the
one-gluon merged results. Comparing different jet merging setups we
confirm that adding a second hard gluon does not change our results
beyond numerical precision, so we do not show it separately. This is
an effect of the large hard scale in the process which renders the
parton shower for the second radiated gluon an excellent
approximation.  Results allowing for one additional quark or gluon jet
we only show when the curves are visibly different from the one-gluon
case. The general agreement of all three merging results shows that
once the double counting from the on-shell states is removed the bulk
of the NLO real emission comes from gluons. Moreover, this gluon
radiation is well described by the {\sc Pythia} parton shower, as long
as the produced particles are heavy~\cite{qcd_radiation}. However,
once an experimental analysis becomes particularly sensitive to the
jet recoil it might pay off to check the parton shower results with a
merged sample~\cite{autofocus,jamie}.

%------------------------------------------------
\begin{figure}[t!]
\begin{center}
\includegraphics[width=0.30\textwidth]{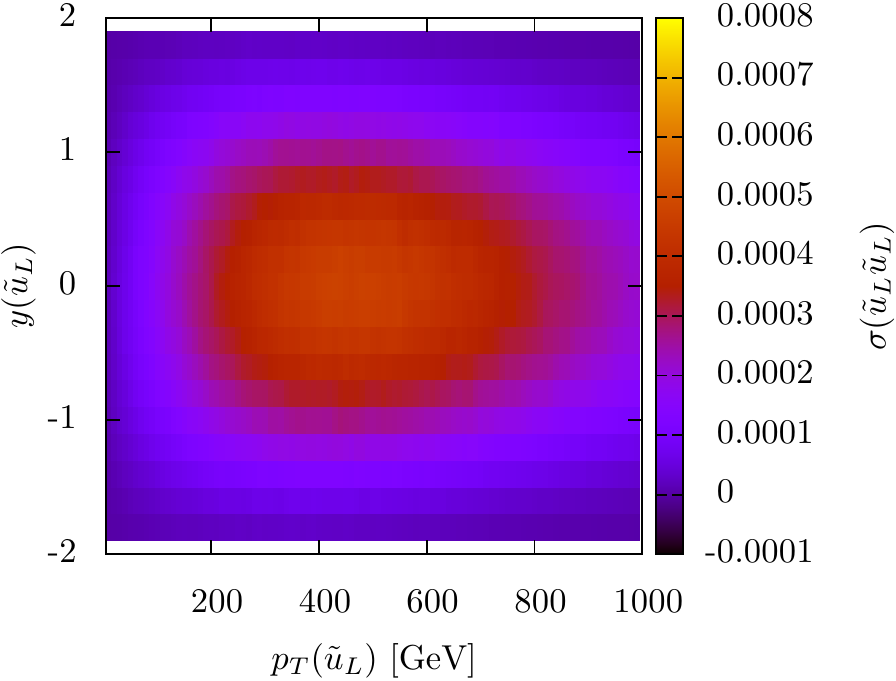}  
\includegraphics[width=0.30\textwidth]{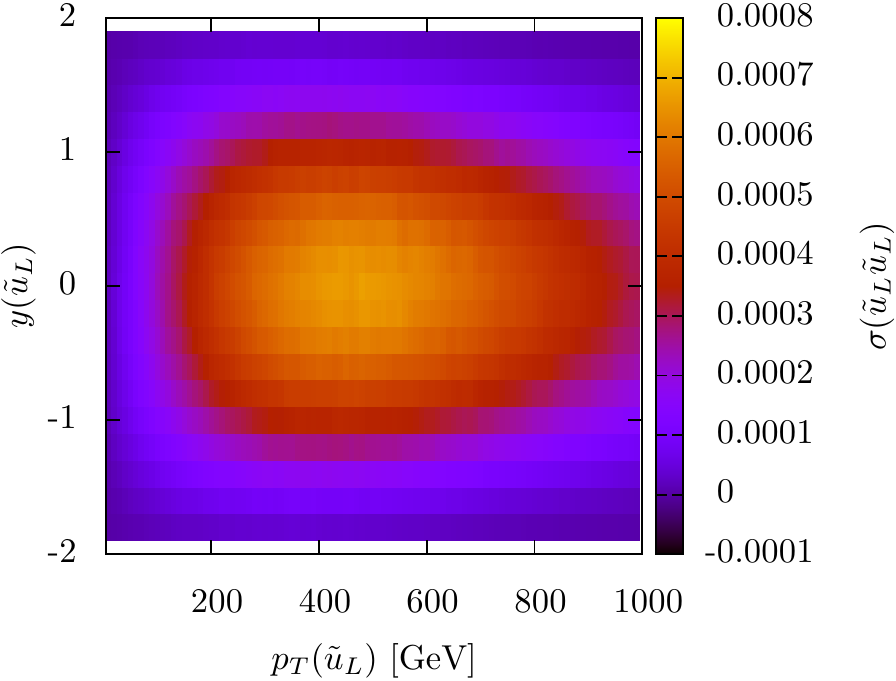} 
\includegraphics[width=0.30\textwidth]{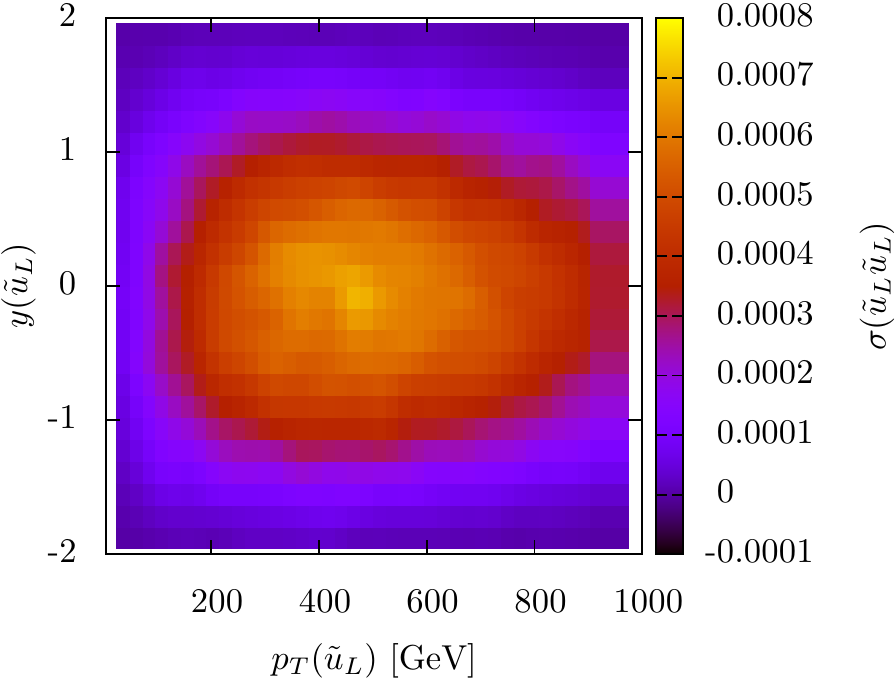}
\end{center}
\caption{Two-dimensional distributions for squark pair production $pp
  \to \sul\sul$ at $\sqrt{S} = 14$ TeV 
  as contour plots in the $p_T(\sul)$-$y(\sul)$ plane.
  The different panels show the results from LO (left), NLO (center),
  and jet merging (right). While the LO result is shown to scale the
  two right histograms are normalized to unity. We use the
  CMSSM~10.2.2 parameters.}
\label{fig:contour}
\end{figure}
%------------------------------------------------

%------------------------------------------------
\begin{figure}[thb]
\begin{center}
\begin{tabular}{ccc}
\includegraphics[width=0.46\textwidth]{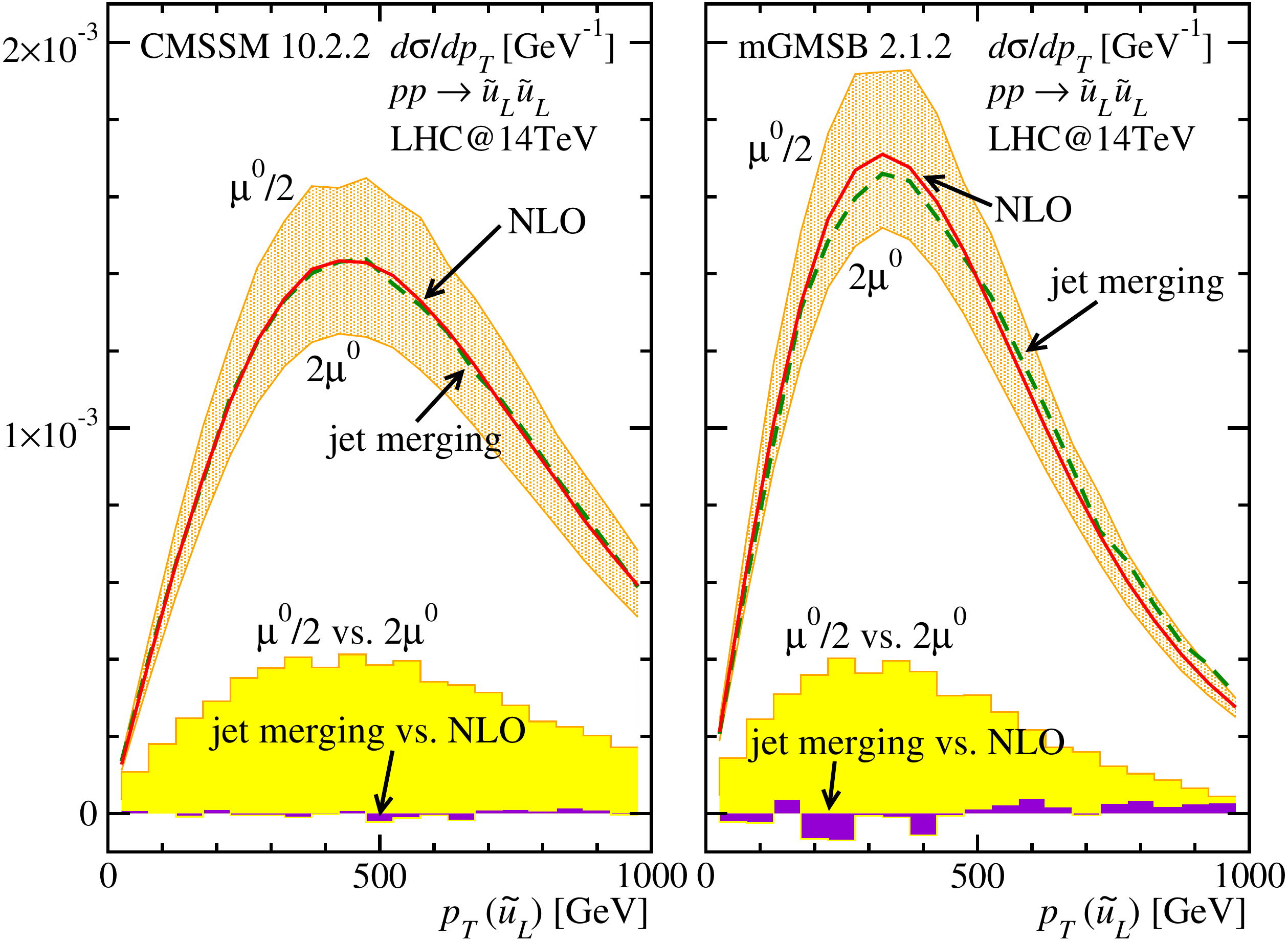} & \hspace{0.4cm}&
\includegraphics[width=0.45\textwidth]{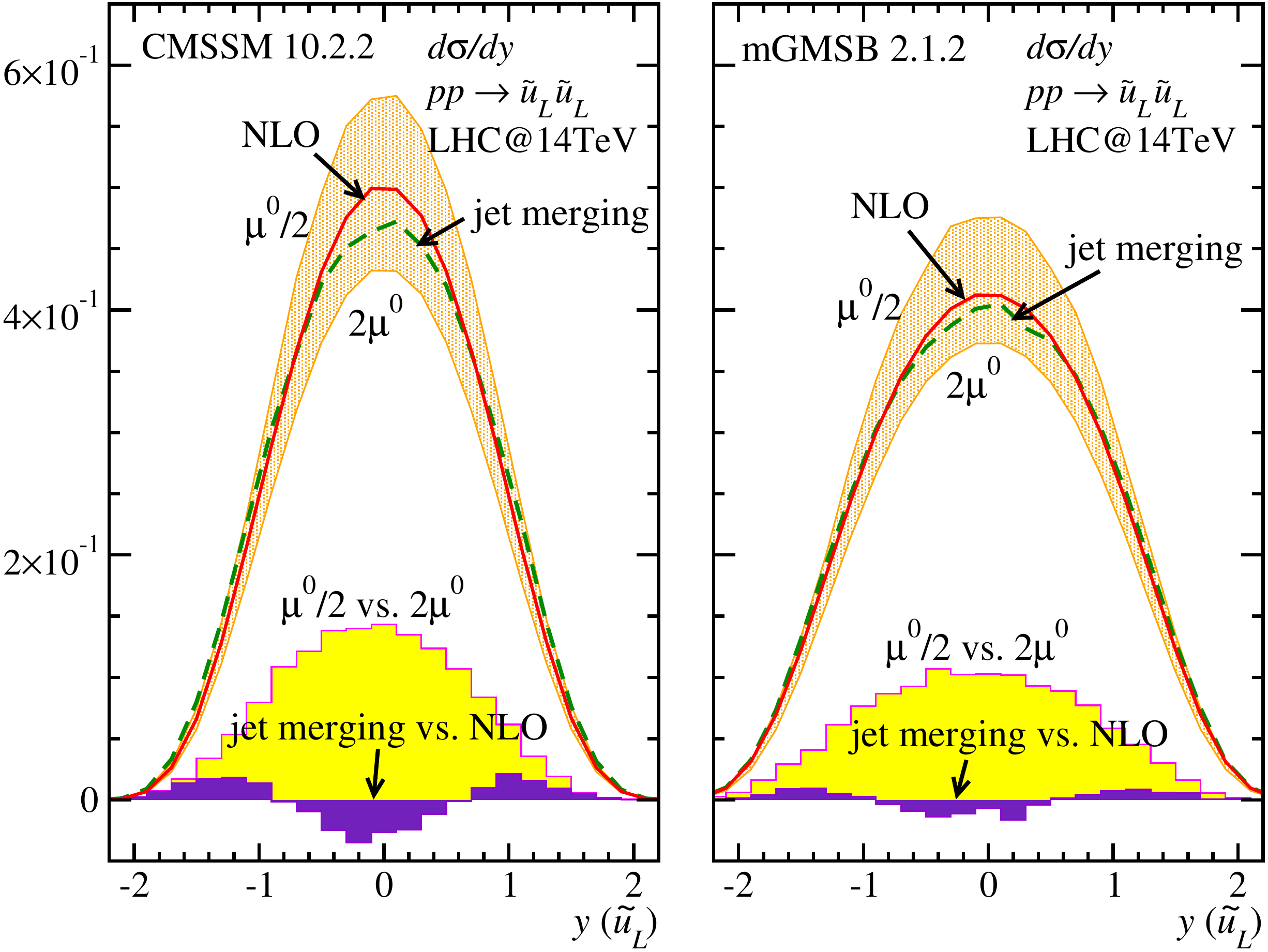} 
 \end{tabular}
\end{center}
\caption{Distributions for squark pair production $pp \to \sul\sul$ as
  a function of the squark transverse momentum (left) and rapidity
  (right). The curves for the central scales we normalize to unity.
  The scale uncertainty curves we normalize to the same central
  value. The yellow area shows the scale uncertainty, \eg
  $d\sigma/dp_T(\mu^0/2)-d\sigma/dp_T(2\mu^0)$, compared to the purple
  area giving $d\sigma^\text{MLM}/dp_T-d\sigma^\text{NLO}/dp_T$. We
  examine the benchmark points CMSSM 10.2.2 and mGMSB~2.1.2.}
\label{fig:uncertainty}
\end{figure}
%-----------------------------------------------

The comparison with the NLO prediction shows that the usual assumption
about the stability of the main distributions is indeed correct. The
normalized distributions from the fixed-order NLO calculation and from
multi-jet merging agree very well. As alluded to above, the multi-jet
merging predictions in turn agree well with the parton shower.  In
spite of the remarkable agreement between both descriptions, some mild
departures are visible. We can essentially understand them as a
fingerprint of the extra recoil jets involved in the matched samples.
For example, in some cases the jet-merging predictions become slightly
harder than the NLO results because they take into account a second
radiated jet. On the other hand, the squarks and gluinos we are
studying are so heavy that it is unlikely that jet radiation makes a
big difference to them.\bigskip

In \mg the generation of any kind of fixed-order distributions, such
as those displayed in Fig.~\ref{fig:distrib}, is completely automated.
This constitutes a substantial improvement for precision BSM
predictions.  Distributions can be computed for a single kinematic
variable, but also two-dimensionally. For example, we show the NLO
phase space dependence on the transverse momentum and the rapidity of
one final-state particle in Fig.~\ref{fig:contour}. The three panels
give LO, NLO, and merged predictions for squark pair production $pp
\to \sul\sul$. The NLO and the merging histograms are normalized to
unity, while the LO distribution is shown to scale. As expected, we do
not find any kind of significant difference between the NLO and the jet merging
results nor any
correlations between the rapidity and transverse momentum.

%%%%%%%%%%%%%%%%%%%%%%%%%%%%%%%%%%%%%%%%%%%%%%%%%%%%%%%%%%%%%%%%%%%%%%%%
\subsection{Scale uncertainties}

%------------------------------------------------
\begin{figure}[b!]
\begin{center}
\begin{tabular}{ccc}
\includegraphics[width=0.46\textwidth]{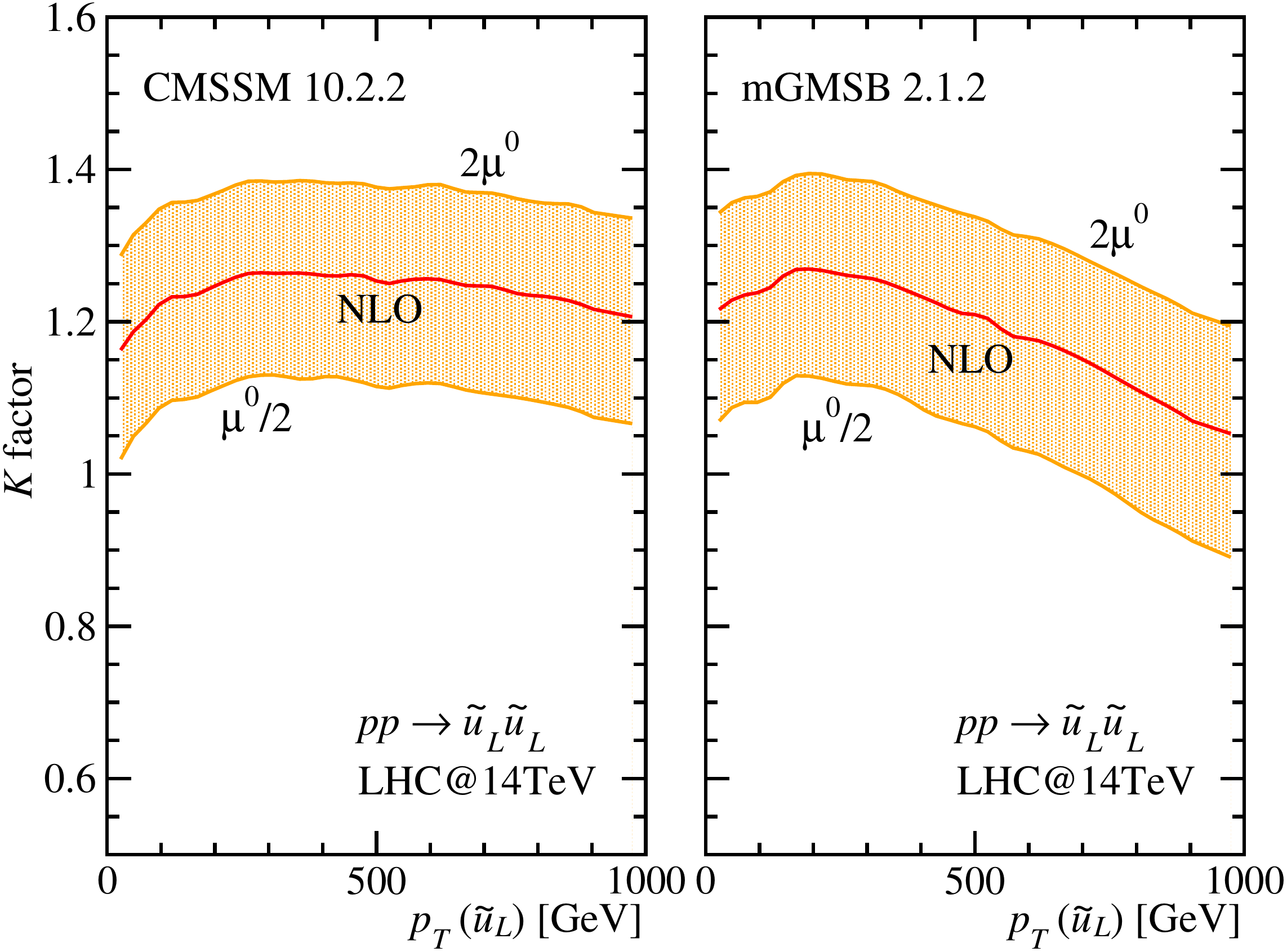}& &
\includegraphics[width=0.45\textwidth]{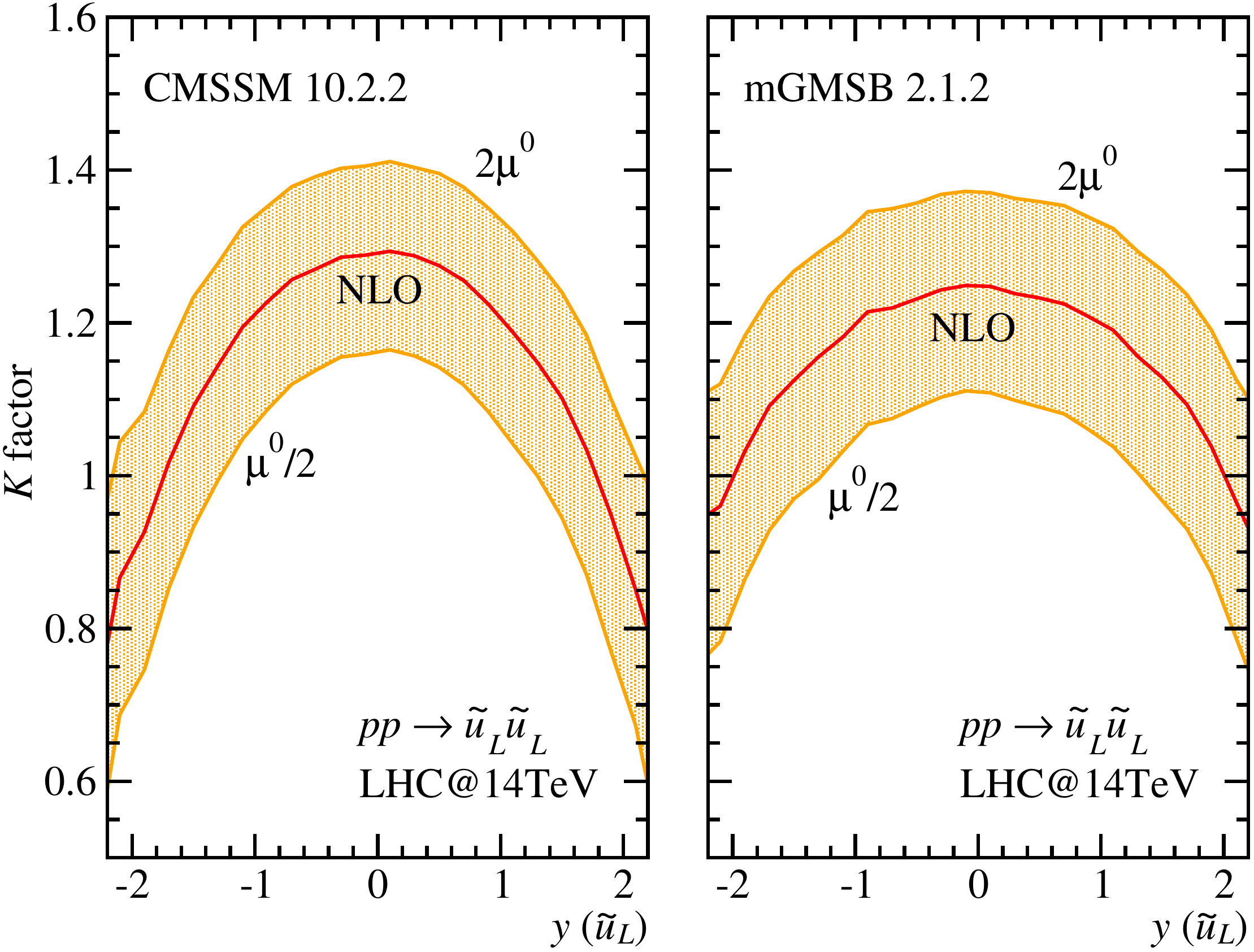}
 \end{tabular}
\end{center}
\caption{$K$ factor as a function of $p_T(\sul)$ and $y(\sul)$ for squark
  production $pp \to \sul\sul$.  The band shows a scale variation
  $\mu^0/2 < \mu < 2\mu^0$.  All MSSM parameters we fix to
  CMSSM~10.2.2 and mGMSB~2.1.2.}
\label{fig:kfac-distrib}
\end{figure}
%------------------------

The stabilization of the (unphysical) dependence with respect to the
choice of the renormalization and factorization scales is a most
prominent feature of NLO calculations. These improvements, which we
have already analyzed for the total cross section, we reexamine now
for the distributions. Again, we study squark pair production $pp \to
\sul\sul$.  In Figure~\ref{fig:uncertainty} we present the squark
transverse momentum and rapidity distributions. 

We first overlay the normalized distributions from the fixed-order NLO
calculation (solid, red line) with the central scale choice
and the jet merging results (dashed, green line) for the CMSSM~10.2.2
and mGMSB~2.1.2 parameter points.  For the NLO curve we compute the
envelope varying the renormalization and factorization scales between
$\mu^0/2$ and $2\mu^0$, keeping the normalization relative to the
central scale choice. This should give a realistic estimate of the
theoretical uncertainty.  

Two differences we show separately: first, the yellow (light) histogram
shows the difference $d\sigma/dp_T(\mu^0/2)-d\sigma/dp_T(2\mu^0)$. It
indicates a theoretical uncertainty of $\mathcal{O}(10\%)$ on the
distribution, with no obvious caveats. In addition, we show the
difference between the central NLO prediction and MLM multi-jet
merging $d\sigma^\text{MLM}/dp_T - d\sigma^\text{NLO}/dp_T$
point-by-point in the purple (dark) histogram. Both comparisons we
repeat for the squark rapidity distributions.  We see that when it
comes to normalized distributions the NLO and MLM multi-jet
merging predictions are in excellent agreement, for example compared
to the sizable NLO scale dependence.\bigskip

A complementary viewpoint in terms of phase space dependent
$K$ factors we display in Fig.~\ref{fig:kfac-distrib}. The NLO
histograms using central scales $\mu^0$ are supplemented by a band
showing a simultaneous renormalization and factorization scale
dependence at NLO.  We confirm that the $K$ factors remain stable and
relatively constant for the transverse momentum and the central
rapidity regime. From the above discussion we know that the slight
change in the $K$ factor over the entire phase space should correspond
to distributions computed using multi-jet merging.  This result we
interpret as a strong argument in favor of the conventional procedure,
where a global $K$-factor or event re-weighting to NLO is applied to
kinematic distributions generated via multi-jet merging.

%%%%%%%%%%%%%%%%%%%%%%%%%%%%%%%%%%%%%%%%%%%%%%%%%%%%%%%%%%%%%%%%%%%%%%%%
\subsection{Non-degenerate squarks}

%------------------------------------------------
\begin{figure}[b!]
\begin{center}
\begin{tabular}{ccc}
\includegraphics[width=0.58\textwidth]{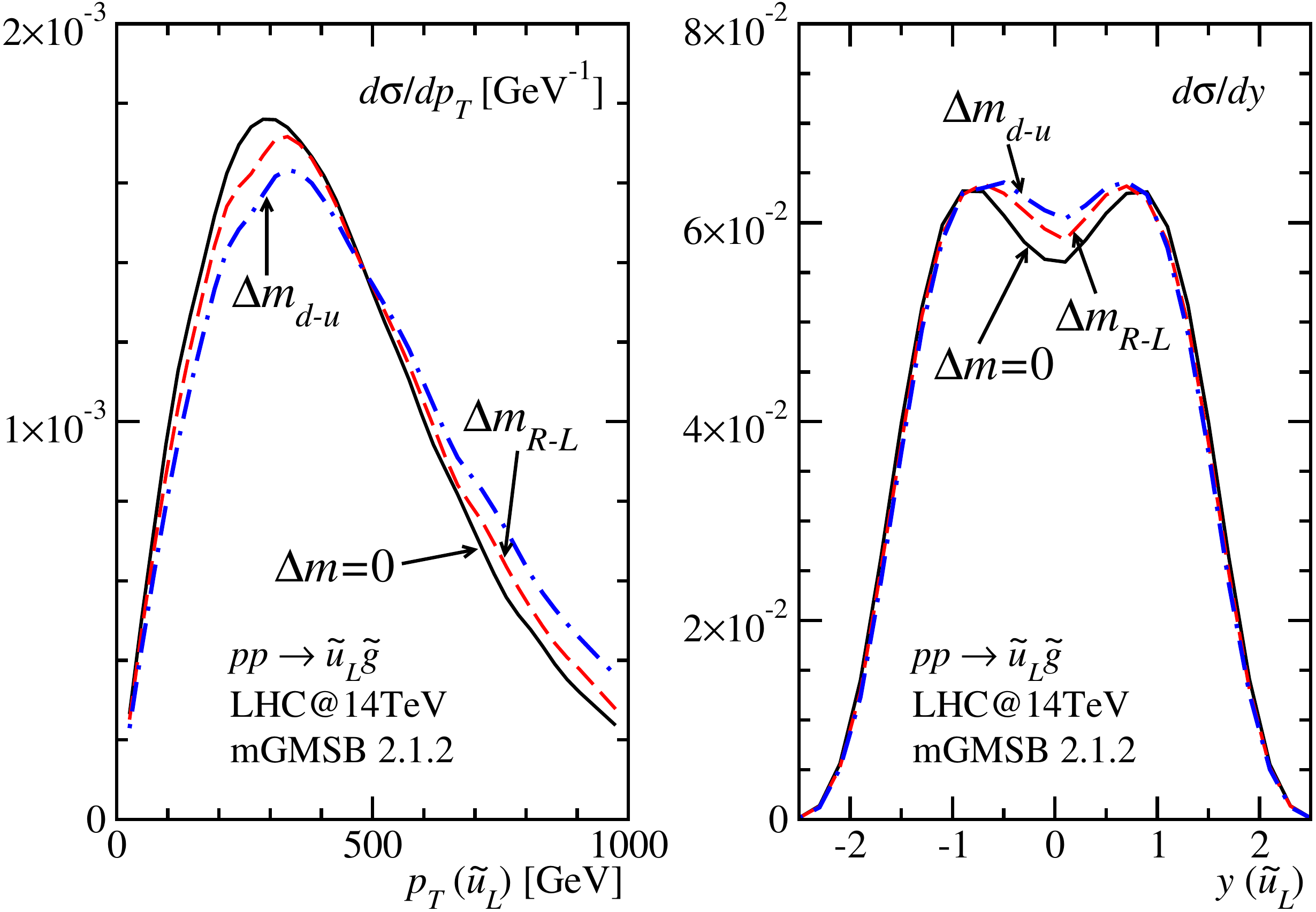}
&
\hspace{2.5cm} & \vspace{-0.2cm}
\includegraphics[width=0.18\textwidth]{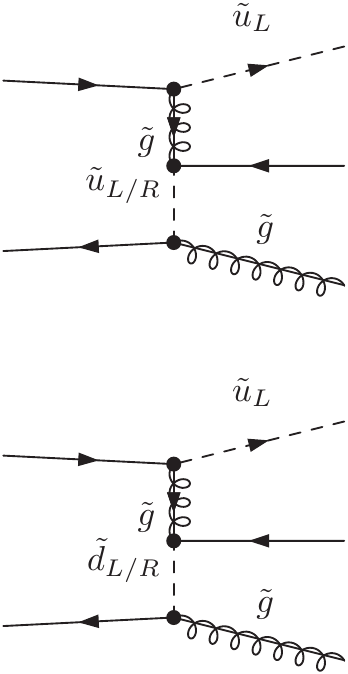}
\end{tabular}
\end{center}
\caption{Normalized transverse momentum (left) and rapidity
  distributions (right) for squark--gluino production $pp\to
  \tilde{u}_L\go$. We assume (i) mass-degenerate squarks with $m_{\sq}
  = 800$~GeV; (ii) a common mass splitting, $\Delta m_{R-L} = 200$~GeV;
  (iii) a common mass splitting, $\Delta m_{d-u} = 200$~GeV.  The
  central MSSM parameters we fix as in mGMSB~2.1.2 benchmark.  The
  Feynman diagrams to the right describe the squark--gluino fusion
  mechanism responsible for the significant differences.}
\label{fig:masses-sqgo-distrib}
\end{figure}
%------------------------------------------------

In Sec.~\ref{sec:rates} we discuss the effect of a general squark mass
spectrum on the different LHC production rates and find that it
largely factorizes from the NLO corrections, \ie the $K$ factors only
change at the per-cent level.  The impact of general squark mass
spectra becomes much more apparent at the distribution level. In
Fig.~\ref{fig:masses-sqgo-distrib} we display the squark transverse
momentum and rapidity distributions. We single out one particular
production channel, $pp \to \sul\go$ and examine the following
representative situations: (i) mass-degenerate squarks, with $m_{\sq}
= 800$~GeV; (ii) a right-left splitting $\Delta m_{R-L} = 200$~GeV
between the right-handed and left-handed squarks; and (iii) a similar
down-up splitting $\Delta m_{d-u} = 200$~GeV. The remaining MSSM
parameters we anchor as in the mGMSB~2.1.2 benchmark point defined in
Table~\ref{tab:sps}. Most importantly, we keep the final-state mass
constant, so the differences between these three scenarios decouple
from the leading phase space effects and instead constitute a genuine
NLO reflect.\bigskip

The finite mass splitting between squarks induces a shift in the
kinematic distributions in the direction of slightly harder and more
central final-state squarks. We can trace this back to the real
emission corrections shown in Fig.~\ref{fig:masses-sqgo-distrib}.
They describe a fusion mechanism where the bulk contribution arises
from internal squark and gluino propagators at very small virtuality,
\ie when these particles are almost on-shell. As a result, they become
particularly sensitive to variations of the squark masses, even if the
final-state squarks masses remain unchanged.

As we can see in Fig.~\ref{fig:masses-sqgo-distrib} the effect of an
$\mathcal{O}(20\%)$ mass splitting between up-type and down-type
squarks saturates the NLO uncertainty on the transverse momentum
distributions. Of course, the squark mass spectrum is not a source of
theory uncertainty which could be captured by the scale
dependence. Therefore, it might be useful to estimate its effect on
LHC analyses independently.

%%%%%%%%%%%%%%%%%%%%%%%%%%%%%%%%%%%%%%%%%%%%%%%%%%%%%%%%%%%%%%%%%%%%%%%%
\section{Summary}
\label{sec:sum}

\mg is a novel approach to the automated computation of total cross
sections and distributions for new heavy particles to next-to-leading
order. It can be used as an add-on to {\sc Madgraph}, making use of
its interfaces to new models as well as to the event generation. In
this paper we present a comprehensive overview of supersymmetric
particle production together with many details of the \mg
implementation. While \mg is not fully public yet, a fully functional
test version can be obtained from the authors upon request.\bigskip

In our application to squark and gluino production we reproduce all
relevant {\sc Prospino} results and extend currently available studies
in several ways:
\begin{itemize}
\item We evaluate NLO corrections to total rates for a completely
  general supersymmetric mass spectrum. For moderate mass splittings
  the leading effects of non-degenerate squarks factorize from the LO
  results, while the effect on $K$ factors stays at the level of few
  per-cent.
\item Instead of identifying the factorization and renormalization
  scales in the estimate of the theoretical uncertainty we vary both
  scales independently. For heavy strongly interacting new particles
  the envelope of all possible scales agrees with a simultaneous scale
  variation.
\item Squark and gluino distributions including the full NLO
  corrections and based on multi-jet merging agree very well within
  the NLO error bands.
\item The effect of non-degenerate spectra on the squark and gluino
  distributions is clearly visible and can exceed the perturbative
  uncertainty already at moderate mass splittings.
\item The composition of the NLO corrections from different classes of
  diagrams and with it the dependence of the $K$ factors on the mass
  of the produced particles is significantly different for
  quark-antiquark vs gluonic initial states, \ie moving from Tevatron
  to LHC.
\end{itemize}
\bigskip

In addition to these specific conclusions we emphasize that with \mg
this kind of study can be easily repeated for any kind of heavy new
particles at the LHC. In the light of the available LHC results this
might be useful not only for general simplified models but also for
specific models outside the usual MSSM model and parameter space.  For
this purpose we include an appendix which provides all necessary
information on the infrared dipole subtraction as well as on a proper
on-shell subtraction as implemented in \mg.\bigskip

\begin{center} 
{\bf Acknowledgments}
\end{center}

We would like to thank Thomas Binoth, who started this project with us
and unfortunately cannot see its completion.  The work presented here
has been in part supported by the Concerted Research action
``Supersymmetric Models and their Signatures at the Large Hadron
Collider'' and the Research Council of the Vrije Universiteit Brussel and by the Belgian
Federal Science Policy Office through the Interuniversity Attraction
Pole IAP VI/11. DG acknowledges support by the International Max Planck
Research School for Precision Tests of Fundamental Symmetries.

\newpage
\appendix

%%%%%%%%%%%%%%%%%%%%%%%%%%%%%%%%%%%%%%%%%%%%%%%%%%%%%%%%%%%%%%%%%%%%%%%%
\section{Supersymmetric dipoles}
\label{sec:cs}

In this appendix we present the unintegrated and integrated dipoles
required for SUSY-QCD calculations~\cite{catani_seymour} including a
phase space constraint~\cite{alpha}. They are implemented as an
independent add-on to the {\sc MadDipole} package~\cite{maddipole} and are
part of the automated \mg framework.\bigskip

There exist two major approaches to remove soft and collinear
singularities: phase space slicing and phase space
subtraction~\cite{lecture}. A simple toy example captures their main
features and highlights the role of an FKS-like phase space
constraint~\cite{alpha}. Let us consider the dimensionally regularized
integral $\int_0^1 dx f(x)/x^{1-\epsilon}$ with $\epsilon>0$. Phase
space slicing based on a small parameter $\alpha$ yields
\begin{alignat}{5}
\int_0^1 dx \; \frac{f\left(x\right)}{x^{1-\epsilon}}
&= \int_\alpha^1 dx \; \frac{f\left(x\right)}{x^{1-\epsilon}}
 + \int_0^\alpha dx \; \frac{f\left(0\right)}{x^{1-\epsilon}}
 + \mathcal{O}\left(\alpha\right)
\notag \\
&= \int_\alpha^1 dx \; \frac{f\left(x\right)}{x}
 + \frac{f\left(0\right)}{\epsilon}
 + f\left(0\right)\log\alpha+\mathcal{O}\left(\alpha;\epsilon\right) \; .
\label{eq:phase_slicing}
\end{alignat}
It requires a choice of small enough $\alpha$ to reach a numerical plateau.
A numerically more stable approach is phase space subtraction, where
the same integral becomes
\begin{alignat}{5}
\int_0^1 dx \; \frac{f\left(x\right)}{x^{1-\epsilon}}
&= \int_0^1 dx \; \frac{f\left(x\right)-f\left(0\right)\Theta\left(x\leq\alpha\right)}
                      {x^{1-\epsilon}}
  + \int_0^\alpha dx \; \frac{f\left(0\right)}{x^{1-\epsilon}}
\notag \\
&= \int_0^1 dx \; 
   \frac{f\left(x\right)-f\left(0\right)\Theta\left(x\leq\alpha\right)}{x}
 + \frac{f\left(0\right)}{\epsilon}
 + f\left(0\right)\log\alpha+\mathcal{O}\left(\epsilon\right) \; .
\label{eq:subtraction_method}
\end{alignat}
Here the divergence is subtracted locally and the final result no
longer depends on $\alpha$. The logarithmic $\alpha$-dependence is
exactly cancelled in the total result.  This feature we can turn into
a sensitive numerical test when varying $0<\alpha\leq1$. For small
values of $\alpha$ we only need to evaluate part of the integrand of
Eq.\eqref{eq:subtraction_method}, which speeds up the
calculation.\bigskip

The toy model of Eq.\eqref{eq:subtraction_method} carries the essence of
the Catani-Seymour subtraction method. Real emission of quarks and
gluons $\left(d\sigma^\text{real}\right)$ leads to IR divergences
after an integration over the emission phase space. Its regularization
relies on a local subtraction term $\left(d\sigma^\text{A}\right)$
which reflects the universality of the soft and collinear limits. The
divergence cancels over the same $n$-particle phase space,
\begin{alignat}{5}
\delta\sigma^\text{NLO}
=\int_{n+1} \; 
 \left(d\sigma_{\epsilon=0}^\text{real}-d\sigma_{\alpha,\:\epsilon=0}^\text{A}
 \right)
+\int_{n} \;
 \left(d\sigma^\text{virtual}+d\sigma^\text{collinear}+\int_1d\sigma_\alpha^\text{A}
 \right)_{\epsilon=0} \; .
\label{eq:cs subtraction}
\end{alignat}
Below, we present the unintegrated dipoles $d\sigma_\alpha^\text{A}$
as well as the integrated dipoles $\int_1 d\sigma_\alpha^\text{A}$
including their $\alpha$ dependence. They are crucial for SUSY-QCD
processes or other NLO QCD predictions beyond the Standard Model. Our
extended set of massive Catani-Seymour dipoles with explicit $\alpha$
dependence has several practical advantages:
\begin{itemize}
\item tuning $\alpha$ we reduce the subtraction phase space and hence
  the number of events for which the real-emission matrix element and
  the subtraction fall into different bins; the so-called
  \emph{binning problem}.
\item choosing $\alpha<1$ we evaluate the subtraction terms only in
  the phase space region where they matter, \ie close to the IR
  divergences.
\item our final result should not depend on $\alpha$. This serves as a
  test for example of the adequate coverage of all the singularities
  or the relative normalization of the two-particle and three-particle
  phase space.
\end{itemize}

In the MSSM gluino and squark interactions induced by the covariant
derivatives $\bar{\go}\slashed{D} \go$, $|D_\mu \sq|^2$ give rise to
new IR divergences which are absent in the Standard Model. The
emission of a soft gluon from these particles requires new final-final
dipoles $D_{ij,k}$ and final-initial dipoles
$D_{ij}^a$. Initial-initial and initial-final configurations can also
have a squark or gluino as spectator, but the dipole only carries
information about the mass of the colored spectator, not about its
spin. This means we can simply use the massive Standard Model
dipoles~\cite{catani_seymour} with an extra SUSY particle in the
final state. To make this Appendix most useful we will firmly stick to
the conventions of Ref.~\cite{catani_seymour}.\bigskip

We start with a collection of formulas for final-final dipoles.  The 
expression for the unintegrated dipole is given by
\begin{alignat}{5}
D_{ij,k}=-\frac{1}{2p_i.p_j}
  \langle...,\widetilde{ij},...,\tilde{k},...|
  \frac{\mathbf{T}_k\mathbf{T}_{ij}}{\mathbf{T}_{ij}^2}
  \mathbf{V}_{ij,k}
 |...,\widetilde{ij},...,\tilde{k},...\rangle \;, 
\end{alignat}
where $|...,\widetilde{ij},...,\tilde{k},...\rangle$ represents the amplitude for the factorized born process,
which in the special case of the SUSY dipoles is made by the removal of the gluon from the diagonal
splitting $\sq(p_{ij})\to\sq(p_j) g(p_i)$. The color matrix $\mathbf{T}_k\mathbf{T}_{ij}/\mathbf{T}_{ij}^2$ acts on the
  born amplitude $|...,\widetilde{ij},...,\tilde{k},...\rangle$ giving the proper color factor.

To
compute the integrated dipoles we integrate over the one-particle
phase space $\left[dp_i(\tilde{p}_{ij},\tilde{p}_k)\right]$ with the
spin average matrices $\langle \mathbf{V}_{ij,k} \rangle$, according to
Eq.(5.22) of Ref.~\cite{catani_seymour}:
\begin{alignat}{5}
\int \left[dp_i\left(\tilde{p}_{ij},\tilde{p}_k\right)\right]
     \frac{1}{\left(p_i+p_j\right)^2-m_{ij}^2}
     \langle \mathbf{V}_{ij,k} \rangle
\equiv 
\frac{\alpha_s}{2\pi}
\frac{1}{\Gamma\left(1-\epsilon\right)}
\left(\frac{4\pi\mu^2}{Q^2}\right)^\epsilon
I_{ij,k}\left(\epsilon\right) \; ,
\end{alignat}
where the squark dipole function, $\langle s|\mathbf{V_{g\sq,k}}|s' \rangle$,
is given by Eq.(C.1) of the same reference,
\begin{alignat}{5}
\frac{\langle s|\mathbf{V}_{g\sq,k}|s' \rangle}
     {8\pi\mu^{2\epsilon}\alpha_sC_F} 
= 
\left[ \frac{2}{1-\tilde{z}_j\left(1-y_{ij,k}\right)}
      -\frac{\tilde{v}_{ij,k}}{v_{ij,k}}
       \left(2+\frac{m_{\sq}^2}{p_ip_j}\right)
\right] \delta_{ss'} 
= \frac{\langle \mathbf{V}_{g\sq,k} \rangle\delta_{ss'}}
       {8\pi\mu^{2\epsilon}\alpha_sC_F} \; .
\end{alignat}
Compared to a massive quark the squark structure is much simpler. This
is because for scalars the labels $s$ and $s'$ are merely tagging the
helicity of the associated quark partners without any effect on the
squark splitting.\bigskip

The integrated dipole $I_{g\sq,k}$ we decompose into an soft or
eikonal part $I^\text{eik}$ and a collinear integral
$I_{g\sq,k}^\text{coll}$ evaluated in $4-2\epsilon$ dimensions,
\begin{alignat}{5}
I_{g\sq,k}\left(\mu_{\sq},\mu_k;\epsilon\right)
&= C_F \left[ 2I^\text{eik}\left(\mu_{\sq},\mu_k;\epsilon\right)
            +I_{g\sq,k}^\text{coll}\left(\mu_{\sq},\mu_k;\epsilon\right)
      \right] \notag \\      
\tilde{v}_{g\sq,k} \; I^\text{eik}  
   &  = \frac{1}{2\epsilon}\log\rho
        - \log\rho\;
          \log\left(1-\left(\mu_{\sq}+\mu_k\right)^2\right)
         -\frac{1}{2}\log^2\rho_{\sq}-\frac{1}{2}\log^2\rho_{k} \notag \\
   &  +\zeta_2+2 \text{Li}_2\left(-\rho\right)
         -2 \text{Li}_2\left(1-\rho\right) 
         - \frac{1}{2} \text{Li}_2\left(1-\rho_{\sq}^2\right)
         - \frac{1}{2} \text{Li}_2\left(1-\rho_k^2\right)  \notag \\ 
I_{g\sq,k}^\text{coll} = \; &
         \frac{2}{\epsilon} 
        -\frac{1}{\epsilon \mu_{\sq}^{2\epsilon}} 
        -\frac{2}{\mu_{\sq}^{2\epsilon}}
      +6 -2\log \left(\left(1-\mu_k\right)^2-\mu_{\sq}^2\right)
     +\frac{4\mu_k\left(\mu_k-1\right)}{1-\mu_{\sq}^2-\mu_k^2} \; .
\label{eq:int_dipole}
\end{alignat}
The rescaled masses $\mu_n$ and the
variables $\rho$ and $\rho_n$ associated with the splitting
$\tilde{ij}\rightarrow i\: j$ and the spectator $k$ are defined in
terms of the final state momenta $p_i$, $p_j$ and $p_k$ as
\begin{alignat}{5}
\mu_n 
& = \frac{m_n}{\sqrt{(p_i+p_j+p_k)^2}}
\notag \\
\rho 
& = \sqrt{\frac{1-\tilde{v}_{ij,k}}{1+\tilde{v}_{ij,k}}}
\qqqquad \text{with} \quad 
\tilde{v}_{ij,k} = 
\frac{\sqrt{\lambda\left(1,\mu_{ij}^2,\mu_k^2\right)}}{1-\mu_{ij}^2-\mu_k^2} 
\notag \\
\rho_n \left(\mu_j,\mu_k\right) 
&=\sqrt{\frac{1-\tilde{v}_{ij,k}+2\mu_n^2/\left(1-\mu_j^2-\mu_k^2\right)}
{1+\tilde{v}_{ij,k}+2\mu_n^2/\left(1-\mu_j^2-\mu_k^2\right)}} 
\qquad \left(n=j,k\right)  \;,
\end{alignat}
with $\lambda$ denoting the K\"{a}llen function
\begin{alignat}{5}
\lambda\left(x,y,z\right)=x^2+y^2+z^2-2xy-2xz-2yz \;.
\end{alignat}
The splitting kinematics we describe using
\begin{alignat}{5}
\tilde{z}_j=1-\frac{p_ip_k}{p_ip_k+p_jp_k}
\qqquad \text{and} \qquad
y_{ij,k}
=\frac{p_ip_j}{p_ip_j+p_ip_k+p_jp_k} 
> y_+=1-\frac{2\mu_k\left(1-\mu_k\right)}{1-\mu_i^2-\mu_j^2-\mu_k^2} \; .
\end{alignat}
Just like for massive quarks there is no collinear singularity, so the
most divergent term in the $I_{g\sq,k}\left(\epsilon\right)$ is a
single $1/\epsilon$ pole.\bigskip 

To include the phase space parameter $\alpha$ into the massive
squark dipole we limit the dipole function to small values of
$y_{ij,k}/y_+$
\begin{alignat}{5}
D_{g\sq,k} \rightarrow 
D_{g\sq,k}\Theta\left(\frac{y_{ij,k}}{y_+} < \alpha\right)
\qqquad \alpha\:\epsilon\:(0,1] \; .
\label{eq:dipole_alpha}
\end{alignat}
For the integrated dipole
$I_{g\sq,k}\left(\epsilon\right)$ we start from
Eq.\eqref{eq:int_dipole} and subtract the finite term including the
same kinematic condition as Eq.\eqref{eq:dipole_alpha}
\begin{alignat}{5}
I_{g\sq,k}\left(\epsilon,\alpha\right) 
&= I_{g\sq,k} \left(\epsilon\right) 
 + \Delta I_{g\sq,k}\left(\alpha\right) 
\notag \\
&= I_{g\sq,k}\left(\epsilon\right) 
 - \frac{2\pi}{\alpha_s}
   \int \left[dp_g\left(\tilde{p}_{g\sq},\tilde{p}_k\right)\right]
       \; \frac{\langle \mathbf{V}_{g\sq,k} \rangle}{2p_{g}p_{\sq}}
       \; \Theta\left(\frac{y_{g\sq,k}}{y^+} > \alpha\right) \; .
\label{eq:I_ijk_alpha}
\end{alignat}
The finite part we can evaluate in four dimensions, because by
definition there exists no divergence in the region $y_{g\sq,k}/y^+ >
\alpha$. The eikonal part
$2/[1-\tilde{z}_{\sq}\left(1-y_{g\sq,k}\right)]$ is the same for
$\langle s|\mathbf{V}_{gQ,k}|s' \rangle$ and $\langle
s|\mathbf{V}_{g\sq,k}|s' \rangle$, so in  Eq.\eqref{eq:I_ijk_alpha} we insert
Eq.\eqref{eq:int_dipole} from our appendix  and  Eq.(A.9) from Ref.~\cite{bevilacqua},

\begin{alignat}{5}
\tilde{v}_{g\sq,k}\;\Delta I^\text{eik}(\alpha)
=& -\text{Li}_2\left(\frac{a+x}{a+x_+}\right)+\text{Li}_2\left(\frac{a}{a+x_+}\right)
      +\text{Li}_2\left(\frac{x_+-x}{x_+-b}\right)-\text{Li}_2\left(\frac{x_+}{x_+-b}\right) \notag \\
  & +\text{Li}_2\left(\frac{c+x}{c+x_+}\right)-\text{Li}_2\left(\frac{c}{c+x_+}\right)
      +\text{Li}_2\left(\frac{x_--x}{x_-+a}\right)-\text{Li}_2\left(\frac{x_-}{x_-+a}\right)    \notag \\
  &  -\text{Li}_2\left(\frac{b-x}{b-x_-}\right)+\text{Li}_2\left(\frac{b}{b-x_-}\right)     
       -\text{Li}_2\left(\frac{x_--x}{x_-+c}\right)+\text{Li}_2\left(\frac{x_-}{x_-+c}\right)    \notag \\     
  &  +\text{Li}_2\left(\frac{b-x}{b+a}\right)-\text{Li}_2\left(\frac{b}{b+a}\right)     
       -\text{Li}_2\left(\frac{c+x}{c-a}\right)+\text{Li}_2\left(\frac{c}{c-a}\right)                    \notag \\     
  &  +\log\left(c+x\right)\log\left(\frac{\left(a-c\right)\left(x_+-x\right)}{\left(a+x\right)\left(c+x_+\right)}\right)     
       -\log\left(c\right)\log\left(\frac{\left(a-c\right)x_+}{a\left(c+x_+\right)}\right)              \notag \\
  &  +\log\left(b-x\right)\log\left(\frac{\left(a+x\right)\left(x_--b\right)}{\left(a+x\right)\left(x_--x\right)}\right) 
       -\log\left(b\right)\log\left(\frac{a\left(x_--b\right)}{\left(a+b\right)x_-}\right)              \notag \\             
  &  -\log\left(\left(a+x\right)\left(b-x_+\right)\right)\log\left(x_+-x\right)        
      +\log\left(a\left(b-x_+\right)\right)\log\left(x_+\right)                                                     \notag \\
  &  +\log\left(d\right)\log\left(\frac{\left(a+x\right)x_+x_-}{a\left(x_+-x\right)\left(x_--x\right)}\right)       
       +\log\left(\frac{x_--x}{x_-}\right)\log\left(\frac{c+x_-}{a+x_-}\right)                             \notag \\    
  &  +\frac{1}{2}\log\left(\frac{a+x}{a}\right)\log\left(a\left(a+x\right)\left(a+x_+\right)^2\right),                             
\end{alignat}
where
\begin{alignat}{5}
a &=\frac{2\mu_k}{1-\mu_{\sq}^2-\mu_k^2} \;, \qqqquad
   b =\frac{2\left(1-\mu_k\right)}{1-\mu_{\sq}^2-\mu_k^2} \;, \\
c &=\frac{2\mu_k\left(1-\mu_k\right)}{1-\mu_{\sq}^2-\mu_k^2} \;, \qqqquad
d =\frac{1}{2}\left(1-\mu_{\sq}^2-\mu_k^2\right) \;, \\
\end{alignat}
and
\begin{alignat}{5}
x_{\pm}=\frac{\left(1-\mu_k\right)^2-\mu_{\sq}^2
      \pm\sqrt{\lambda\left(1,\mu_{\sq}^2,\mu_k^2\right)}}{1-\mu_{\sq}^2-\mu_k^2}\;.
\end{alignat}
The collinear part is different for squarks, so we supplement its form
in Eq.\eqref{eq:int_dipole} by
\begin{alignat}{5}
\Delta I_{g\sq,k}^\text{coll}\left(\alpha\right)
= - \frac{C_F}{2\pi^2}
  \left[ \frac{\left(1-\mu_k\right)^2-\mu_{\sq}^2}{1-\mu_{\sq}^2-\mu_k^2}
         \left(1-\alpha\right)
        +\log\alpha
  \right] \; .
\end{alignat}
\bigskip

Following the same logic we tackle the final-initial dipoles. The
final-initial dipole function is given by Eq.(C.3) of
Ref.~\cite{catani_seymour},
\begin{alignat}{5}
\langle \mathbf{V}_{g\sq}^a \rangle
= 8\pi\mu^{2\epsilon}\alpha_sC_F
  \left( \frac{2}{2-x_{g\sq,a}-\tilde{z}_{\sq}}-2-\frac{m_{\sq}^2}{p_{g}p_{\sq}}\right)
\; .
\end{alignat}
The integrated dipole function $I_{g\sq}^a$ becomes
\begin{alignat}{5}
I_{g\sq}^a\left(x;\epsilon\right)
= C_F
  \left[ \left(J_{g\sq}^a\left(x,\mu_{\sq}\right)\right)_+
        + \delta \left(1-x\right) 
          \left( J_{g\sq}^{a;S}\left(\mu_{\sq};\epsilon\right)
                +J_{g\sq}^{a;NS}\left(\mu_{\sq}\right)
          \right)
  \right]
+\mathcal{O}\left(\epsilon\right) \; ,
\end{alignat}
with the three contributions $I_{g\sq}^a$ 
\begin{alignat}{5}
\left[J_{g\sq}^a\left(x,\mu_{\sq}\right)\right]_+ 
&= -2 \left( \frac{1+\log\left(1-x+\mu_{\sq}^2\right)}{1-x} \right)_+
   +\left(\frac{2}{1-x}\right)_+\log\left(2+\mu_{\sq}^2-x\right)
\notag \\
J_{g\sq}^{a;S}\left(\mu_{\sigma};\epsilon\right) 
&= \frac{1}{\epsilon^2}
   -\frac{\pi^2}{3}
   -\frac{1}{\mu_{\sq}^{2\epsilon}} \; 
    \left( \frac{1}{\epsilon^2}
          +\frac{1}{\epsilon}
          +\frac{\pi^2}{6}+2
    \right)
   -\frac{\log\left(1+\mu_{\sq}^2\right)}{\epsilon} 
   +\frac{2}{\epsilon}+4-\frac{\pi^2}{6}
\notag \\
J_{g\sq}^{a;NS}\left(\mu_{\sq}\right) 
&= \frac{\pi^2}{3}
   -2\text{Li}_2\left(\frac{1}{1+\mu_{\sq}^2}\right)
   -2\text{Li}_2\left(-\mu_{\sq}^2\right)
   -\frac{1}{2}\log^2\left(1+\mu_{\sq}^2\right) \; .
\end{alignat}
In analogy to the final-final case of Eqs.\eqref{eq:dipole_alpha} and
\eqref{eq:I_ijk_alpha} we introduce a phase space cutoff 
\begin{alignat}{5}
D_{g\sq}^a 
& \rightarrow D_{g\sq}^a\Theta\left(\alpha-1+x_{g\sq,a}\right)
\notag \\
\Delta I_{g\sq}^a\left(\alpha\right) 
&= - C_F \; \frac{\Theta\left(1-\alpha-x\right)}{1-x}
           \left[ -2+2\log\left(1+\frac{1}{1+\mu_{\sq}^2-x}\right) \right] \; ,
\end{alignat}
where the kinematic variable $x_{ij,a}$ is given by
\begin{alignat}{5}
x_{ij,a}=\frac{p_ap_i+p_ap_j-p_ip_j+\dfrac{m_{ij}^2-m_i^2-m_j^2}{2}}
             {p_ap_i+p_ap_j}\; .
\end{alignat}
\bigskip

\begin{figure}[t]
\includegraphics[width=0.4\textwidth]{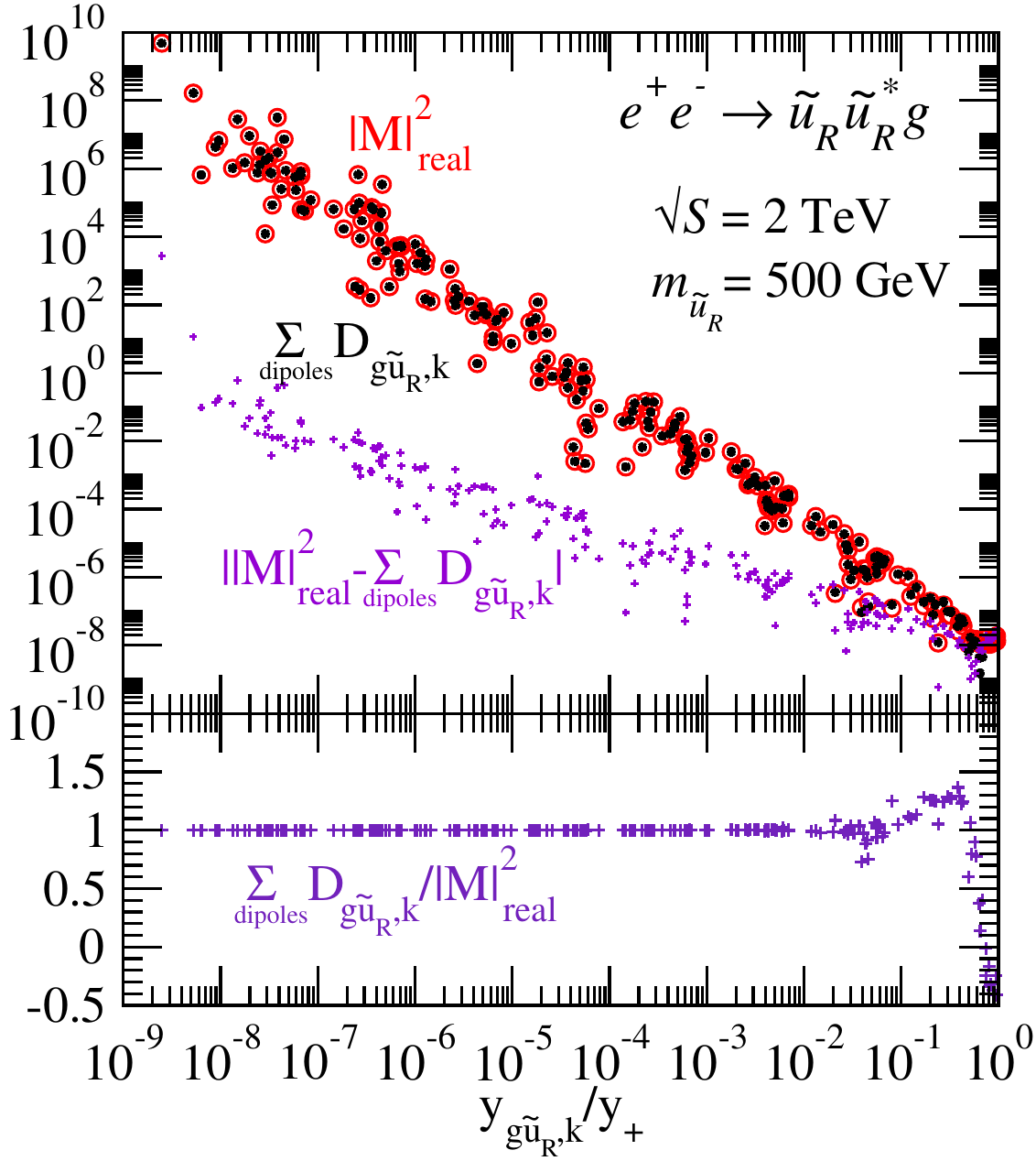}
\hspace*{0.1\textwidth}
\includegraphics[width=0.4\textwidth]{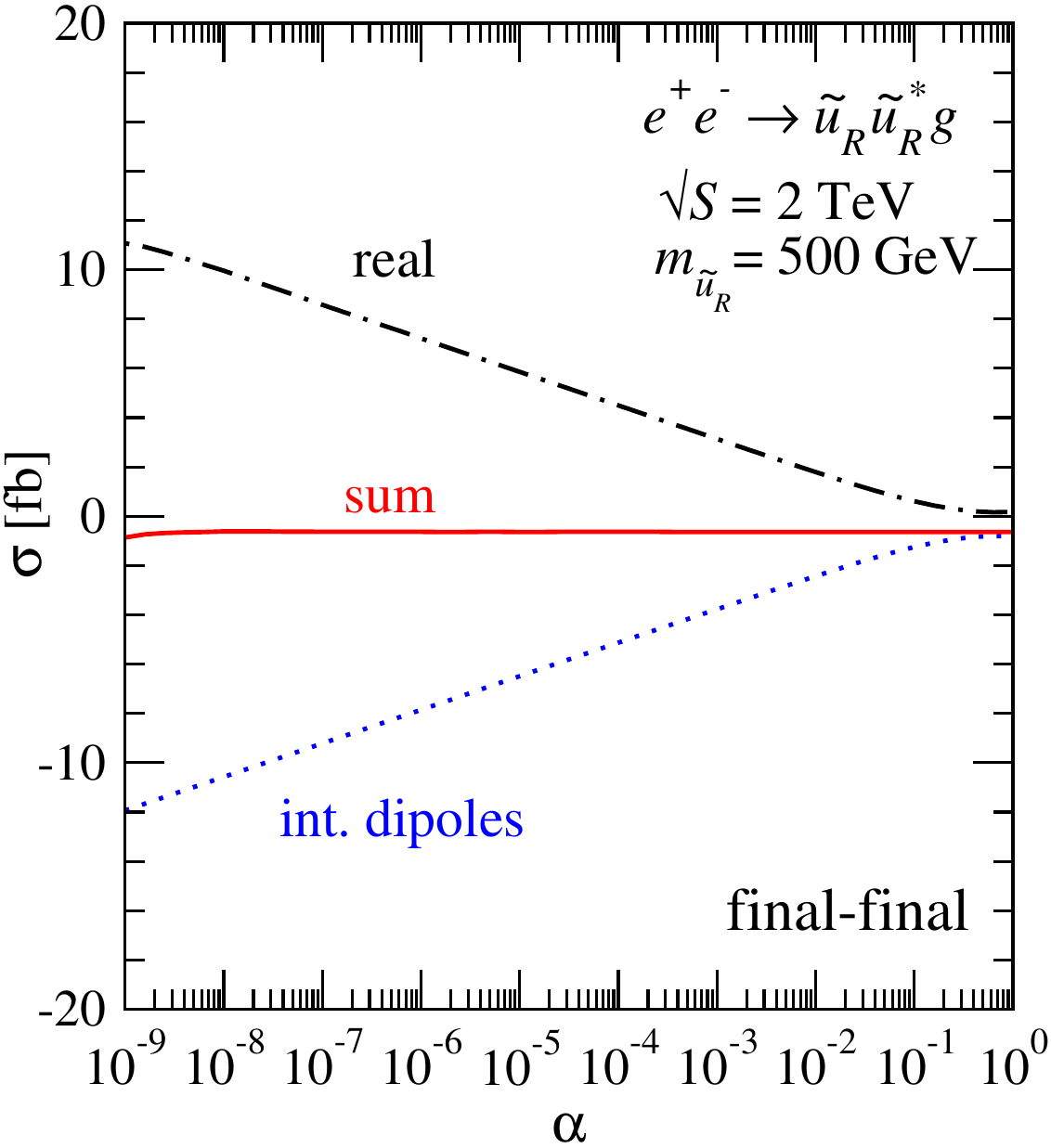}
\caption{Left: real emission matrix element (red circles) and the
  dipole subtraction (black crosses inside) towards the soft limit
  $y_{g\sq,k}\rightarrow0$. Right: $\alpha$ dependence for final-final
  squark dipoles.}
\label{fig:Local-dipole-subtraction-plot}
\end{figure}

As an example for numerous numerical tests of our dipole
implementation, we discuss soft gluon emission off the hard process
$e^+ e^- \rightarrow \sur \sur^*$.  In the left panel of
Fig.~\ref{fig:Local-dipole-subtraction-plot} we show how the dipole
subtraction cancels the IR divergence locally, \ie point by point. The
numerical agreement of the real emission matrix element with the
dipole subtraction term improves for softer gluons. In the soft limit
both terms grow as $1/E_g^2$.  Even though we find
$|\mathcal{M}_\text{real}^2-\sum\limits_{\rm dipoles}\, D_{g\sq,k}| \sim 1/E_g$ the phase space
factor $E_g dE_g$ cancels this dependence.

%------------------------------------------------
%------------------------------------------------

In the right panel of Fig.~\ref{fig:Local-dipole-subtraction-plot} we
show the $\alpha$ dependence for the final-final squark dipole.  Both,
the real emission and the integrated dipole, depend separately on
$\alpha$. Their sum is numerically stable over many orders of
magnitude and down to $\alpha=10^{-9}$.

%%%%%%%%%%%%%%%%%%%%%%%%%%%%%%%%%%%%%%%%%%%%%%%%%%%%%%%%%%%%%%%%%%%%%%%%
\section{On-shell subtraction}
\label{sec:os}

%------------------------------------------------
\begin{figure}[b]
\includegraphics[width=0.9\textwidth]{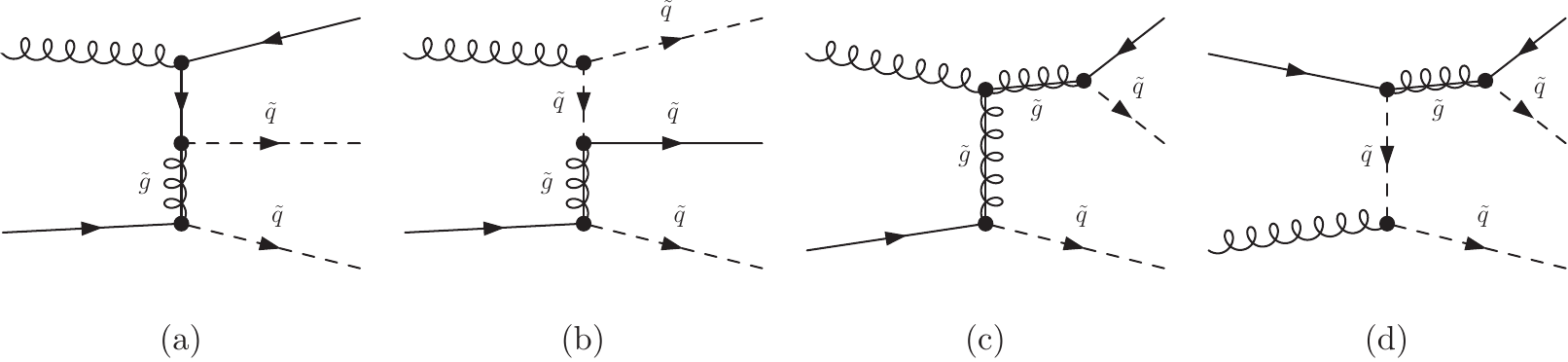} 
\caption{Sample diagrams for the real-emission corrections to squark
  pair production with an additional quark in the final state.}
\label{fig:os_diagrams}
\end{figure}
%------------------------------------------------

From single top production it is well known that when including NLO
corrections we have to avoid double counting of diagrams which are
attributed to different physics processes.  As an example we consider
real emission corrections to squark pair production $pp\rightarrow
\sq\sq$: the partonic sub-channels with an additional quark in the
final state $qg\rightarrow\sq\sq\bar{q}$ display a peculiar behavior
which we illustrate in Fig.~\ref{fig:os_diagrams}.  The diagrams (a)
and (b) are part of the genuine NLO corrections to squark pair
production. In contrast, the diagrams (c) and (d) can be interpreted in
two ways:
\begin{alignat}{5}
 qg &\rightarrow \sq\go^{(*)} \rightarrow \sq\sq\bar{q} 
\qqqquad &&\text{squark pair production} 
\notag \\
 qg &\rightarrow \sq\go \times \text{BR}(\go\rightarrow \sq\bar{q}) 
\qqqquad &&\text{squark--gluino production}
\end{alignat}
The first interpretation simply assumes NLO corrections to the hard
process $pp(qg)\rightarrow\sq\sq$ and is generally valid for on-shell and
off-shell gluinos. The second interpretation points to the LO process
for $qg \rightarrow \sq \go$ followed by the branching
$\text{BR}(\go\rightarrow \sq \bar{q})$ and implicitly assumes an
on-shell gluino. For a mass hierarchy $m_{\go}>m_{\sq}$ we can
therefore separate the two assignments into off-shell and on-shell
gluinos. This distinction avoids double counting and is the basis of
our on-shell subtraction scheme. Approaches to tackle this problem
include
\begin{itemize}
\item a slicing procedure which separates the phase space related to
  the on-shell emissions and removes the on-shell divergence by
  requiring $|\sqrt{s_{\sq\bar{q}}}-m_{\go}|>\delta$~\cite{Belyaev}.
  Phase space methods of this kind do not offer a cancellation of the
  $\delta$ dependence and do not act locally in phase space. Moreover,
  as a pure phase space approach it does not allow for a proper
  separation into finite, on-shell and interference contributions
  which is crucial for a reliable rate prediction. 
\item diagram removal where the resonant diagrams are removed
  by hand. In lucky cases this method might work in the limit
  $\Gamma/m \ll 1$~\cite{mcnlo}, but it ignores any kind of
  interference contributions which do not actually have to vanish in
  narrow width limit. This scheme is theoretically poorly motivated in
  many ways.
\item local on-shell subtraction in the so-called {\sc Prospino}
  scheme~\cite{squarkgluinoNLO,on-shell} which under the name `diagram
  subtraction' is also used in the single-top computation of {\sc
    Mc@nlo}.  This method used in \mg.
\end{itemize}   

\begin{figure}[t]
\includegraphics[width=0.14\textwidth]{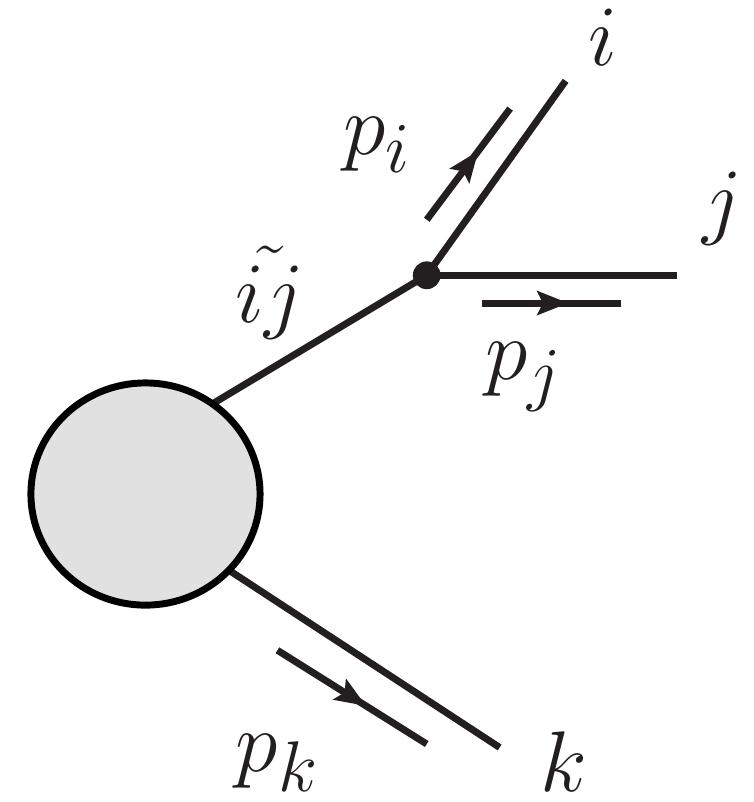} 
\caption{Kinematic variables for the on-shell subtraction.}
\label{fig:os_kinematics}
\end{figure}

To define the on-shell subtraction we split the contributions of the
matrix element in two parts: the first piece concerns the resonant
diagrams (c) and (d) and is denoted as $\mathcal{M}_\text{res}$, while
the second piece represents the non-resonant (remnant) diagrams (a) and (b) as
$\mathcal{M}_\text{rem}$. Note that this separation is defined at the
amplitude level and not based on the amplitude squared. The full
matrix element squared becomes
\begin{alignat}{5}
|\mathcal{M}|^2=
  |\mathcal{M}_\text{res}|^2 
+ 2\text{Re} (\mathcal{M}_\text{res}^*\mathcal{M}_\text{rem}) 
+ |\mathcal{M}_\text{rem}|^2.
\label{eq:matrix_element}
\end{alignat}
The divergent propagator in $\mathcal{M}_\text{res}$ we regularize as a
Breit-Wigner propagator
\begin{alignat}{5}
\frac{1}{p_{ij}^2-m_{ij}^2} \rightarrow 
\frac{1}{p_{ij}^2-m_{ij}^2+im_{ij}\Gamma_{ij}} \; ,
\end{alignat}
where $m_{ij}$ is the mass of the mother particle in the splitting
$\widetilde{ij}\rightarrow i\,j$, as shown in
Fig.~\ref{fig:os_kinematics}, and $\Gamma_{ij}$ is a regulator.\bigskip

%------------------------------------------------
%------------------------------------------------

As explained above, a possible double counting is limited to the
on-shell configuration in $|\mathcal{M}_\text{res}|^2$.  To remove it
we define a local subtraction term $d\sigma^\text{OS}$ and include it
in complete analogy to the Catani-Seymour dipole subtraction
Eq.\eqref{eq:cs subtraction}, such that the total cross section is
given by
\begin{alignat}{5}
\delta\sigma^\text{NLO}
=\int_{n+1} 
 \left( d\sigma_{\epsilon=0}^\text{real}
       -d\sigma_{\alpha,\:\epsilon=0}^\text{A}
       -d\sigma_{\epsilon=0}^\text{OS}
 \right)
+\int_{n}
 \left(d\sigma^\text{virtual}
      +d\sigma^\text{collinear}
      +\int_1 d\sigma_\alpha^\text{A}\right)_{\epsilon=0} \; .
\label{eq:os subtraction}
\end{alignat}
The extra subtraction term $d\sigma^\text{OS}$ is
$|\mathcal{M}_\text{res}|^2$ with its momenta remapped to the on-shell
kinematics,
\begin{alignat}{5}
d\sigma^\text{OS}
= \Theta(\hat{S}-(m_{ij}+m_k)^2) \;
  \Theta(m_{ij}-m_i - m_j) \;
  \dfrac{\dfrac{1}{(p_{ij}^2-m_{ij}^2)^2 + m_{ij}^2\Gamma_{ij}^2}}
        {\dfrac{1}{m_{ij}^2\Gamma_{ij}^2}} \;
%  \frac{\mathcal{BW}(s_{ij})}{\mathcal{BW}(m_{ij}^2)} \; 
  |\mathcal{M}_\text{res}|^2 \Bigg|_\text{remapped} \; .
\label{eq:dsigmaos}
\end{alignat}
The kinematic configuration is depicted in
Fig.~\ref{fig:os_kinematics}.  The two step functions in
Eq.\eqref{eq:dsigmaos} ensure that the partonic center-of-mass energy
is sufficient to produce the intermediate on-shell particle and that
it can decay on-shell into the two final state particles.  The ratio
of the Breit-Wigner functions ensures that the subtraction has the
same profile as the original $|\mathcal{M}_\text{res}|^2$ over the
entire phase space. In the small width limit this ratio reproduces the
delta distribution which factorizes the $2 \to 3$ diagrams into
$\sigma \times \text{BR}$.

Note that this method works with a mathematical regulator
$\Gamma_{ij}$ which can be related to the physical width as in the
{\sc Mc@nlo} implementation; alternatively we can go into the
well-defined limit $\Gamma_{ij} \ll m_{ij}$ used in the original {\sc
  Prospino} implementation.\bigskip

%------------------------------------------------
\begin{figure}[ht]
\includegraphics[width=0.5\textwidth]{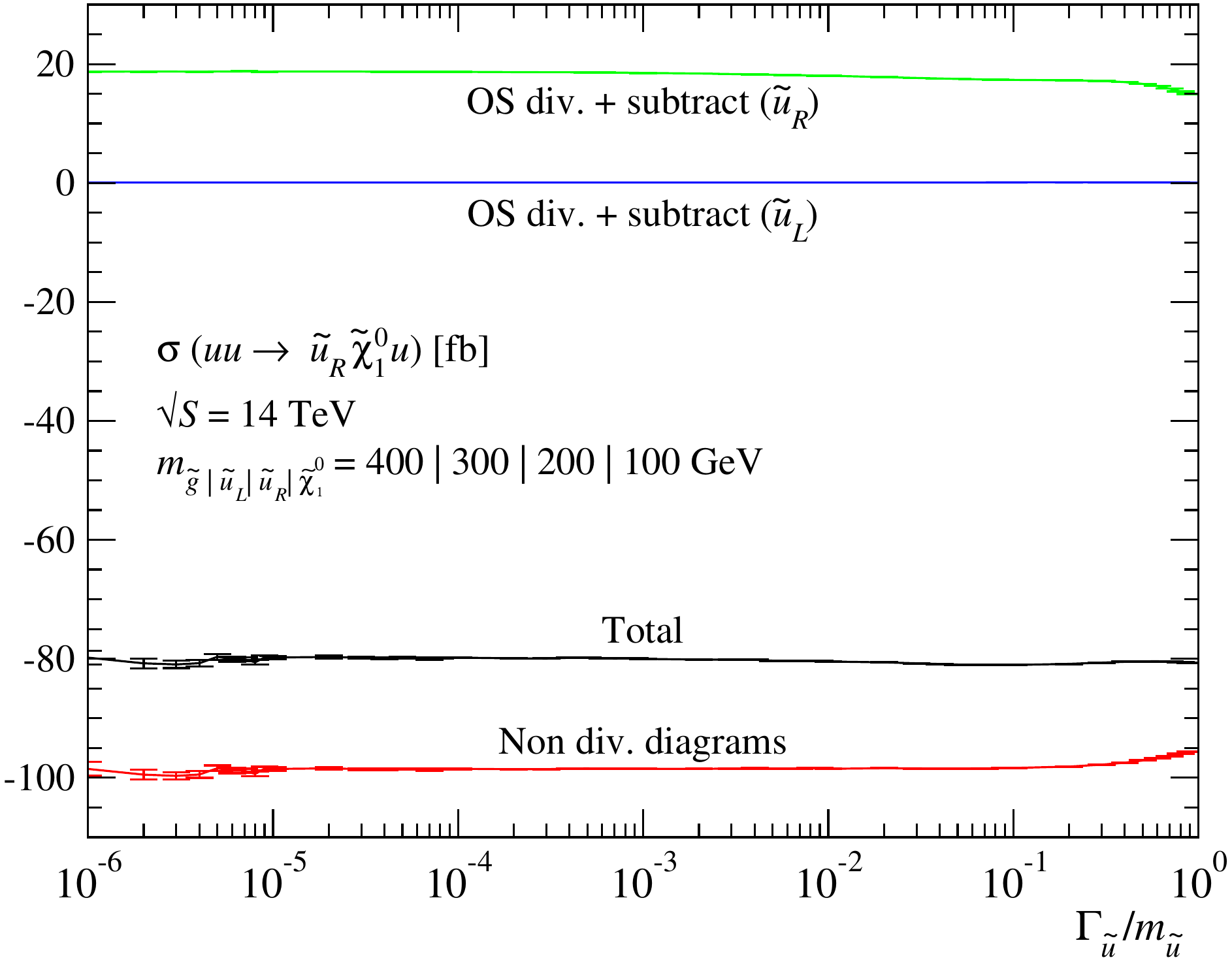} 
\caption{NLO contributions from intermediate on-shell particles in $u
  u \to \sur \nz{1}+X$ production as a function of
  $\Gamma_{\su}/m_{\su}$. The squark width acts as a cutoff in the
  {\sc Prospino} subtraction scheme~\cite{squarkgluinoNLO,on-shell}.
  The masses are chosen to illustrate all different resonant channels;
  virtual corrections are not included.}
\label{fig:onshell}
\end{figure}
%------------------------------------------------

In summary, this on-shell subtraction implemented in \mg exhibits
several attractive features when it comes to prediction of total and
differential cross sections. First, it subtracts all on-shell
divergences point-by-point over the entire phase space. This means
that all distributions are automatically safe.  Second, it preserves
gauge invariance at least in the narrow-width limit.  Third, it takes
into account spin correlations, because it includes the full
$2\rightarrow3$ matrix element. Fourth, it keeps track of the
interference of the resonant and non-resonant terms, $2 \text{Re}
(\mathcal{M}_\text{res}^*\mathcal{M}_\text{rem})$, which can be
numerically sizable. Finally, Fig.~\ref{fig:onshell} shows that it
smoothly interpolates between a finite width $\Gamma_{ij}/m_{ij} \sim
0.1$ and the narrow-width limit $\Gamma_{ij}/m_{ij} \rightarrow
0$.

%%%%%%%%%%%%%%%%%%%%%%%%%%%%%%%%%%%%%%%%%%%%%%%%%%%%%%%%%%%%%%%%%%%%%%%%
\section{One-loop amplitudes}
\label{sec:virtual}

The virtual corrections module of \mg is based on the
Feynman-diagrammatic approach defined in Ref~\cite{golem}.  It
calculates the virtual corrections to any $2\to2$ process such that it
can be applied to the Standard Model, MSSM, or any other
renormalizable theory.

The module uses {\sc Qgraf}~\cite{qgraf}, {\sc Form}~\cite{form}, {\sc
  Maple}, and the {\sc Golem95} integral library~\cite{golem_lib} to
produce {\sc Fortran90} code that calculates the virtual cross section
for a $2\to2$ process, given a set of phase space points.  It also
produces analytical {\sc Maple} output in the form of partial
amplitudes.  Each virtual diagram is broken down according to its color
structure, helicity, and scalar integrals.  This allows for a careful
test before the numerical calculation is even started.  The approach
can be divided into four main steps:
\begin{enumerate}
\item generate the tree level diagrams, counterterms, and one-loop
  diagrams with {\sc Qgraf} and translate the output into {\sc Form}
  code suitable for symbolic manipulation. The analytic structures
  keep track of external wave functions, vertex couplings and internal
  propagators, color factors, Lorentz structure, and the overall sign
  from external fermions.
\item map the analytic structures onto partial amplitudes using a
  basis in color, helicity and tensor structures using the
  spinor-helicity formalism.
\item apply an analytical reduction of tensor integrals to scalar loop
  integrals based on the {\sc Golem} reduction
  scheme~\cite{golem,golem_lib}.
\item collect all results and insert the renormalization constants
  into the counterterms.  The final output is available as
  analytical partial amplitudes in {\sc Maple} and as numerical {\sc
    Fortran90} output. The latter is implemented into the {\sc
    Madgraph} structure.
\end{enumerate}
These four blocks are coordinated and run by the {\sc Perl} script
\emph{run\_golem.pl}. We will describe them in detail below.\bigskip

In the first step {\sc Qgraf} generates all Born diagrams,
counterterms, and QCD one-loop corrections for a hard process
specified in a {\sc MadGraph}-like file \emph{proc\_card.dat}.  The
topological rules {\sc Qgraf} obtains from modified {\sc MadGraph}
model files. In addition to the familiar {\sc MadGraph} options, we
include novel functionalities specific to a NLO calculation; for
instance, the flag \emph{nlotype} in the process card generates either
pure QCD (gluon mediated), or full SUSY-QCD virtual corrections. This
division relies on a constrained set of propagators within {\sc
  Qgraf}. Note that SUSY-QCD corrections also include loop diagrams
which do not involve either gluons or gluinos, \eg mediated by the
four-squark vertex.  Therefore, {\sc Qgraf} first includes all loop
diagrams with gluons, gluinos and squarks as well as Faddeev-Popov ghosts. The
number of loop diagrams is reduced later by checking the order of
$\alpha_s$.

Counterterms are generated automatically by {\sc Qgraf} via tree-level
diagrams containing place-holders for all renormalization
constants. These renormalization constants depend on $\mathcal{O}(\alpha_s)$
corrections to a set of two-point functions involving the different colored
particles present within a given model. We provide them as a set
of separated (model-dependent)
libraries, implemented as {\sc Maple} list files.

Additional topological constraints, \eg requiring only gluonic
$t$-channel contributions, or only self-energy corrections, can be
included via \emph{run\_golem.pl}. At this stage the code groups
topologically equivalent structures and applies loop filtering
techniques \eg removing diagrams which are trivially zero. The {\sc
  Perl} script also assigns the overall sign to each diagram, because
standard {\sc Qgraf} is not valid for Majorana fermions.

Once the {\sc Qgraf} output is filtered by \emph{run\_golem.pl}, the
remaining set of diagrams is processed in {\sc Form} to apply Feynman
rules. Assigning the correct fermion flow is crucial for diagrams with
Majorana particles~\cite{majoranaflip}.\bigskip

The second step of \mg treats the color and helicity structure of the
{\sc Qgraf} and {\sc Form} output. For the QCD structure of each
Feynman diagram {\sc Form} uses a color-flow
decomposition~\cite{colorflow}. Each external gluon is matched with an
adjoint generator $T^a_{ij}$, which means we can re-write the gauge
structure using delta functions in color-space, and factorize it from
the remaining amplitude. This way the color flow within the amplitude
becomes more apparent.

The spin structure of each diagram is also manipulated in {\sc Form}.
Using the spinor-helicity formalism~\cite{helicity,Dixon:1996wi} the
amplitude is projected onto a set of helicity amplitudes. Each fermion
pair and vector boson is re-written as massless spinor products, of
which we take the traces.  This way each diagram is expressed in the
Mandelstam variables $s$, $t$ and $u$, with a spinor product
prefactor. Massive spinors require a helicity projection in the
direction of an auxiliary reference vector which we choose to be the
light-like momentum of one of the other external particles
$k_i$. Whenever this is not allowed we instead use $k_5 =
(E_1+E_2)(1,1,0,0)/2$. In this case the kinematic structure is
no longer defined by the usual Mandelstam variables, so we also
include $s_{15}$, $s_{25}$, $s_{35}$, $s_{45}$, and
$i\epsilon_{\mu\nu\rho\sigma}k_1^\mu k_2^\nu k_3^\rho
k_5^\sigma$. This additional reference vector seriously impacts the
tensor reduction described below and slows down the amplitude
generation and evaluation.\bigskip

In step three of \mg we simplify the loop diagrams using the {\sc
  Golem} approach~\cite{golem,golem_lib}. For a fully analytical
reduction of tensor integrals to a linear combination of scalar
integrals we rely on a combination of {\sc Form} and {\sc Maple}.  All
one-loop integrals are regularized dimensionally; internal momenta and
gamma matrices are split into four-dimensional and
$(-2\epsilon)$-dimensional components, with the latter only
contributing at $\mathcal{O}(\epsilon)$~\cite{Reiter:2009kb}. Tensor
loop integrals we simplified using a Passarino-Veltman
reduction~\cite{passvelt}. This reduces an $N$-point tensor integral
of rank $r$ to a scalar $N$-point integral plus a series of integrals
with fewer external legs and reduced rank. The final result we can
express in terms of known scalar integrals ($D_0,C_0,B_0,A_0$) in
$4-2\epsilon$ dimensions.  Their divergence structure is simple: IR
poles arise purely from $D_0$ and $C_0$, UV poles arise purely from
$B_0$ and $A_0$. The only exception to this rule is the scalar
two-point function
\begin{alignat}{5} 
B_0(0,0,0) 
= \int\,\frac{d^D\,q}{(2\pi)^{D}}\, \frac{1}{q^2\,(q+p)^2}
= \frac{i}{16\pi^2} \; 
  \frac{(4\pi)^\epsilon}{\Gamma(1-\epsilon)} \;
  \left(\frac{1}{\epsilon_\text{UV}}-\frac{1}{\epsilon_\text{IR}}\right) \; ,
\label{eq:beta0}
\end{alignat}
which in this schematic notation is IR and UV divergent. Its finite
part vanishes, but in our calculation we need to keep track of its IR
and UV poles separately.  As mentioned above, four-point tensor
integrals with $k_5 \cdot l$ 
in the numerator ($l$ standing for the internal loop momentum) cannot be simplified
using the Passarino-Veltman approach and are kept as un-reduced form
factors to be numerically processed by {\sc Golem95}.\bigskip

In the final step we collect all partial amplitudes for a given
process using {\sc Maple}. Two analytical output files contain all
information about the Born amplitude and the renormalized virtual
amplitude:
\begin{itemize}
\item \emph{AMP\_TREE.mapout} lists the total non-zero Born
  amplitudes, sorted by diagram, helicity, and color representation.
  If the flag \emph{nlosymsimp} is enabled in the {\sc MadGraph}
  process card, the helicity amplitudes are tested for the possible
  symmetry $\mathcal{M}^{\{\lambda\}} = \mathcal{M}^{\{\lambda'\}*}$,
  where $\{\lambda'\}$ is a different helicity from $\{\lambda\}$.
  Only the minimal set of helicity amplitudes is kept, along with a
  note of which helicities are conjugates of which.  This greatly
  reduces the size of the output for pure QCD or QED processes.
\item \emph{AMP\_LOOP.mapout} lists all finite loop amplitudes as
  kinematic coefficients sorted by diagram, helicity, color
  representation, and type (scalar integral, form factor, or number).
  In the same format it also lists the counterterms after the
  renormalization constants have been inserted. The simplification
  flag \emph{nlosymsimp} is also applied to the loop amplitudes.
\end{itemize}

For a numerical evaluation we do not rely on this analytic
output. Instead, \mg writes several {\sc Fortran90} routines for the
computation of the virtual corrections:
\begin{itemize}
\item \emph{libcoeffs\_all.so} and \emph{libcoeffs\_all\_tree.so}
  contain the amplitude coefficients for the virtual corrections and
  Born amplitudes. For size reasons we generate a separate library for
  each partial amplitude. Each library we pre--compile before linking
  them dynamically and launching them at runtime.
\item \emph{golem(k,mu,amplitude\_array)} takes the external
  four-momenta $k_{1,2,3,4}$ and the renormalization scale $\mu$ and
  returns
  \begin{alignat}{5}
  \emph{amplitude\_array} 
  = \left( a_0,
          \frac{a_{-1}}{\epsilon},
          \frac{a_{-2}}{\epsilon^2},|\mathcal{M}|_\text{Born}^2,a_\text{UV}
    \right) \;.
  \end{alignat}
  The different $a_j$ are defined through
  the interference term between Born and virtual amplitude,
  \begin{alignat}{5}
  |\mathcal{M}|_\text{1-loop}^2 
  = 2 \text{Re}\left(\mathcal{M}_\text{born}^*\mathcal{M}_\text{virt} \right) 
  = a_0+\frac{a_{-1}}{\epsilon}+\frac{a_{-2}}{\epsilon^2} \; ,
  \end{alignat}
  and correspond to the finite contribution ($a_0$), and the
  coefficients of the single ($a_{-1}$) and double ($a_{-2}$) IR
  poles.  The Born term is included for comparison with {\sc
    MadGraph}, and $a_\text{UV}$ returns the numerical value of the UV
  pole which is zero after proper renormalization. All results are
  averaged over initial state colors and helicities.
\item \emph{virtual\_corrections.f90} contains the subroutine
  \emph{golem(k,mu,amplitude\_array)}, which calls the integral
  library {\sc Golem95} and the coefficient libraries
  \emph{libcoeffs*.so}. It calculates the fully renormalized matrix
  element at one-loop level.
  The three values $a_0$, $a_{-1}$, $a_{-2}$ are combined with the integrated dipoles
  from the real emission corrections for the complete NLO corrections
  to a $2\to2$ process. 
The cancellation of the single and double poles is automatically 
checked in our numerical implementation.
\end{itemize}

As is appropriate for an automized NLO tool we have undertaken an
exhaustive program of checks to ensure the robustness and reliability
of our \mg.  We have calculated the (SUSY)-QCD one-loop corrections
for a large set of $2\to 2$ processes both in the Standard Model and the MSSM,
covering all representative possibilities of spins, color charges,
interaction patterns and topologies.  The cancellation of the UV, IR
and OS divergences, as well as the gauge invariance of the overall
result, can be confirmed numerically, for some specific cases also
analytically. The finite renormalized one-loop amplitudes we have
systematically compared with {\sc FeynArts}, {\sc FormCalc} and {\sc
  LoopTools}~\cite{feynarts}.

%%%%%%%%%%%%%%%%%%%%%%%%%%%%%%%%%%%%%%%%%%%%%%%%%%%%%%%%%%%%%%%%%%%%%%%%
\section{Renormalization}
\label{sec:cts}

As discussed in the previous appendix, we automatically generate the
ultraviolet counter terms using the tree-level amplitude from {\sc
  Qgraf}. As an input we express all renormalization constants in
terms of two-point functions as a separate library. The current \mg
implementation fully supports renormalized QCD effects for the
Standard Model, the MSSM, sgluons~\cite{madgolem_sgluons}, 
and other new physics models.  To document our notation
we give all relevant expressions for the renormalization of
supersymmetric QCD here.\bigskip

The renormalization constants we define through the relation between
the bare and the renormalized fields, masses and the coupling
constant:
\begin{alignat}{5}
\Psi^{(0)} \to Z^{1/2}_{\Psi}\,\Psi 
\qqqquad 
m_\Psi^{(0)}  \to m_\Psi + \delta\,m_\Psi
\qqqquad 
g_s^{(0)} \to g_s + \delta\,g_s \; .
\label{eq:defRC}
\end{alignat}
The field $\Psi = q,\sq,g,\go$ denotes all strongly interacting MSSM
fields. We express the SUSY-QCD counter terms to vertices and propagators
in Table~\ref{tab:cts}.\bigskip

%------------------------------------------------
\begin{table}[t]
\begin{center}
\begin{tabular}{cl} 
\hline & \\ & \\
\myrbox{\includegraphics[width=0.15\textwidth]{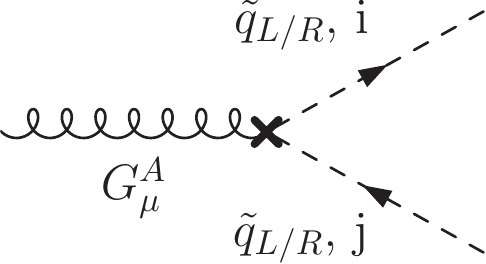}} & 
  $-i\,g_s\,T^A_{ij}\,\left[
    \delta\,g_s + \dfrac{\delta\,Z_{\sq_{L/R,i}}+\delta\,Z_{\sq_{L/R, j}} + \delta\,Z_G}{2} \right]\,
    \sq_{L/R, i}\,(p_i + p_j)^\mu\,G^A_\mu\,\sq_{L/R, j}$  
\\ & \\ 
\myrbox{\includegraphics[width=0.15\textwidth]{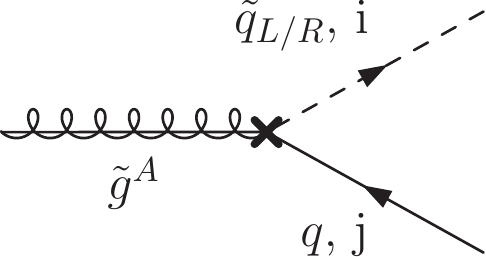}} &  
  $\mp \, i\,g_s\,\sqrt{2}\,T^A_{ij}\,
  \left[\delta\,g_s + \dfrac{\delta\,Z_{\sq_{L/R, i}}+\delta\,Z_{q_{j}} +\delta\,Z_{\go}}{2} 
        + \delta_\text{SUSY}\right]\,
\go^A\,
  P_{L/R}\,q_j\,\sq_{L/R, i}$
 \\ & \\ 
\myrbox{\includegraphics[width=0.15\textwidth]{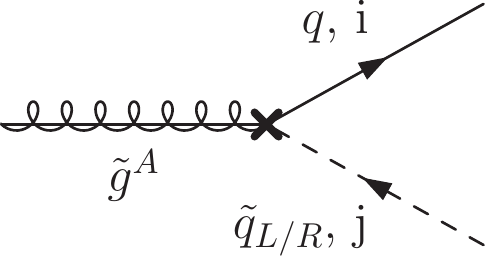}} &  
  $\pm\, i\,g_s\,\sqrt{2}\,T^A_{ij}\,
  \left[\delta\,g_s + \dfrac{\delta\,Z_{\sq_{R/L, j}} + \delta\,Z^{*}_{q_i}+\delta\,Z_{\go}}{2}
      + \delta_\text{SUSY}\right]\,
\bar{q}_i\, 
  P_{L/R}\,\go^A\,\sq_{R/L, j}$
 \\ & \\ 
\myrbox{\includegraphics[width=0.15\textwidth]{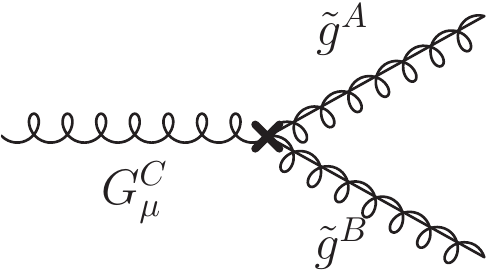}} &  
  $-g_s\,f^{ABC}\,\left[\delta\,g_s + \delta\,Z_{\go} + \dfrac{\delta\,Z_G}{2} \right] 
  \go^A\gamma^\mu\,\go^B\,G^C_{\mu}$
 \\ & \\ 
\myrbox{\includegraphics[width=0.15\textwidth]{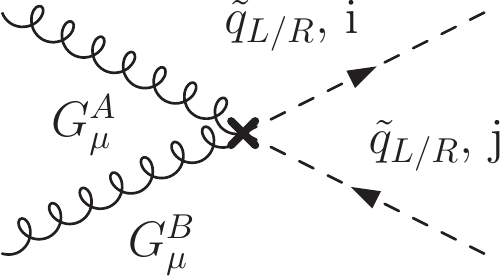}} &
  $i\,g_s\,\{T^A\,T^B\}_{ij}\,
  \left[\delta\,g_s + \delta\,Z_{G} + \dfrac{\delta\,Z_{\sq_{L/R,i}}+\delta\,Z_{\sq_{L/R, j}}}{2} \right] 
  \, \sq_{L/R, i}\,\sq_{L/R, j}\,G_\mu^A\,G^{B\mu}$
  \\ & \\
\myrbox{\includegraphics[width=0.12\textwidth]{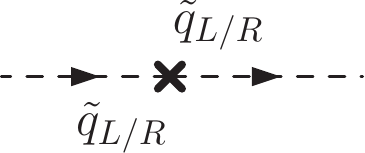}} &
  $p^2\,\delta\,Z_{\sq_{L/R}} 
   - m^2_{\sq_{L/R}}\,\delta\,Z_{\sq_{L/R}}
   - \delta\,m^2_{\sq_{L/R}}\, 
$
  \\ & \\
\myrbox{\includegraphics[width=0.12\textwidth]{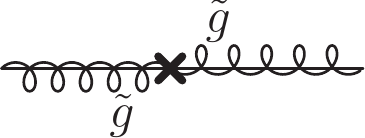}} &
  $\slashed{p}\,\delta\,Z_{\go} - m_{\go}\,\delta\,Z_{\go} - \delta\,m_{\go}$
  \\ & \\
\hline
\end{tabular}
\end{center}
\caption{Strong interaction counter terms for the MSSM.  The finite
  supersymmetry-restoring counter term $\delta_\text{SUSY}$ is given
  in Eq.\eqref{eq:susyrestoration}.}
\label{tab:cts}
\end{table}
%------------------------------------------------

The actual counter terms, presented below, we include in a separate
library. The strong coupling constant we renormalize in the $\msbar$
scheme and explicitly decouple all particles heavier than the bottom
quark. This zero-momentum subtraction
scheme~\cite{squarkgluinoNLO,Berge:2007dz,decoup} leaves us with the
renormalization group running of $\alpha_s$ to light colored particles
only. It corresponds to the measured value of the strong coupling, for
example in a combined fit with the parton densities. Its
renormalization constant reads
\begin{alignat}{5}
\delta\,g_s = 
- \frac{\alpha_s}{4\pi} \;
  \frac{\beta^L_0+\beta^H_0}{2} \, \frac{1}{\tilde{\epsilon}}
- \frac{\alpha_s}{4\pi} \,
\left( \frac{1}{3} \, \log \frac{m^2_t}{\mu_R^2}
      + \log \frac{m^2_{\go}}{\mu_R^2}
      + \frac{1}{12} \, \sum_\text{squarks} \,
        \log \frac{m^2_{\sq_j}}{\mu_R^2}
\right) \; .
\label{eq:alphas_ct}
\end{alignat}
The UV divergence appears as $1/\tilde{\epsilon} \equiv
(4\pi)^{\epsilon}/\Gamma(1-\epsilon) = 1/\epsilon - \gamma_E +
\log(4\pi) + \mathcal{O}(\epsilon)$.  Both light ($L$) and heavy
($H$) colored particles contribute to the beta function
\begin{alignat}{5}
\beta_0 = 
\beta_0^L + \beta_0^H = 
\left[\frac{11}{3}\,C_A - \frac{2}{3}\,n_f \right]
+ 
\left[- \frac{2}{3}  -\frac{2}{3}\,C_A - \frac{1}{3}\,(n_f+1) \right] \; . 
\end{alignat}
\mg sets the number of active flavors to $n_f = 5$.  The $SU_C(3)$
color factors are $C_F = 4/3$ and $C_A = 3$.\bigskip

The analytic form of all renormalization constants we reduce down to
one-point and two-point scalar one-loop functions, which we handle by
means of the standard 't Hooft-Veltman dimensional
regularization scheme in $4-2\epsilon$ dimensions.  The field and mass
renormalization constants we compute from the one-loop self-energies
which involve virtual gluons and gluinos.  All fields are renormalized
on-shell. In addition, for the gluon field we subtract the heavy modes
consistently with our $g_s$ renormalization scheme. The underlying
Slavnov-Taylor identities link the corresponding finite counter terms
as $\delta\,Z_G = -2\,\delta\,g_s$.\bigskip

In addition, we need to pay attention to dimensional regularization
breaking supersymmetry through a mismatch of two gaugino and the
$2-2\epsilon$ gauge vector degrees of freedom~\cite{Martin:1993yx}. As
a result, the Yukawa coupling $\hat{g}_s$ appearing in the $q\sq\go$
vertex departs from $g_s$. We restore supersymmetry by hand, forcing
$\hat{g}_s = g_s$. The corresponding finite counter term can be
computed using dimensional reduction,
\begin{alignat}{5}
\frac{\hat{g}_s}{g_s} = 
\frac{\alpha_s}{4\pi} \; 
\left(\frac{2}{3}n_f - \frac{3}{2}\,C_F \right) 
\qquad \Rightarrow \qquad 
\delta_\text{SUSY} =  \frac{4}{3}\,\frac{\alpha_s}{4\pi} \; .
\label{eq:susyrestoration}
\end{alignat}
\bigskip

Finally, we quote the analytical expressions for the field and mass
renormalization. For the scalar one-point and two-point functions we
adopt the notation of Ref.~\cite{oneloop}. The corrections to the
massless quarks including the non-chiral SUSY contributions are
\begin{alignat}{5}
\delta\,Z_{q_{L/R}} 
= -\frac{\alpha_s}{4\pi}\,C_F\,
    &\left[B_0(0,0,0) + B_0(0,\mgo^2,m^2_{\sq_{L/R}}) \right. \notag \\ 
& \left.
+ (\mgo^2 - m^2_{\sq_{L/R}})\,B_0'(0,\mgo^2,m^2_{\sq_{L/R}})
+ (\mgo^2 - m^2_{\sq_{R/L}})\,B_0'(0,\mgo^2,m^2_{\sq_{R/L}})\right] \; .
\label{eq:quarkf_sqcd}
\end{alignat}
The corresponding squark fields ($\tilde{q}_{s=L/R}$) and mass are renormalized as
\begin{alignat}{5}
\delta\,Z_{\sq_s\sq_s} 
&= \frac{\alpha_s}{2\pi}\,C_F\,
   \left[\,B_0(\msqs^2,0,\msqs^2)
   +\msqs^2\,B^{'}_0(\msqs^2,0,\msqs^2) - B_0(\msqs^2, \mgo^2,0) 
   +(\mgo^2-\msqs^2)\,B^{'}_0(\msqs^2, \mgo^2,0)
   \right] \notag \\
\delta\,m_{\sq_s} 
&= -\frac{\alpha_s}{4\pi}\,C_F\,
    \left[ 4\msqs^2 + 3 A_0(\msqs^2)+ 2\,A_0( \mgo^2) 
         + 2\,(\mgo^2 - \msqs^2)\,B_0(\msqs^2, \mgo^2,0)
    \right] \; .
\label{eq:squarkmass}
\end{alignat}
The gluon wave function renormalization, linked to the counter term
for the strong coupling, is
\begin{alignat}{5}
\delta\,Z_G 
&= \frac{\alpha_s}{4\pi}\,\left(\beta^L_0\, + \beta^H_0\right)\,\frac{1}{\tilde{\epsilon}}
  +\frac{\alpha_s}{2\pi}\,
   \left[  
 \frac{1}{3}\,\log \frac{m^2_t}{\mu^2} +
\log \frac{m^2_{\go}}{\mu^2}
         +\frac{1}{12}\,\sum_\text{squarks} \,\log \frac{m^2_{\sq}}{\mu^2} 
   \right] \; .
\label{eq:gluonfield_ct}
\end{alignat}
Finally, the gluino field and mass renormalization constants are
\begin{alignat}{5}
\delta\,Z_{\go} 
&= \frac{\alpha_s}{4\pi}\,C_A\,
   \left[1 + 4\,m^2_{\go}\,B'_0(m^2_{\go},0,m^2_{\go}) - \frac{A_0(m^2_{\go})}{m^2_{\go}} 
   \right] 
\notag \\
&+ \frac{\alpha_s}{8\pi\mgo^2} \sum_\text{light (s)quarks}\,
   \left[ A_0(\msq^2)-(\mgo^2+\msq^2)\,B_0(\mgo^2,0,\msq^2)
        - 2\,\mgo^2\,(\mgo^2-\msq^2)\,B'_0(\mgo^2,0,\msq^2) \right]
\notag \\
&+ \frac{\alpha_s}{8\pi\,\mgo^2} \sum_\text{heavy (s)quarks}\,
   \left[ 2\mgo^2\,(m^2_{\sq}-m_q^2-\mgo^2)\,B'_0(\mgo^2,m_{q}^2,m^2_{\sq_s}) 
        + (m_q^2-m^2_{\sq}-\mgo^2)\,B_0(\mgo^2,m_q^2,m^2_{\sq}) \right.
\notag \\
&\left. \phantom{haaallllooooooooooooooo} + A_0(m^2_{\sq}) - A_0(m_q^2) \right]
\notag \\
&+\frac{\alpha_s}{\pi}\, \sum_\text{heavy (s)quarks}\,
\mgo\,m_q\,R^{q}_{s1}\,R^{q}_{s2}\,B'_0(\mgo^2,m_{q}^2,m^2_{\sq_s}) 
\notag \\
\delta\,m_{\go} 
&= -\frac{\alpha_s}{4\pi}\,C_A\,\mgo\,\left[1+3\frac{A_0(\mgo^2)}{\mgo^2} \right]
   + \frac{\alpha_s}{8\pi\mgo}\,\sum_\text{light (s)quarks}
    \,\left[ A_0(\msq^2)+(\mgo^2-\msq^2)\,B_0(\mgo^2,0,\msq^2) \right]
\notag \\
&+ \frac{\alpha_s}{8\pi\,\mgo}\,\sum_\text{heavy (s)quarks}\,
   \left[A_0(m^2_{\sq})-A_0(m_q^2)
        -(m^2_{\sq}-m_{q}^2-\mgo^2)\,B_0(\mgo^2,m_q^2,m^2_{\sq}) 
   \right] 
\notag \\ 
&- \frac{\alpha_s}{2\pi}\,\sum_\text{heavy (s)quarks}\,
   m_q\,R^{q}_{s1}\,R^{q}_{s2}\,B_0(\mgo^2,m_{q}^2,m^2_{\sq_s})
\label{eq:gluinoCT}.
\end{alignat}
The sum over heavy squarks covers all squark flavors corresponding to
heavy quarks. We usually consider the bottom quark massless, which
means that only the two stop eigenstates feel top mass
effects. However, the bottom/sbottom loops can be trivially moved from
the light to the heavy category. The stop mass eigenstates
$\tilde{t}_{1,2}$ are related to the electroweak interaction bases
through a rotation with $R = \pm 1$.

%%%%%%%%%%%%%%%%%%%%%%%%%%%%%%%%%%%%%%%%%%%%%%%%%%%%%%%%%%%%%%%%%%%%%%%%

%%%%%%%%%%%%%%%%%%%%%%%%%%%%%%%%%%%%%%%%%%%%%%%%%%%%%%%%%%%%%%%%%%%%%%%%

\end{document}